\documentclass[fleqn,usenatbib]{mnras}

\usepackage{newtxtext,newtxmath}

\usepackage[T1]{fontenc}

\DeclareRobustCommand{\VAN}[3]{#2}
\let\VANthebibliography\thebibliography
\def\thebibliography{\DeclareRobustCommand{\VAN}[3]{##3}\VANthebibliography}

\usepackage{graphicx}	
\usepackage{amsmath}	
\usepackage{xspace}
\usepackage{comment}
\usepackage{gensymb}

\usepackage{physics}
\usepackage{bm}

\usepackage{siunitx}

\usepackage{threeparttable} 
\newlength{\abovecaptionskip}%
\newcommand{\cm}{\text{cm}}
\newcommand{\erg}{\text{erg}}
\newcommand{\Msun}{\text{M}_\odot}
\newcommand{\nH}{n_\text{H}}
\newcommand{\K}{\text{K}}
\newcommand{\pc}{\text{pc}}
\newcommand{\kpc}{\text{kpc}}
\newcommand{\Mpc}{\text{Mpc}}
\newcommand{\Gpc}{\text{Gpc}}
\newcommand{\yr}{\text{yr}}
\newcommand{\eV}{\text{eV}}

\newcommand{\kms}{\text{km}\,\text{s}^{-1}}

\newcommand{\vsn}{\Delta v_\text{SN}}
\newcommand{\tagn}{\Delta T_\text{AGN}}
\newcommand{\betabh}{\beta_\text{BH}}
\newcommand{\vjet}{v_\text{jet}}

\newcommand{\bahamas}{BAHAMAS}
\newcommand{\flamingo}{FLAMINGO}
\newcommand{\swift}{\textsc{Swift}\xspace}
\newcommand{\sphenix}{\textsc{sphenix}\xspace}
\newcommand{\healpix}{\textsc{HEALPix}\xspace}
\newcommand{\namaster}{\textsc{NaMaster}\xspace}

\newcommand{\FastDF}{\textsc{fastdf}\xspace}
\newcommand{\panphasia}{\textsc{panphasia}\xspace}
\newcommand{\monofonIC}{\textsc{monofonIC}\xspace}
\newcommand{\vr}{\textsc{VELOCIraptor}\xspace}

\title[The FLAMINGO project]{The FLAMINGO project: cosmological hydrodynamical simulations for
large-scale structure and galaxy cluster surveys}

\author[J. Schaye et al.]{
Joop Schaye,$^{1}$\thanks{E-mail: schaye@strw.leidenuniv.nl}
Roi Kugel,$^{1}$
Matthieu Schaller,$^{2,1}$
John C. Helly,$^{3}$
Joey Braspenning,$^{1}$
Willem Elbers,$^{3}$\newauthor
Ian G. McCarthy,$^{4}$
Marcel P. van Daalen,$^{1}$
Bert Vandenbroucke,$^{1}$
Carlos S. Frenk,$^{3}$
Juliana Kwan,$^{4}$\newauthor
Jaime Salcido,$^{4}$
Yannick M. Bah\'e,$^{1,5}$
Josh Borrow,$^{3,6}$
Evgenii Chaikin,$^{1}$
Oliver Hahn,$^{7,8}$
Filip Hu\v{s}ko,$^{3}$\newauthor
Adrian Jenkins,$^{3}$
Cedric G. Lacey$^{3}$
and Folkert S. J. Nobels$^{1}$
\\
$^{1}$Leiden Observatory, Leiden University, PO Box 9513, 2300 RA Leiden, the Netherlands\\
$^{2}$Lorentz Institute for Theoretical Physics, Leiden University, PO box 9506, 2300 RA Leiden, the Netherlands\\
$^{3}$Institute for Computational Cosmology, Department of Physics, University of Durham, South Road, Durham, DH1 3LE, UK\\
$^{4}$Astrophysics Research Institute, Liverpool John Moores University, Liverpool L3 5RF, UK\\
$^{5}$Institute of Physics, Laboratory of Astrophysics, Ecole Polytechnique F\'ed\'erale de Lausanne (EPFL), Obervatoire de Sauverny, 1290 Versoix, Switzerland\\
$^{6}$Department of Physics, Kavli Institute for Astrophysics and Space Research, Massachusetts Institute of Technology, Cambridge, MA 02139, USA\\
$^{7}$Department of Astrophysics, University of Vienna, T\"urkenschanzstrasse 17, 1180 Vienna, Austria\\
$^{8}$Department of Mathematics, University of Vienna, Oskar-Morgenstern-Platz 1, 1090 Vienna, Austria
}

\date{Accepted XXX. Received YYY; in original form ZZZ}

\pubyear{2022}

\begin{document}
\label{firstpage}
\pagerange{\pageref{firstpage}--\pageref{lastpage}}
\maketitle

\begin{abstract}
We introduce the Virgo Consortium's \flamingo\ suite of hydrodynamical simulations for cosmology and galaxy cluster physics. To ensure the simulations are sufficiently realistic for studies of large-scale structure, the subgrid prescriptions for stellar and AGN feedback are calibrated to the observed low-redshift galaxy stellar mass function and cluster gas fractions. The calibration is performed using machine learning, separately for each of \flamingo's three resolutions. This approach enables specification of the model by the observables to which they are calibrated. The calibration accounts for a number of potential observational biases and for random errors in the observed stellar masses. The two most demanding simulations have box sizes of 1.0 and 2.8~Gpc on a side and baryonic particle masses of $1\times10^8$ and $1\times10^9\,\Msun$, respectively. For the latter resolution the suite includes 12 model variations in a 1~Gpc box. There are 8 variations at fixed cosmology, including shifts in the stellar mass function and/or the cluster gas fractions to which we calibrate, and two alternative implementations of AGN feedback (thermal or jets). The remaining 4 variations use the unmodified calibration data but different cosmologies, including different neutrino masses. The 2.8~Gpc simulation follows $3\times10^{11}$ particles, making it the largest ever hydrodynamical simulation run to $z=0$.  Lightcone output is produced on-the-fly for up to 8 different observers. We investigate numerical convergence, show that the simulations reproduce the calibration data, and compare with a number of galaxy, cluster, and large-scale structure observations, finding very good agreement with the data for converged predictions. Finally, by comparing hydrodynamical and `dark-matter-only' simulations, we confirm that baryonic effects can suppress the halo mass function and the matter power spectrum by up to $\approx20$ per cent.
\end{abstract}




\section{Introduction}
The standard model of cosmology allows us to compute the evolution of the universe and the growth of structure in it starting shortly after inflation. Among its ingredients are the physics of the standard model of particle physics, whose constituents include baryonic matter, photons and neutrinos, and a distribution of initial density perturbations. However, to fit the cosmological data two additional, ad-hoc contributions to the energy content are required, namely non-baryonic dark matter and dark energy \citep[for a review see e.g.][]{Workman2022}. 

In the simplest version of the model, referred to as LCDM (or $\Lambda$CDM), the universe is spatially flat; the dark matter is assumed to be non-interacting (apart from gravitational interactions) and cold (i.e.\ the free streaming scale is negligible for practical purposes); the dark energy is a cosmological constant (i.e.\ the dark energy has an equation of state $P= w \rho c^2$ with $w=-1$); there are three neutrino flavours and the sum of the neutrino masses is equal to the minimum allowed by ground-based neutrino oscillation experiments (0.06~eV; e.g.\ \citealt{Esteban2020}); there are no primordial tensor metric fluctuations; and the primordial density perturbations are adiabatic, Gaussian and their power spectrum is described by a power law. This model has only 7 free parameters. One of these, the current temperature of the cosmic microwave background (CMB) radiation, is measured very accurately from its spectrum \citep{Fixsen2009}. This leaves only 6 parameters that need to be measured more precisely, which include the amounts of cold dark matter (CDM) and baryons, the normalisation and slope of the primordial power spectrum of density perturbations, the optical depth to photon scattering due to reionization (which is determined by galaxy formation physics) and a final parameter which must depend on the normalisation of the expansion history and can therefore be thought of as the Hubble constant. 

The standard model of cosmology can reproduce an impressively diverse set of observations that span a wide range of length and time-scales \citep[for a review see e.g.][]{lahav2022}. Examples are the abundances of the light elements; anisotropies in the CMB; standard rods like the baryon acoustic oscillations (BAO) in the distributions of galaxies and intergalactic hydrogen; geometric probes such as the Alcock-Paczynski effect; the distance-redshift relation of standard candles like supernovae of type Ia; the age-redshift relation of cosmic chronometers; and last but not least the growth of structure as measured e.g.\ through the redshift evolution of the abundance of galaxy clusters, galaxy clustering and associated redshift distortions, cosmic shear and CMB lensing, the Ly$\alpha$ forest, and the Sunyaev-Zel'dovich effect (SZE). 

Although the overall agreement of the model with observations is impressive, there is some tension between different data sets (see \citealt{Abdalla2022} for a recent review of the problems and proposed solutions). The Hubble constant measured from the local distance ladder, particularly the value of $H_0 = 73 \pm 1~\kms\,\Mpc^{-1}$ measured by \citet{Riess2022} from supernovae Ia calibrated using Cepheid stars (which in turn are calibrated using Gaia parallaxes), is significantly greater than the $H_0 = 67.4 \pm 0.5~\kms\,\Mpc^{-1}$ inferred from the CMB by \citet{Planck2020cosmopars} assuming the LCDM model. Another area of tension concerns the clumpiness of the matter distribution. For example, CMB anisotropies imply $S_8 \equiv \sigma_8(\Omega_\text{m})^{0.5} = 0.83 \pm 0.01$ \citep{Planck2020cosmopars}, whereas cosmic shear measurements and a variety of other low-redshift probes of large-scale structure (LSS) find $S_8 = 0.77 \pm 0.02$ \citep[e.g.][]{Heymans2021, Abbott2022}. At present it is unclear whether these tensions require an extension of the base model or whether they are due to underestimated or unidentified systematic errors.

A diverse set of surveys is about to map the LSS down to smaller scales than was possible before, which will greatly reduce the statistical uncertainties on the cosmological parameters. However, their scientific potential can only be realized if the model predictions are at least as accurate and precise as the observations. While predictions for the CMB are thought to be sufficiently robust, the same is not true for the growth of structure and its observational manifestations on scales $\lesssim 10$~Mpc. On these scales baryonic matter cannot  be assumed to trace the CDM. On scales of $1 \lesssim \lambda \lesssim 10$~Mpc the baryons are predicted to be distributed more smoothly than the CDM, mainly due to their redistribution by galactic winds driven by feedback from star formation and particularly active galactic nuclei (AGN), while on smaller scales the baryons are predicted to cluster more strongly than the CDM due to their ability to radiate away their binding energy, which allows them to condense into galaxies \citep[e.g.][]{VanDaalen2011}. This prediction from hydrodynamical simulations is confirmed by halo models and enhanced dark matter only (DMO) simulations that use the observed hot gas and stellar content of galaxies and clusters as input \citep[e.g.][]{Schneider2019,Debackere2020,Giri2021}. The emerging picture is that the low baryon fractions of groups and low-mass clusters are closely related to the baryonic suppression of the matter power spectrum on relatively large ($1 \lesssim \lambda \lesssim 10$~Mpc) scales \citep[e.g.][]{Semboloni2011,Semboloni2013,VanDaalen2020, Arico2021,Salcido2023}. If these baryonic effects are not accounted for, then they will, for example, result in catastrophic systematic errors on upcoming cosmic shear surveys \citep[e.g.][]{Semboloni2011,Eifler2015,Huang2019,Lu2021,Martinelli2021}. 

Hydrodynamical simulations \citep[for a recent review see][]{Vogelsberger2020} offer a number of key advantages over DMO-based techniques. First, they can provide much more detail, e.g.\ galaxy colour gradients, galaxy shapes, and intrinsic alignments between galaxies, which may all bias cosmic shear results, and spatially-resolved X-ray and SZE (virtual) observations of clusters. Second, they self-consistently model the gravitational back-reaction onto the dark matter due to the redistribution of baryons. Third, they self-consistently model the relations between different physical processes and galaxy properties, e.g.\ the fact that the dynamical friction experienced by a satellite of a given total mass will depend on its stellar mass, or that the star formation activity of a central galaxy may correlate with the distribution of gas around it. Fourth, they predict not only the properties of the galaxies, but also the 3-D distribution, kinematics, temperature, and chemical composition of the gas. This enables direct comparisons with more types of data, such as diffuse X-ray emission, the SZE, dispersion measures, as well as their cross correlations with galaxy clustering, cosmic shear, and CMB lensing. Even if they cannot yet replace DMO simulations or DMO-based semi-analytic or semi-empirical models in parameter inference, hydrodynamical simulations are needed to validate those methods' assumptions and to calibrate the corrections for baryonic effects that they apply. 

The necessity for cosmological simulations to model baryonic effects blurs the line between the fields of LSS and galaxy formation and demands a new approach. This development constitutes a challenge for the modellers, but it also represents an opportunity, as probes of LSS can also be used to advance our understanding of the formation and evolution of galaxies and clusters of galaxies.

Predictions for observables that are directly sensitive to the distribution of baryons are crucial, because the ab initio predictive power of the simulations is limited. Without additional observational constraints, it is, for example, impossible to predict the effect of outflows driven by AGN feedback with the accuracy needed for upcoming experiments. 

One approach is to brute force the problem by running a very large number of hydrodynamical simulations that vary all the relevant subgrid parameters and then look for predictions that depend on cosmology but are insensitive to the uncertain subgrid physics, or vice versa \citep[e.g.][]{Villaescusa2021,Ni2023}. However, computational expense then dictates the use of volumes that are too small for most of the commonly used probes of cosmology. Another approach is to calibrate the subgrid physics explicitly, which requires a choice of calibration target. For example, the EAGLE simulations of galaxy formation \citep{Schaye2015,Crain2015} were calibrated to the local galaxy stellar mass function (SMF) as well as the relations between galaxy mass and size, and between galaxy and supermassive black hole (BH) mass, while the Illustris-TNG simulations \citep{Pillepich2018TNGmethod} included additional constraints such as the cosmic star formation history, the intragroup medium, the mass-metallicity relation and galaxy quenching. Because the TNG model was calibrated for the resolution of the TNG100 simulation, the lower-resolution $(205~\Mpc/h)^3$ TNG300 \citep{Springel2018} and $(500~\Mpc/h)^3$ MillenniumTNG \citep[MTNG;][]{Pakmor2022} simulations do not fit the calibration data as well, but are more useful for cosmology due to their larger volumes. 

For observational cosmology the most relevant calibration targets are arguably the SMF, because it constrains the galaxy-halo connection, which is, for example, critical for galaxy clustering, and the gas and baryon fractions of groups and clusters, because those correlate with the degree to which feedback processes suppress the matter power spectrum on large scales. The $(400~\Mpc/h)^3$ \bahamas\ simulations \citep{McCarthy2017,McCarthy2018} were calibrated to match these two observables, both at $z\approx 0$. Because the SMF and gas fractions are not precisely known, \bahamas\ included a strong and a weak AGN feedback model, which skirted the observational error bars on the cluster gas fractions. The ANTILLES suite \citep{Salcido2023} uses a similar approach but includes a much larger number of models spanning a wider range of subgrid parameter values, though in a much smaller volume of $(100~\Mpc/h)^3$.

Other suites of hydrodynamical simulations with volumes $\gg (10^2\,\Mpc)^3$ that were run to $z=0$, such as the cosmo-OWLS \citep{LeBrun2014} and Magneticum \citep{Dolag2016} projects, did not have an explicit calibration strategy. For cluster physics zooms of haloes selected from very large volume but low-resolution (DMO) simulations are often used, both as stand alone samples \citep[e.g.][]{Hahn2017,Cui2018,Cui2022,Tremmel2019,Henden2020,Pellissier2023} and to complement large-volume hydrodynamical runs by extending their mass range, e.g.\ the MACSIS sample \citep{Barnes2017MACSIS} for \bahamas\ and the Hydrangea \citep{Bahe2017Hydrangea} and C-EAGLE \citep{Barnes2017C-EAGLE} samples for EAGLE. 

\begin{figure*}
    \centering
    \includegraphics[width=\textwidth]{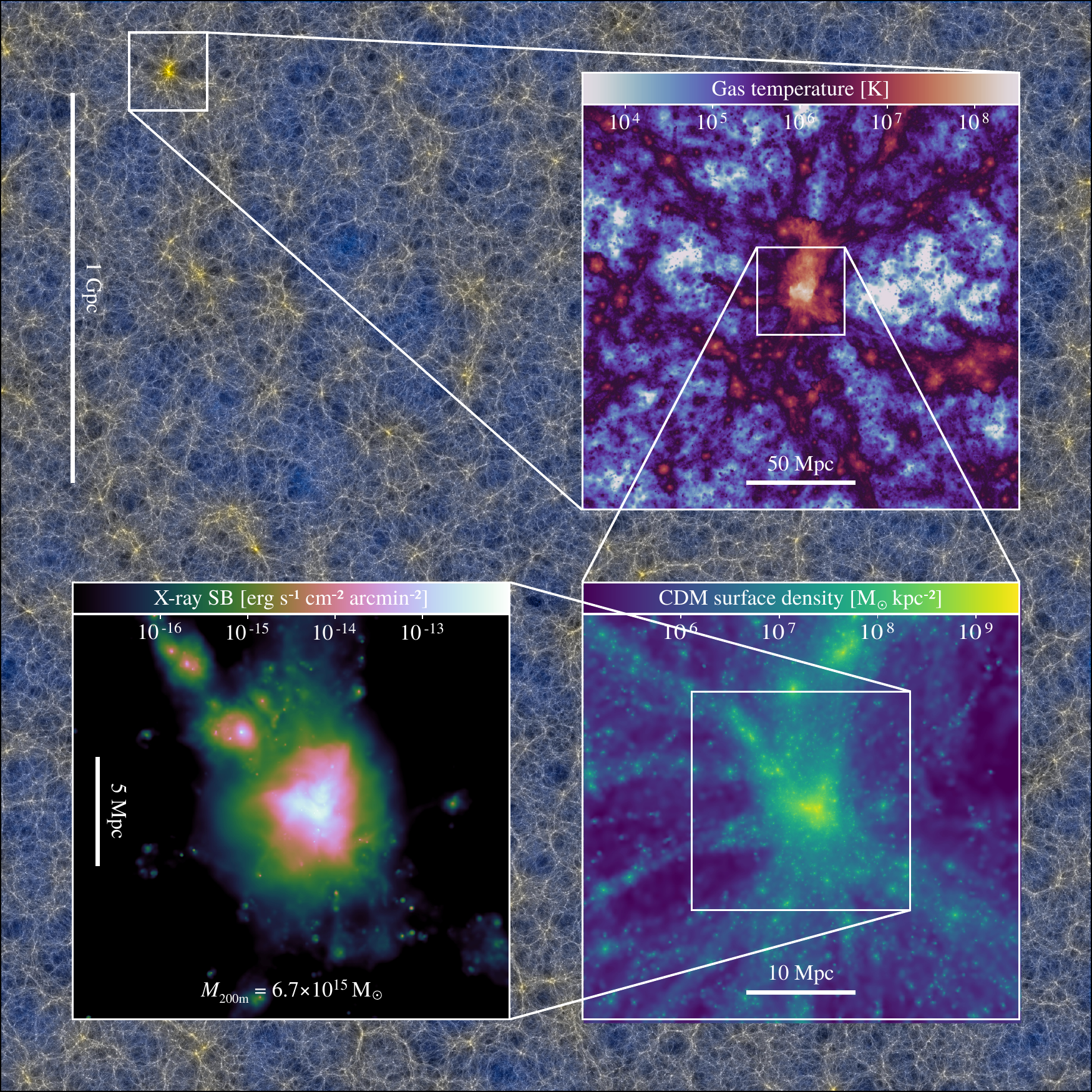}
    \caption{A projection through a 40~Mpc thick slice through the fiducial, intermediate-resolution simulation with box side length 2.8~Gpc (run L2p8\_m9 in Table~\ref{tab:simulations}) at $z=0$. The luminosity of the background image gives the CDM surface density whilst the colour encodes the surface density of massive neutrinos, both on a logarithmic scale (see Fig.~\ref{fig:cdm_nu_slice} for a side-by-side comparison of images of the CDM and neutrino surface densities with color bars for each). The insets show three consecutive zooms centred on the most massive halo (total mass $M_\text{200m} = 6.7\times 10^{15}\,\Msun$). \emph{First inset}: a projection of a $200\times 200 \times 40~\Mpc^3$ sub-volume containing the cluster, showing the mass-weighted gas temperature along the line of sight. \emph{Second inset}: CDM surface density in a $40\times 40\times 20~\Mpc^3$ region. \emph{Final inset}: X-ray surface brightness in the 0.5-2~keV band in a $20\times 20\times 20~\Mpc^3$ region, computed from the $z=0$ snapshot but placing the cluster at $z=0.025$.}    \label{fig:zoom-in}
\end{figure*}

\begin{figure}
    \centering
	  \includegraphics[width=\columnwidth]{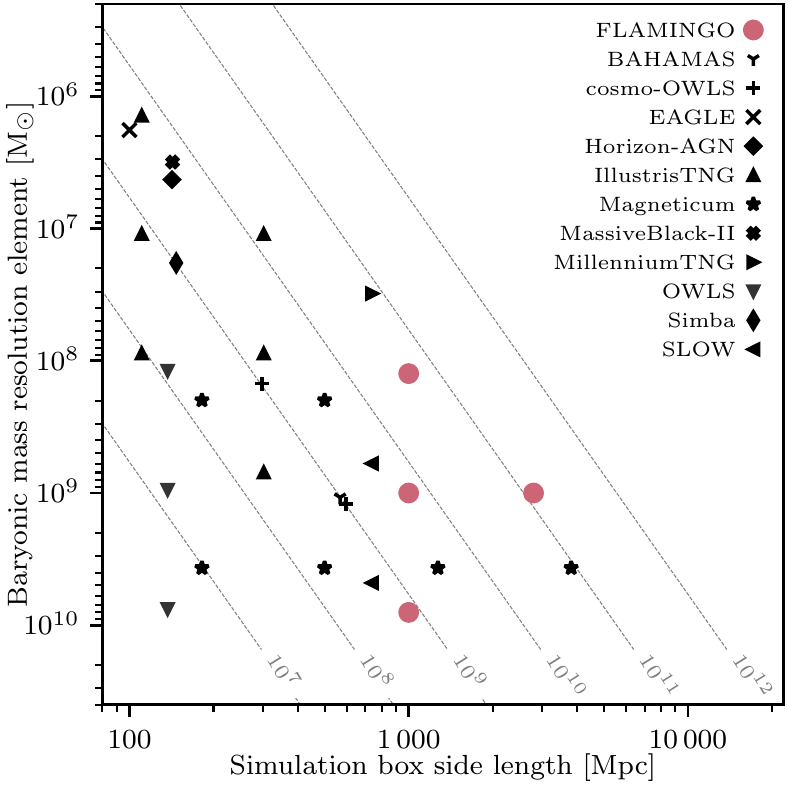}
     \vspace{-0.5cm}
    \caption{Comparison of the resolutions (baryonic particle mass or target cell mass; resolution increases along the y-axis) and box sizes of the \flamingo\ runs (filled red circles) with cosmological, hydrodynamical simulations from the literature that include radiative cooling, use a box size of at least $(100~\Mpc)^3$, and were run down to $z=0$. The grey diagonal lines indicate the total number of baryonic resolution elements. The simulations from the literature shown are \bahamas\ \citep{McCarthy2017}, cosmo-OWLS \citep{LeBrun2014}, EAGLE \citep{Schaye2015}, Horizon-AGN \citep{Dubois2014}, IllustrisTNG \citep{Springel2018}, Magneticum \citep{Dolag2016}, MassiveBlack-II \citep{Khandai2015}, MillenniumTNG \citep{Pakmor2022}, OWLS \citep{Schaye2010}, SIMBA \citep{Dave2019}, and SLOW \citep{Dolag2023}.}
    \label{fig:simulation_sizes}
\end{figure}

Here we present the \flamingo\footnote{\flamingo\ is a project of the Virgo consortium for cosmological supercomputer simulations. The acronym stands for \textbf{F}ull-hydro \textbf{L}arge-scale structure simulations with \textbf{A}ll-sky \textbf{M}apping for the \textbf{I}nterpretation of \textbf{N}ext \textbf{G}eneration \textbf{O}bservations.} project, which improves on \bahamas\ and other large-volume hydrodynamical simulations in many respects. Below we list some of \flamingo's key features. 

First, \flamingo\ uses three different resolutions that are all calibrated to the same data. The two flagship runs use volumes of $(2.8~\Gpc)^3$ and $(1~\Gpc)^3$ and baryonic particle masses of $1\times 10^9\,\Msun$ (which we will refer to as intermediate/m9 resolution) and $1\times 10^8\,\Msun$ (high/m8 resolution), respectively. While the former has the same resolution as \bahamas, its volume is more than two orders of magnitude larger. To highlight the dynamic range captured by this simulation, Fig.~\ref{fig:zoom-in} zooms in from the full simulation box with the large-scale structure of the cosmic web to a single massive galaxy cluster and its internal structure. The color scale of the background image encodes the density of neutrinos, while the intensity encodes the CDM density, which can clearly be seen to be modulated on smaller scales than the neutrino density. The consecutive zooms in the three insets show, respectively, the gas temperature, CDM surface density and X-ray surface brightness.  

Second, the simulations use very large numbers of particles, up to $3\times 10^{11}$, which is the largest number of resolution elements for any existing cosmological hydrodynamical simulation run to $z=0$. See Fig.~\ref{fig:simulation_sizes} for a comparison with simulations from the literature. 

Third, the calibration of the subgrid physics is not done by trial and error, but systematically using machine learning (Gaussian process emulators), also accounting for observational errors and biasing \citep{FlamingoCal}. 

Fourth, massive neutrinos are modelled using particles with a new, self-consistent and efficient implementation, the `$\delta f$'method \citep{Elbers2021}, that was designed to reduce shot noise. 

Fifth, besides the fiducial simulations, \flamingo\ includes eight astrophysics variations, all in $(1~\Gpc)^3$ volumes with intermediate resolution. These models are each calibrated using our emulators by shifting the calibration data according to the allowed error bars. Model variations are then no longer expressed only as variations of particular subgrid parameters that have no direct connection with observables, like the subgrid AGN heating temperature that was varied in the cosmo-OWLS and \bahamas\ projects. Instead, the variations can be specified in terms of the data that they are calibrated to. For example, we include runs where the observed stellar masses or cluster gas fractions have been shifted by different multiples of the expected systematic errors. Multiple subgrid parameters are then adjusted to accomplish the new fits. However, having models that span the uncertainties in the observables used for calibration may not be sufficient to quantify the uncertainty in the predictions for observations that were not considered in the calibration. A different model, in particular a different implementation of AGN feedback, may result in different predictions even when the model is calibrated to the same data. \flamingo\ therefore includes two simulations that use jet-like AGN feedback instead of the fiducial thermal AGN feedback, but that are calibrated to the same data. 

Sixth, besides the fiducial cosmology, which assumes cosmological parameter values from the Dark Energy Survey year three (3x2pt plus external constraints) for a spatially flat universe with $\sum m_\nu c^2 = 0.06~\eV$; \citep{Abbott2022}, \flamingo\ includes four (spatially flat) cosmology variations that each use the fiducial calibration data and were run in $(1~\Gpc)^3$ volumes with intermediate resolution. The variations are the \citet{Planck2020cosmopars} cosmology, one with $\sum m_\nu c^2 = 0.06~\eV$ and two with $\sum m_\nu c^2 = 0.24~\eV$, and a model with a lower amplitude of the power spectrum, as preferred by many LSS surveys \citep{Amom2023}.

Seventh, we produce on-the-fly full-sky lightcone output for up to eight different observers, including \healpix\ maps for gravitational lensing, X-ray emission, the thermal and kinematic SZE, and dispersion measures. 

Eigth, we use a new hydrodynamics code (\swift; \citealt{SWIFT_release}) with improved subgrid models. 

Nineth, we use 3-fluid initial conditions with separate transfer functions for CDM, baryons and neutrinos, perturbing particle masses rather than positions to suppress discreteness noise \citep{Hahn2020,Hahn2021,Elbers2022a}.

This paper serves to introduce the \flamingo\ project, document the simulation methods, describe the simulation suite and the calibration strategy, provide a comparison with some key observables, and to report some first results on the effects of baryonic physics on structure formation. 

The paper is organized as follows. In Section~\ref{sec:simulation_methods} we discuss the simulation methods. This section is meant as a reference and can be skipped by readers not interested in the technical details. The calibration of the subgrid galaxy formation physics as well as the observational bias factors is detailed in \citet{FlamingoCal} and summarized in Section~\ref{sec:calibration}. Section~\ref{sec:simulations} summarizes the numerical, cosmological and subgrid parameters for the simulation runs and presents some visuals. We compare with the calibration data in Section~\ref{sec:comparison_calibration_data} and with a selection of other observables in Section~\ref{sec:comparison_other_data}, including the cosmic star formation history (\S\ref{sec:sfh}), galaxy properties (\S\ref{sec:galaxy_properties}), cluster scaling relations (\S\ref{sec:cluster_properties}), and the cross-correlation of the thermal SZE and CMB lensing signals (\S\ref{sec:lss}). We investigate baryonic effects on the halo mass function and the matter power spectrum in Sections~\ref{sec:hmf} and \ref{sec:power_spectra}, respectively, and we conclude in Section~\ref{sec:conclusions}. A summary of the on-the-fly generation of lightcone data as well as the associated data products (particle data and \healpix\ maps) can be found in Appendix~\ref{sec:lightcones}. 

\section{Simulation methods}\label{sec:simulation_methods}
This section discusses the gravity and hydrodynamics solver (\S\ref{sec:swift}), the implementation of neutrinos (\S\ref{sec:neutrinos}), the subgrid models for unresolved processes (\S\ref{sec:subgrid}), the initial conditions (\S\ref{sec:ICs}), and the (sub)halo finding (\S\ref{sec:VR}).

\subsection{The gravity and hydrodynamics solver \swift}\label{sec:swift}
The simulations were performed using \swift \citep{SWIFT_release}, a fully open-source coupled cosmology, gravity, hydrodynamics, and galaxy formation code\footnote{Publicly available, including the version used for these simulations, at \url{www.swiftsim.com}}. \swift uses task-based parallelism within compute nodes and interacts between compute nodes via \texttt{MPI} using non-blocking communications resulting in excellent scaling up to $>10^5$ compute cores \citep{Schaller2016, SWIFT_release}. The short- and long-range gravitational forces are computed using a 4$^{\mathrm{th}}$-order fast multipole method \citep{Greengard1987, Cheng1999, Dehnen2014} and a particle-mesh method solved in Fourier space, respectively, following the force splitting approach of \cite{Bagla2003}. The accuracy of the gravity solver is mainly controlled by an adaptive acceptance criterion for the fast multipole method similar to the one proposed by \cite{Dehnen2014}.

The equations of hydrodynamics are solved using the smoothed-particle hydrodynamics (SPH) method \citep[for a review, see][]{Price2012}. In particular, the \flamingo\ simulations use the SPHENIX flavour of SPH \citep{Borrow2022sphenix} that was specifically designed for galaxy formation simulations. The particle smoothing is done using a \cite{Wendland1995} C2 kernel with the resolution parameter $\eta=1.2348$\footnote{This corresponds to 58 weighted neighbours or, equivalently, to 48 neighbours with a cubic spline kernel.}. The SPHENIX scheme uses a density-energy formulation of the equations of motion combined with artificial viscosity and conduction terms. Viscosity and conduction limiters are included in the solver to prevent spurious energy losses across feedback-generated shocks or radiative cooling events. \swift also uses the mechanism limiting the time step of inactive neighbour particles of \cite{Durier2012} to properly evolve the fluid even in the most extreme shocks. 

Time integration is performed using a standard leapfrog scheme with the individual time steps of the particles set by the minimum of the gravity time step ($\Delta t = \left(0.025 \epsilon/a\right)^{1/2}$), where $\epsilon$ is the gravitational softening length and $a$ the acceleration, and the Courant–Friedrichs–Lewy (CFL) condition for hydrodynamics with parameter value $0.2$. All particles use the same fixed, but time evolving, gravitational softening length (Table \ref{tab:simulations}) and we impose a floor on the gas particle smoothing length at $0.01$ of the gravitational softening length.

\subsection{Treatment of neutrinos}\label{sec:neutrinos}

The baseline cosmology for the \flamingo{} simulations includes a single massive neutrino species with a mass of \SI{0.06}{\eV} and two massless species, representing the minimum neutrino mass scenario under the normal ordering \citep{Esteban2020,Salas2021}. The \flamingo{} suite also includes variations with larger neutrino masses (see \S\ref{sec:simulations} and Table~\ref{tab:cosmologies}). A key requirement for our implementation of massive neutrinos is therefore that small and large neutrino masses should both be treated accurately. To accomplish this task without exceeding the memory and time constraints imposed by the size and scope of \flamingo{}, we make use of the recently proposed $\delta f$ method \citep{Elbers2021}.

Massless neutrinos are included in the calculation of the Hubble expansion rate and in the initial conditions, but are otherwise treated as a smooth component. Massive neutrinos are included at both the background and perturbation levels using the $\delta f$ method. This method uses particles to capture the full non-linear evolution of the neutrino phase-space distribution, but statistically weights the particle contributions by comparing the known phase-space density, which is manifestly conserved by the symplectic leapfrog integration scheme of \swift, with the phase-space density expected at the background level. This minimizes the level of shot noise in the neutrino density field and thus significantly reduces the required number of simulation particles. The suppression of shot noise is particularly strong at early times, eliminating the spurious back-reaction on the CDM and baryon components that otherwise results. 

As the centre of expansion for the multipoles in \swift is the centre of mass of each tree node and not the geometric centre of the node (see \citealt{SWIFT_release} for details), we cannot allow for negatively weighted particles to dominate as this would push the centre of expansion outside of the cell. This would in turn lead to large errors in the calculation of the gravity forces. We therefore treat neutrinos as ordinary massive particles in short-range interactions and only apply the $\delta f$ weighting scheme in the mesh-based long-range gravity calculation. This choice nevertheless ensures that the back-reaction on large scales is eliminated and that non-linear neutrino interactions are not neglected. The weights are also used in the power spectrum calculation and preserved for post processing, since they reveal additional information about the phase-space distribution and make it possible to probe neutrino clustering on smaller scales \citep{Elbers2021,Adamek2022}.

Massive neutrinos are relativistic at early times. We account for this by using relativistic velocities for the neutrino particles. Relativistic corrections to the acceleration are negligible in the time frame of the simulations ($z\leq31$) and are not included to preserve symplecticity of the leapfrog integration scheme \citep{Elbers2022c}. Further modifications are needed to account for the presence of neutrinos in the initial conditions, as discussed in Section \ref{sec:ICs}.

\subsection{Subgrid prescriptions} \label{sec:subgrid}
Like any hydrodynamical simulation, \flamingo\ relies on subgrid prescriptions to model unresolved physical processes. These prescriptions build on those developed for the OWLS \citep{Schaye2010} project, which were also used for cosmo-OWLS \citep{LeBrun2014} and \bahamas\ \citep{McCarthy2017}, and some of which were developed further for the EAGLE project \citep{Schaye2015}. For \flamingo\ the models were ported from the code \textsc{gadget} \citep{Springel2005Gadget} used for these previous projects to the \swift\ code used for \flamingo. 

Below we will summarise the subgrid models used for radiative cooling and heating (\S\ref{sec:cooling}), star formation and the interstellar medium (ISM) (\S\ref{sec:SF}), stellar mass loss (\S\ref{sec:mass_loss}), stellar energy feedback (\S\ref{sec:SF_feedback}), BHs (\S\ref{sec:BHs}), and AGN feedback (\S\ref{sec:AGN}). We will mention significant differences with respect to \bahamas.

\subsubsection{Radiative cooling and heating}
\label{sec:cooling}
Radiative cooling and heating rates are implemented element-by-element and are taken from \citet[their fiducial model UVB\_dust1\_CR1\_G1\_shield1]{Ploeckinger2020}. They used the spectral synthesis and radiative transfer code \textsc{cloudy} \citep[version 17.01]{Ferland2017} to tabulate the rates as a function of density, temperature, chemical composition, and redshift. The gas is assumed to be in ionisation equilibrium, and exposed to the CMB, the evolving meta-galactic UV/X-ray background radiation given by a modified version of the \citet{Faucher2020} model, and, at high densities, also to a diffuse interstellar radiation field. The gas and dust column densities used to account for self-shielding and to compute the intensity of the interstellar radiation field scale with the density and temperature like the local Jeans column density. We do not allow cooling below a temperature of 100~K.

\begin{figure}
	\includegraphics[width=\columnwidth]{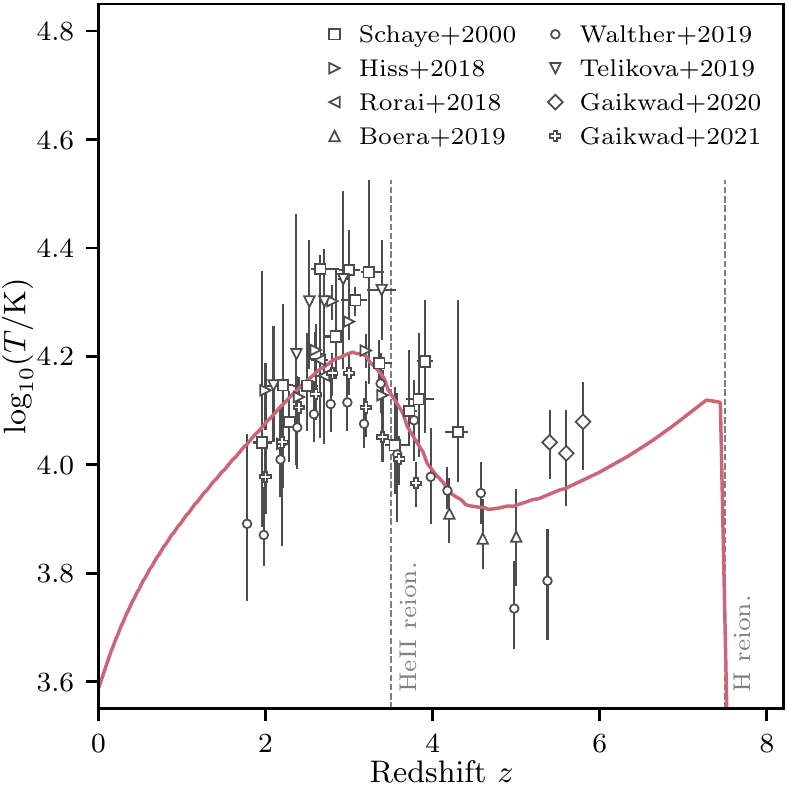}
    \caption{The temperature of gas at the cosmic mean density as a function of redshift. The peaks at $z\approx 7$ and 3 are due to H and He reionization, respectively. The thermal evolution is in good agreement with observations of the Ly$\alpha$ forest. Data points are based on measurements of absorption line widths as a function of strength \citep{Schaye2000,Rorai2018,Hiss2018,Telikova2019,Gaikwad2020}, on the small-scale cut-off in the flux power spectrum \citep{Boera2019,Walther2019}, or on both types of methods \citep{Gaikwad2021}.}\label{fig:thermal_evolution}
\end{figure}

At $z> 3$ ($z>7.2$) \citet{Ploeckinger2020} attenuate the \citet{Faucher2020} spectrum above the \ion{H}{i} (\ion{He}{ii}) ionisation energies using \ion{H}{i} (\ion{He}{ii}) column densities tuned to match the effective photo-ionisation and photo-heating rates that \citet{Faucher2020} finds to be needed to match observations of the optical depth seen by the CMB and observations of the intergalactic medium. In the models used to compute the cooling rates, hydrogen and helium reionize at $z=7.8$ and $z=3.5$, respectively. To account for the extra heat due to spectral hardening and non-equilibrium effects, we inject an extra 2~eV per hydrogen atom at H reionization. We also inject 2~eV per hydrogen atom at $z=3.5$, spread over a Gaussian redshift interval with $\sigma(z)=0.5$, to account for the later reionization of \ion{He}{ii}. Fig.~\ref{fig:thermal_evolution} shows that this yields a thermal evolution of the IGM that agrees with observations. 

Compared with the rates of \citet{Wiersma2009Cooling} used by \bahamas, the \citet{Ploeckinger2020} rates use a newer version of \textsc{cloudy}, a more recent model for the background radiation, a lower redshift of reionization, and they account for self-shielding, the presence of dust, cosmic rays and an interstellar radiation field. 

A new addition is the treatment of rapidly cooling gas. In \bahamas\ and our other \textsc{gadget} simulations we computed the entropy that a particle is expected to reach at the end of its time step based on its current radiative cooling rate and then adjusted the time derivative of the entropy such that the particle would gradually drift to this final entropy over the course of its time step. However, this is inappropriate if the cooling time is short compared with the time step. Therefore, if the internal energy\footnote{While the \textsc{gadget} simulations used entropy as an independent variable, in \flamingo's hydrodynamics solver \sphenix\ the internal energy is used instead.} is expected to change by more than one third, then we immediately set it to the value that we estimate it will reach at the end of the time step. 

\subsubsection{Star formation and the pressure of the interstellar medium}
\label{sec:SF}
Since \flamingo\ lacks the resolution to model the multiphase ISM, we follow \citet{Schaye2008} and impose a density-dependent lower limit on the pressure corresponding to the temperature 
\begin{equation}
    T_\text{EoS}(\nH) = 8000~\K~\left (\frac{\nH}{10^{-1}\,\cm^{-3}}\right )^{1/3},
\end{equation}
onto gas with proper hydrogen number density $\nH > 10^{-4}\,\cm^{-3}$ and overdensity greater than 10. This relation, which is often referred to as an `equation of state', corresponds to a constant Jeans mass of $\sim 10^7\,\Msun$, which is, however, unresolved by the \flamingo\ simulations. The temperatures of gas near $T_\text{EoS}$ should be interpreted as representative of the pressure of a multiphase medium and should therefore not be used to compute observables.

During the simulation, gas particles are converted into collisionless star particles. Gas particles with proper density\footnote{Due to a bug, in all intermediate-resolution simulations except for the Jet models, particles with metallicity equal to precisely zero used a threshold of $\nH^* = 10\,\cm^{-3}$. Tests show that this only has significant effects on galaxies with fewer than 10 star particles, where it artificially suppresses the stellar-to-halo mass ratio. Hence, the bug compromised our ability to fit the SMF to even lower masses in this, in any case, poorly resolved regime. The error was fixed before running the low- and high-resolution simulations and the simulations using jet-like AGN feedback. \label{fn:SF_bug}} $\nH > \nH^* = 10^{-1}\,\cm^{-3}$, overdensity $>100$ and pressure at most 0.3~dex above the temperature $T_\text{EoS}$ are eligible for star formation. The proper density threshold is motivated by the models of \citet{Schaye2004} for the transition from the warm, atomic interstellar gas phase to the cold, molecular phase, though we neglect the predicted (weak) dependence on metallicity. The overdensity threshold ensures intergalactic gas at very high redshift does not form stars. The temperature ceiling prevents star formation in high-temperature gas.

The low-resolution runs do not sample the density distribution to sufficiently high values to form enough stars if the above star formation criteria are used, even in the absence of feedback. These simulations therefore use lower thresholds of $\nH^* = 10^{-3} \,\cm^{-3}$ and overdensity $>10$ combined with a temperature ceiling of $T < 10^5\,\K$. 

Gas particles are stochastically converted into star particles using the pressure-dependent star formation rate (SFR) of \citet{Schaye2008}, 
\begin{equation}
    \dot{m}_* = m_\text{g} A (1~\Msun\,\pc^{-2})^{-n} \left (\frac{\gamma}{G} f_\text{g} P\right )^{(n-1)/2},
    \label{eq:SFlaw}
\end{equation}
where $m_\text{g}$ is the gas particle mass, $\gamma=5/3$ is the ratio of specific heats, $G$ is the gravitational constant, and $P$ is the pressure. This relation is derived from the observed Kennicutt-Schmidt surface density star formation law, 
\begin{equation}
    \dot{\Sigma}_\ast = A \left (\frac{\Sigma_\text{g}}{1~\Msun\,\pc^{-2}} \right )^n ,
    \label{eq:KSlaw}
\end{equation}
under the assumption of vertical, hydrostatic equilibrium with a gas fraction of unity, $f_\text{g}=1$. We use the values $A =1.515\times 10^{-4}\,\Msun\,\yr^{-1}\,\kpc^{-2}$ and $n=1.4$ measured by \citet{Kennicutt1998}, where the former has been converted to the value for a \citet{Chabrier2003} stellar initial mass function (IMF). Note that equation~(\ref{eq:SFlaw}) implies a minimum possible nonzero SFR (attained for a single gas particle with density $\nH = \nH^*$) of $\dot{m}_* \approx 0.4~\Msun\,\yr^{-1} (m_\text{g}/10^9\,\Msun)$. 

\subsubsection{Stellar mass loss}
\label{sec:mass_loss}
Star particles are treated as simple stellar populations with a \citet{Chabrier2003} IMF for zero age main sequence masses between 0.1 and 100~$\Msun$. Time-dependent stellar mass loss by stellar winds from massive stars, asymptotic giant branch stars, core-collapse supernovae, and type Ia supernovae are implemented as in \cite{Wiersma2009Chemo} with the modifications described in \S4.4 of \citet{Schaye2015}. 

Briefly, we track the abundances of the elements H, He, C, N, O, Ne, Mg, Si, and Fe. These abundances are used to compute the corresponding element-by-element radiative cooling rates, along with those of Ca and S, whose abundances we assume to track that of Si with mass ratios of 0.094 and 0.605, respectively. Mass loss occurs when a star leaves the main sequence. At each time step\footnote{To limit the computational expense, mass loss is executed after 10 time steps for stellar particles with ages $> 100$~Myr.}, we therefore compute the (pre-main sequence) mass range of stars that leave the main sequence using the mass- and metallicity-dependent stellar lifetimes of \citet{Portinari1998}. The mass (and momentum) released by stars in this mass range is transferred from the star particle to its gaseous neighbours using SPH-weighting (but the weights are computed using the gas particles' initial mass; see \citealt{Schaye2015}) according to the stellar yields tabulated by\footnote{As in \citet{Wiersma2009Chemo}, the massive star yields of C, Mg and Fe are multiplied by factors of 0.5, 2 and 0.5, respectively. } \citet{Marigo2001}, \citet{Portinari1998} and \citet{Thielemann2003}. In contrast to \bahamas\ and EAGLE, we force the time steps of star particles with ages $\le 45$~Myr to be no longer than 1~Myr in order to ensure that the evolution of massive stars is sufficiently well sampled. 

Supernova type Ia (SNIa) rates per unit formed stellar mass are taken from \citet{Schaye2015}, who found that the following function results in an evolving cosmic SNIa rate density that agrees with observations, 
\begin{equation}
    \dot{N}_\text{SNIa} = \nu \frac{e^{-t/\tau}}{\tau} ,
    \label{eq:SNIa}
\end{equation}
where $t$ is the age of the stellar population, $\nu = 2\times 10^{-3}\,\Msun^{-1}$ and $\tau=2$~Gyr. Motivated by the idea that SNIa require a compact stellar remnant, and in contrast to \citet{Schaye2015}, we set the rate to zero for ages below 40~Myr, which corresponds to the lifetime of an $8~\Msun$ star. However, this does not have a significant impact on the cosmic SNIa rate.

While stellar mass loss reduces the masses of star particles, it increases the masses of gas particles, thus causing more metal-rich particles to be more massive. To keep the masses of baryonic particles similar, we split particles whose mass exceeds $4 m_\text{g}$ into two equal mass particles, where $m_\text{g}$ is the mean, initial gas particle mass. Note that this was not done in \bahamas.

\subsubsection{Stellar energy feedback}
\label{sec:SF_feedback}
Massive stars and supernovae inject energy into their surroundings. We implement stellar energy feedback kinetically by stochastically kicking SPH neighbours of young star particles, as in \citet{DallaVecchia2008}. We assume that stars with masses between 8 and $100~\Msun$ each inject $10^{51}\,\erg$ at the end of their main sequence lifetime and that a fraction $f_\text{SN}$ of this energy couples to the ISM. For our IMF the fraction $f_\text{SN}=1$ then corresponds to an energy budget of $1.18\times 10^{49}\,\erg\,\Msun^{-1}$. Besides the energy fraction $f_\text{SN}$, the feedback prescription uses a second parameter, the target wind velocity $\vsn$. The values of the stellar feedback parameters $f_\text{SN}$ and $\vsn$ are assumed to be constant and are calibrated as described in \S\ref{sec:calibration}.

In contrast to \bahamas, we do not impose the fixed time delay of 30~Myr between star formation and feedback from \citet{DallaVecchia2008}. Instead, we follow \citet{Richings2016} and evaluate the probabilities for feedback at each time step according to the energy associated with massive stars leaving the main sequence during that time step. 

Differently from \citet{DallaVecchia2008} and \bahamas, we do not simply increase the velocity of SPH particles by the wind velocity, because this violates energy conservation if those particles have non-negligible velocities relative to that star particle. We instead use the method\footnote{As in \citet{Chaikin2023}, we do not allow a gas particle to be kicked more than once per time step. However, while in \citet{Chaikin2023} unused energy is saved in a reservoir, we inject it thermally at the end of the time step. As `kick collisions' are rare, we expect resulting differences to be negligible.} of \citet{Chaikin2023}, which conserves linear momentum, angular momentum and energy by kicking particles in pairs in random but opposite directions. In this scheme the actual post-kick velocity of the wind particles relative to the star will differ somewhat from the target kick velocity if the particles are moving with respect to the star or if the two target particles have different masses. 

While \citet{DallaVecchia2008} and \bahamas\ used mass-weighting\footnote{All SPH neighbours are given a probability proportional to their mass.} to select the neighbours to be kicked, we use the method of \citet{Chaikin2022isotropic} to ensure the energy distribution is statistically isotropic. As demonstrated by \citet{Chaikin2022isotropic}, this difference is important because mass-weighting is biased to higher densities, which results in stronger cooling losses. 

Note that contrary to what is done in e.g.\ the IllustrisTNG \citep{Pillepich2018TNGmethod} and SIMBA \citep{Dave2019} cosmological simulations, we do not decouple wind particles from the hydrodynamics. As demonstrated in \citet{DallaVecchia2008}, such decoupling has major consequences and results in a qualitatively different feedback prescription. Decoupled winds also underestimate the energy required because of the neglect of thermal losses and work done during the opening up of the channels through the ISM that the decoupled models implicitly assume to exist.

As in \bahamas, the energy from SNIa is injected thermally at every time step, assuming each supernova provides $10^{51}\,\erg$. As discussed in e.g.\ \citet{DallaVecchia2012}, such a `thermal dump' results in only small temperature increases of the gas, which means it is subject to excessively large radiative losses and has very little effect.

\subsubsection{Black holes}
\label{sec:BHs}
The origin of supermassive BHs is still unclear \citep[e.g.][]{Volonteri2021BHreview}, but proposed mechanisms for their seeding such as direct collapse from metal-free gas in low-mass haloes, population III stellar remnants, and mergers inside star clusters are all unresolved by our simulations. Following \citet{DiMatteo2008} and \citet{Booth2009}, we therefore place seed BHs in haloes that are sufficiently massive and do not yet contain a BH. Starting from time $a=1/(1+z) = 0.05$, after every $\Delta\log_{10}a = 1.00751$ we run a friends-of-friends (FoF) halo finder with linking length 0.2 times the mean inter-particle distance. The minimum halo mass for seeding is set to $2.757\times 10^{11}\,\Msun ~(m_\text{g}/1.07\times 10^9\,\Msun)$. If the FoF halo were to consist purely of dark matter, then this would correspond to 49 particles. If the halo contains baryonic particles, then the number of particles is larger because baryonic particles are less massive than CDM particles. We use a BH seed mass of $10^5\,\Msun$. However, for the low-resolution simulations the seed mass had to be increased by a factor of 100 because those runs lack the resolution to follow the rapid growth phase of the BHs \citep[e.g.][]{Bower2017}. The BH seed is placed at the position of the densest gas particle in the halo (if the halo does not contain gas then it is not seeded), which is converted into a collisionless BH particle of the same mass as the progenitor gas particle. 

Because the BH mass can be much smaller than the particle mass, BH particles carry a subgrid BH mass that is initially set equal to the seed mass. As in \citet{Springel2005BHs}, BH processes such as gas accretion and BH mergers are computed using the subgrid BH mass, while gravitational forces are computed using the particle mass.

For comparison, \bahamas\ used the \citet{Booth2009} values of a minimum halo mass for seeding equal to 100 dark matter particles and a BH seed mass of $10^{-3}\,m_\text{g}$. \citet{McCarthy2017} found that the minimum halo mass for seeding has a non-negligible effect on the SMF, though its effect can be largely compensated for using other subgrid parameters. 

Massive BHs are subject to significant dynamical friction, which causes them to sink to the center of the galaxy and limits their excursions thereafter. Cosmological simulations lack the resolution to capture this dynamical friction for two reasons. First, to experience dynamical friction, the BH must be much more massive than the surrounding particles. For \flamingo\ this condition is violated across the entire range of BH masses. Second, scattering of particles with small impact parameters is important for dynamical friction, but is not captured because the gravitational forces are softened and because of poor sampling. To compensate for the inability to simulate dynamical friction directly, we follow \citet{Springel2005BHs} and \citet{Booth2009} and reposition the BHs by hand. \citet{Bahe2022} have recently demonstrated that BH repositioning is critical for simulations like \bahamas, and hence for \flamingo, as well as for much higher-resolution simulations like EAGLE. Without repositioning, BH growth through gas accretion and mergers is dramatically reduced and AGN feedback is consequently ineffective. 

At each time step, we move the BH to the location of the SPH neighbour within three gravitational softening lengths that has the lowest gravitational potential, provided it is lower than at the location of the BH. Note that the velocity of the BH is not explicitly altered when it is repositioned. While \bahamas\ allowed repositioning onto star and dark matter particles, we only consider gas particles in order to reduce the computational cost of neighbour finding (the gaseous neighbours need to be found in any case in order to compute the BH accretion rate). 
In contrast to \bahamas, we do not impose an upper limit on the velocity of gas particles on to which the BHs can be repositioned. As discussed in \citet{Bahe2022}, such a restriction is unnecessary and at the resolution of \bahamas\ it strongly reduces the efficiency of AGN feedback, though it is possible the effect would be reduced if repositioning onto stellar or dark matter particles were allowed and the effect can be compensated with changes to the subgrid parameters. 

When computing the gravitational potential for the purpose of repositioning, we should subtract the contribution of the BH itself in order to prevent the BH from becoming trapped by its own local potential well. This subtraction was done neither in \bahamas\ nor in any other simulation we are aware of. We also neglected to do so for nearly all our intermediate-resolution simulations, but did implement it for the high- and low-resolution simulations, as well as the intermediate-resolution Jet models, after recognizing the problem. To test its effect, we repeated a (400~Mpc)$^3$ intermediate-resolution simulation that did include the subtraction of the BH's contribution to the potential. For massive galaxies ($M_\ast \gtrsim 10^{12}\,\Msun$) this roughly doubled the fraction of quenched galaxies while for active galaxies the specific SFRs decreased by a factor of a few. However, for intracluster gas and lower-mass galaxies we did not find any significant differences. It should thus be kept in mind that for high-mass galaxies the quenched fractions in the intermediate-resolution \flamingo\ simulations with our fiducial implementation of AGN feedback may be artificially low. The failure to subtract the BH's own potential for the purpose of repositioning may also have caused excessive star formation in brightest cluster galaxies in earlier simulations.

The prescription for the merging of nearby BHs is taken from \citet{Bahe2022}. BHs are merged if they are separated by less than 3 gravitational softening lengths, $r< 3\epsilon$, and if their relative velocity satisfies $\Delta v < \sqrt{2Gm_\text{BH}/r}$, where $m_\text{BH}$ is the mass of the most massive of the two BHs and $r$ is their separation. This differs from \bahamas, which used the \citet{Booth2009} criterion $r< h$ and $\Delta v < \sqrt{Gm_\text{BH}/h}$, where 
$h$ is the SPH smoothing length of the most massive BH. The new criterion, which we think is more appropriate because merging is a gravitational process, leads to somewhat more rapid merging, particularly in low-density environments. When two BHs merge, we conserve momentum, subgrid BH mass, and particle mass and the particle carrying the lower-mass BH is removed from the simulation.

As in \citet{Springel2005BHs}, BHs are assumed to accrete at a modified Bondi-Hoyle rate limited by the Eddington rate. The modified Bondi-Hoyle rate is given by 
\begin{equation}
  \dot{m}_\text{accr} = \alpha \frac{4\pi G^2 m_\text{BH}^2\rho}{(c_\text{s}^2+v^2)^{3/2}},   
  \label{eq:BHaccr}
\end{equation}
where $\rho$ and $c_\text{s}$ are the gas density and the speed of sound of the ambient medium, $v$ is the velocity of the BH with respect to its environment, and the coefficient $\alpha$ is a boost factor described below (unmodified Bondi-Hoyle accretion corresponds to $\alpha=1$). The Eddington rate is 
\begin{equation}
    \dot{m}_\text{Edd} = \frac{4\pi G m_\text{BH} m_\text{p}}{\epsilon_\text{r} \sigma_\text{T} c},
    \label{eq:EddLimit}
\end{equation}
where $m_\text{p}$ is the proton mass, $\sigma_\text{T}$ is the Thomson cross section for electron scattering, $c$ is the speed of light, and $\epsilon_\text{r} = 0.1$ is the assumed radiative efficiency, i.e.\ the fraction of the accreted rest mass energy that is converted into light. 

Because the simulations generally do not resolve the Bondi radius, $r_\text{B} = Gm_\text{BH}/c_\text{s}^2 \approx 4~\kpc\ (m_\text{BH}/10^8\,\Msun) (c_\text{s}/10~\kms)^{-2}$, and because the simulations do not model the multiphase ISM, it is justified and necessary to boost the Bondi-Hoyle accretion rate. Following \citet{Booth2009}, we multiply the Bondi-Hoyle rate by the factor
\begin{equation}
    \alpha = 
    \begin{cases}
        1 & n_\text{H} < n_\text{H}^\ast \\
        \left (\frac{n_\text{H}}{n_\text{H}^\ast}\right )^{\betabh} & n_\text{H} \ge n_\text{H}^\ast ,
    \end{cases}
    \label{eq:alpha}
\end{equation}
where $\nH^\ast = 0.1~\cm^{-3}$ is the threshold for star formation and $\betabh$ is a parameter that is calibrated (see \S\ref{sec:calibration} and Table~\ref{tab:parameters}). As discussed by \citet{Booth2009}, at low densities the accretion rate should not be boosted for two reasons. First, for sufficiently high BH masses and low temperatures we can resolve Bondi-Hoyle accretion and for densities $\nH < \nH^\ast$ we do not expect the gas to contain a cold phase \citep{Schaye2004}. Second, if the large boost factors that are needed at high densities are applied everywhere, then for typical BH masses the Bondi-Hoyle rate only falls below the Eddington rate for extremely low densities, $\nH \ll \nH^\ast$, which forces AGN feedback to reduce the density of the ISM and the inner CGM to unrealistically low values in order to regulate BH growth through feedback. 

While the subgrid BH mass is smaller than its host particle's mass, the growth of the subgrid BH does not require the transfer of mass from the surrounding gas particles. If the BH subgrid mass exceeds the BH particle mass, then we transfer the difference in mass from the BH's SPH neighbours to the BH particle using the method of \citet{Bahe2022}. The mass (and momentum) taken from each neighbour is proportional to that neighbour's contribution to the SPH density at the location of the BH. Neighbours with mass less than half the initial baryonic particle mass are excluded (a limit that is not reached in practice). This method of letting the BH `nibble' on its neighbours differs from the method of \citet{Booth2009} used in \bahamas, where entire gas particles were stochastically swallowed by the BH. The new method leads to a smaller relative difference between the BH's particle and subgrid masses, particularly when the latter is similar to the mass of gas particles.
 
 Gas accretion increases the BH subgrid mass as $\dot{m}_\text{BH} = (1-\epsilon_\text{r}) \dot{m}_\text{accr}$ \citep{Booth2009} and decreases the BH particle mass by $\epsilon_\text{r}\dot{m}_\text{accr}$ \citep{Bahe2022}. We note that the correction to the particle mass, which accounts for the loss of rest mass to radiation, was neglected in \bahamas. 

\subsubsection{AGN feedback}
\label{sec:AGN}

In our fiducial model AGN feedback is implemented thermally as in \citet{Booth2009}, but in some of our intermediate-resolution runs we instead use anisotropic, kinetic feedback in order to enable tests of the sensitivity of the results to the implementation of AGN feedback. For simplicity we employ only a single mode of AGN feedback in each run. 

\subsubsection*{Fiducial model: thermal injection}
Although our fiducial model does not include jets, the outflows emerging from our thermal AGN feedback prescription are anisotropic because they naturally take the path of least resistance, and in clusters AGN feedback episodes result in buoyantly rising bubbles of high-entropy gas \citep[e.g.][]{Nobels2022}.  

A fraction $\epsilon_\text{f}$ of the energy available in the time step, $\epsilon_\text{r} \dot{m}_\text{accr} c^2 \Delta t$, is assumed to couple to the gas surrounding the BH. In order to prevent numerical overcooling, i.e.\ the overestimate of the radiative cooling rate due to the underestimate of the post-shock gas temperature that results from having to heat at least the mass of one gas particle \citep[e.g.][]{DallaVecchia2012}, we do not inject feedback energy at every time step. Instead, we store the energy in the subgrid reservoir of the BH particle until it suffices to increase the temperature of $n_\text{heat}$ gas particles by $\tagn$.
To ensure the feedback is well sampled, we adopt the BH time step limiter of \citet{Bahe2022}. Unless it would result in a time step smaller than $10^5~\yr$, we limit the time step of each BH particle such that its energy reservoir will not increase by more than the energy needed to heat one particle if the BH continues to accrete at its current rate.

As demonstrated by \citet{Booth2009,Booth2010}, because self-regulation determines the amount of energy that AGN provide for a given galaxy-scale gas accretion rate, the coupling efficiency, $\epsilon_\text{f}$, effectively determines the BH mass of galaxies that are regulated by AGN feedback, but other galaxy properties are insensitive to its value. If $\epsilon_\text{f}$ is increased, then the BH has to accrete less gas in order to inject the same amount of energy. As in \citet{Booth2009} and \bahamas, we set $\epsilon_\text{f}=0.15$, which means a fraction $\epsilon_\text{f}\epsilon_\text{r} = 0.015$ of the accreted rest-mass energy is used for feedback. We will show in Section~\ref{sec:BHmass} that this value results in good agreement with the observed $z=0$ relation between BH and stellar mass. 

As in (Cosmo-)OWLS, we set $n_\text{heat} = 1$ whereas \bahamas\ used $n_\text{heat}=20$. Hence, at a fixed mass resolution, $m_\text{g}$, and for a fixed $\tagn$, AGN events in \flamingo\ are an order of magnitude less energetic than in \bahamas. However, \citet{McCarthy2017} found that reducing $n_\text{heat}$ to unity only slightly increases the SMF and cluster gas fractions. In contrast to our previous work, we do not use mass-weighting to select the SPH neighbour that receives the energy. Instead, we inject the feedback energy into the gas particle nearest to the BH. This choice minimizes the occurrence of feedback at large distances. As demonstrated by \citet{Chaikin2022isotropic} for the case of stellar feedback, for small values of $n_\text{heat}$ selecting the nearest particles yields a nearly isotropic distribution of energy and gives results that are nearly indistinguishable from their isotropic scheme.

Finally, $\tagn$ is a parameter of the model that we calibrate (see \S\ref{sec:calibration} and Table~\ref{tab:parameters}).

\subsubsection*{Model variation: kinetic, jet-like injection}
For accretion rates that are not extremely small compared to the Eddington rate, our fiducial, thermal implementation can be thought to represent the effect of radiatively-driven winds from geometrically thin accretion discs \citep{Shakura1973}. However, when the accretion rate drops below some critical fraction of the Eddington rate ($\dot{m}_\text{BH}/\dot{m}_\text{Edd}\sim 10^{-2}$), accretion discs are expected to be radiatively-inefficient, advection-dominated, geometrically thick \citep{Narayan1995} and to be efficient at launching collimated jets \citep[e.g.][]{Yuan2014} whose power (or efficiency) depends on the dimensionless BH spin $a=J_\mathrm{BH}c/(m_\mathrm{BH}^2G) \in[-1,\hspace{0.3mm}1]$, where $J_\mathrm{BH}$ is the BH angular momentum \citep{Tchekhovskoy2010,Narayan2022}, and that are directed along the BH spin axis (i.e.\ the direction of its angular momentum vector). 

Our `Jet' simulations employ a simplified version of the BH spin and AGN jet implementation of \citet{Husko2022}. However, for simplicity, and to maximize the difference with respect to our fiducial model, we use the jet mode (with its accompanying accretion disc subgrid physics) for all accretion rates, and we employ a constant jet feedback efficiency of 0.015 for consistency with the fiducial, thermal model. This efficiency results in BH masses that are consistent with observations (see Fig.~\ref{fig:bh}). Given this choice, the BH spins obtained using the aforementioned model are used only to determine the direction that the jets are launched in.

 The model accounts for the following processes when evolving BH spin: 1) gas accretion; 2) jet spindown (negligible in this case due to the small jet efficiencies we have assumed); 3) BH-BH mergers \citep[as in][]{Rezzolla2008}; and 4) Lense-Thirring torques that mediate the angular momentum transfer between the disc and the BH. If the accretion rate is high, then these torques can cause the BH spin to be redirected towards the angular momentum of the outer regions of the accretion disc on Myr time-scales \citep{King2005}. At low accretion rates (matching the assumptions used here), the redirection is instead much slower. For the BH spin axis to change appreciably, the BH therefore needs to accrete a large fraction of its current mass \citep[see appendix~B of][]{Husko2022}, or the spin evolution needs to be dominated by major BH-BH mergers, which are relatively rare. Given these considerations, redirection is expected to occur on Gyr rather than Myr time-scales, at least in galaxy clusters.
 
 We assume that the direction of the angular momentum of the outer accretion disc (whose size is up to $10^5$ gravitational radii, $\sim\,$pc scales) is the same as that of the gas in the BH smoothing kernel, which we acknowledge to be a strong assumption given the low resolution of the simulations. While this may imply that the direction of the BH spin vector (and thus the jets) is not entirely realistic, \citet{Husko2022} showed that jet redirection is unimportant for long-term feedback effects, as long as it occurs relatively rarely. To be more precise, provided the jets do not redirect more frequently than the typical duration of a jet episode ($\lesssim 10^2$ Myr), their effects are insensitive to their direction and thus the time-scale of redirection (which we expect to be $\sim\,$1~Gyr here). The most important point is that the jets do not redirect frequently (e.g.\ on  Myr time-scales), since that would, at the resolutions employed in this project, correspond more to an isotropic kinetic wind (as in e.g.\ MTNG) than jets.

The jets are launched by kicking gas particles from within the BH's SPH kernel. We use a constant target jet velocity, $\vjet$. Every time the BH's feedback energy reservoir exceeds $2\times(1/2)m_\mathrm{g}\vjet^2$, two particles are kicked. We choose the two particles closest to the BH spin axis (in terms of angular distance), and their velocities are increased along unit vectors chosen randomly from within two cones with $7.5\degree$ opening angles around the BH spin axis (one on each side of the BH spin axis, for each particle). Note that as was the case for stellar feedback (see \S\ref{sec:SF_feedback}), energy conservation implies that the actual magnitude of the velocity increase can be different from the target velocity, depending on the initial particle velocity.

AGN jets can have velocities approaching the transrelativistic regime on the scales resolved by our simulations. However, due to the low mass resolution, jets using such high velocities would be extremely poorly sampled. Thus, $\vjet$ is treated as a subgrid parameter of the model whose value we calibrate (see \S\ref{sec:calibration} and Table~\ref{tab:parameters}) and whose role is analogous to that of $\tagn$ for the case of thermal AGN feedback.

\subsection{Initial conditions}\label{sec:ICs}

Initial conditions (ICs) for purely gravitational $N$-body simulations of dark matter are commonly set up with higher-order Lagrangian perturbation theory (LPT), which is known to be significantly more accurate than the first-order Zel'dovich approximation \citep{Scoccimarro1998,Sirko2005,Crocce2006}. However, doing so for simulations with multiple fluid components with distinct transfer functions, such as hydrodynamical simulations and simulations with neutrinos, is non-trivial. For \flamingo{}, we made use of several recent theoretical developments that enable multi-fluid third-order Lagrangian perturbation theory (3LPT) ICs that accurately reproduce the relative growth of the individual fluid components. These developments were incorporated in the \monofonIC{} code \citep{Hahn2020,Michaux2021}. For \flamingo{}, a modified version of \monofonIC{} was used that implements the effects of massive neutrinos\footnote{\url{https://github.com/wullm/monofonic}}.

We use the prescriptions for 3-fluid ICs with CDM, baryons, and massive neutrinos outlined in \citet{Elbers2022a}, which builds on the 2-fluid formalism of \citet{Rampf2021} and \citet{Hahn2021}. CDM and baryon particles are set up in a two-stage process. First, the combined mass-weighted CDM + baryon fluid is initialized with single-fluid 3LPT, accounting for the presence of neutrinos. This single fluid is then split into separate components with distinct transfer functions by perturbing the masses and velocities in accordance with the first-order compensated mode. \citet{Hahn2021} showed that discreteness errors can be suppressed by perturbing particle masses rather than displacements, thereby eliminating spurious growth of the compensated mode (see also \citealt{Bird2020,Liu2023}).

The underlying Gaussian random fields were chosen from subregions of \panphasia\ to facilitate future zoom-in resimulations \citep{Jenkins2013}, see appendix~\ref{sec:panphasia} for details. To limit cosmic variance without compromising the ability to do zooms, we used partially fixed ICs \citep{Angulo2016}, setting the amplitudes of modes with $(kL)^2 < 1025$ to the mean variance, where $k$ is the wavenumber and $L$ is the side length of the simulation box. A paired simulation with inverted phases was run for the $L=1~\Gpc$ DMO fiducial cosmology simulation to enable further limiting of the cosmic variance \citep{Angulo2016}. The starting redshift, $z=31$, was chosen to be as late as possible to limit discreteness errors and reduce computational cost, but before shell-crossing, such that LPT remains valid, and before BH seeding and star formation are initiated. The linear power spectra and transfer functions were computed with \textsc{class} \citep{lesgourgues11,lesgourgues11b}. Given that \flamingo\ includes simulations with side lengths of $L=2.8~\Gpc$ and greater, relativistic effects become a factor on the largest scales. The ICs were therefore set up in $N$-body gauge \citep{Fidler2015,Fidler2017}. In the absence of radiation and neutrinos, this is enough to ensure that the relativistic fluid equations coincide with the usual Newtonian equations solved by $N$-body codes at first order.

Since our simulations include massive neutrinos, we also need to account for their presence \citep{Zennaro2017,Aviles2020}. At first order, massive neutrinos change the transfer functions and introduce a scale-dependence in the growth factors \citep[e.g.][]{Lesgourgues2006}. This is taken into account in a generalized back-scaling procedure \citep{Zennaro2017}, the purpose of which is to correct for the remaining differences between the relativistic and Newtonian fluid systems. This is accomplished by starting with the desired result: the $z=0$ transfer functions computed by \textsc{class}, and evolving them back to $z=31$ with a Newtonian fluid approximation implemented in the \textsc{zwindstroom} code \citep{Elbers2022a}. In practice, this amounts to a small boost in the initial power spectrum on large scales. The simulations correctly model the expansion history, including the effects of massive neutrinos, an amount of radiation corresponding to a CMB temperature of $T=2.7255~\K$, and an effective number of relativistic species $N_\text{eff}=3.046$ at high redshift. Accordingly, the same is done in the back-scaling calculation. Neutrinos also change the growth rate of matter perturbations relative to the geometric expansion, which feeds back into the higher-order LPT solutions. While the full third-order theory requires successive expensive convolutions \citep{Aviles2020}, the effects can be captured to high accuracy by scale-independent correction factors obtained from an all-order recursive solution in the small-scale limit \citep{Elbers2022a}, which were included. Finally, the neutrino particles themselves also require ICs. It is understood that neutrino perturbations are suppressed relative to dark matter perturbations, such that neutrinos can still be treated linearly at $z=31$. To take into account linear perturbations to the neutrino phase-space distribution function, we integrated neutrinos from $z=10^9$, when all modes of interest were outside the horizon, down to $z=31$ using the \FastDF{} code \citep{Elbers2022c}. This represents a substantial improvement over the Zel'dovich approximation, which neglects higher moments of the neutrino distribution and underestimates the power spectrum on large scales \citep{Elbers2022c}.

\subsection{Structure finding}\label{sec:VR}

The identification of haloes and substructures in the outputs of the simulations was performed using the \vr subhalo finder \citep{Elahi2019vr}. We summarize the procedure here. In a first phase, haloes are identified in configuration space using a 3D FoF algorithm with a linking length $l=0.2$ of the mean interparticle separation \citep{Davis1985}. The FoF search includes all particle types except neutrinos. Within each halo, we then search for substructures that are dynamically distinct from the mean background halo. This is achieved by performing an iterative 6D FoF search, in phase space and again including all particle types except neutrinos, on each halo individually with a velocity-space linking length set to the velocity dispersion of the original 3D FoF object and a real-space linking length set to 0.1 times the one used in the initial 3D FoF search, i.e.\ $0.1\times 0.2=0.02$. The most prominent object thus found (the one that is the most distinct from the background halo) is labelled as the central and the remaining objects as satellites. Once these substructures have been identified, we clean them up via an unbinding procedure which removes particles that are not gravitationally bound to the object. The most bound particle in each cleaned object is then used to define its centre. 

Finally, we note that all the galaxy and cluster measurements presented here were computed using the Spherical Overdensity and Aperture Processor (SOAP), a tool that we developed for the \flamingo\ project. SOAP takes the (sub)halo centers and particle membership determined by \vr as input, and computes a large number of (sub)halo properties for a range of apertures, which can be 3D or projected, include or exclude other substructures and unbound particles, and whose sizes can be specified as physical radii or mean internal overdensities.

\section{Calibration of subgrid parameters and observational biases}\label{sec:calibration}
Subgrid prescriptions for unresolved physical processes necessarily involve choices and free parameters. Some parameter values are chosen based on theoretical considerations. An example is the star formation threshold density, which is motivated by the radiative transfer models of \citet{Schaye2004} for the atomic to molecular phase transition. Others are fixed based primarily on numerical considerations, e.g.\ the equation of state imposed on the ISM discussed in Section~\ref{sec:SF}. 

Some parameters can be taken directly from observations. This is for example the case for the parameters $A$ and $n$ appearing in the pressure-based star formation law (eq.~\ref{eq:SFlaw}), which are, respectively, the normalisation and the slope of the Kennicutt-Schmidt surface density law (eq.~\ref{eq:KSlaw}). Other parameters are fit to specific observations that directly constrain the corresponding subgrid process. This is for example the case for the normalisation, $\nu$, and time delay, $\tau$, appearing in the SNIa delay function (eq.~\ref{eq:SNIa}), which are fit to the observed cosmic SNIa rate density assuming the observed cosmic star formation history. A less intuitive example is the efficiency of AGN feedback, $\epsilon_\text{f}$, which, in galaxies regulated by AGN feedback, determines the masses of BHs but has little effect on everything else and is chosen to reproduce the observed relation between stellar mass and BH mass for massive galaxies \citep{Booth2009,Booth2010}.  

However, there are also subgrid parameters that affect multiple observables of primary interest and that are not directly constrained by specific observations. In those cases we have to choose observables to calibrate to and we require a method to set the values of said subgrid parameters. We have chosen to calibrate to two observables that are particularly important for the goals of \flamingo: the present-day galaxy stellar mass function (SMF) and the gas mass fraction in clusters of galaxies. The SMF is important because it constrains the relation between stellar mass and halo mass, where the former is observed and the latter determines the clustering properties. The gas fraction in groups and clusters is important because it largely determines the baryonic suppression of the matter power spectrum \citep[e.g.][]{Semboloni2011,Semboloni2013,McCarthy2018,Schneider2019,Debackere2020,VanDaalen2020,Arico2021,Delgado2023,Salcido2023} and of the cluster mass function \citep[e.g.][]{Velliscig2014,Cui2014,Debackere2021}. We wish to calibrate our galaxy formation model to these observations by varying as few parameters as possible. 

From previous simulation work \citep[e.g.][]{McCarthy2017} we know that the SMF and gas fractions are most sensitive to stellar and AGN feedback, respectively, though these processes are not independent \citep{Booth2013}. After some experimentation, we found it necessary and sufficient to vary two parameters related to stellar feedback, the energy budget ($f_\text{SN}$) and the kick velocity ($\vsn$), and two parameters related to BHs, the AGN heating temperature ($\tagn$) and the logarithmic slope of the density dependence of the BH accretion rate boost factor ($\betabh$). For the model variations using AGN jet feedback the parameter $\tagn$ is replaced by the jet velocity, $\vjet$. For the low-resolution runs we do not use stellar feedback, because the mass range in which stellar feedback dominates remains unresolved, thus leaving only the two BH parameters. 

Below we summarize our calibration method and results. For more details, motivation, discussion and additional results, including the covariance between the different parameters, we refer to the companion paper by \citet{FlamingoCal}. 

\subsection{Emulation}
In previous projects such as EAGLE and \bahamas\ we calibrated the subgrid parameters through trial and error, mostly be systematically varying one parameter at a time. For \flamingo\ we instead use a more systematic and quantitative Bayesian approach that has already been applied to the semi-analytic model GALFORM \citep[][]{Bower2010,Rodrigues2017}. After settling on the set of parameters to vary and the ranges over which to vary them (i.e.\ our priors), which we based on physical arguments and a small number of test runs, we employ machine learning to fit the subgrid parameters to the calibration data. We use Gaussian process emulators \citep[e.g.][]{GPbook} trained on 32-node Latin hypercubes of simulations. The dimensionality of the hypercube equals the number of subgrid parameters and the side lengths are our priors. The 32 nodes are distributed quasi-randomly throughout the hypercube such that the minimum distance between the nodes is maximized. A separate hydrodynamical simulation is run for each node. We then build a different emulator for each observable based on all 32 simulations. The SMF emulator takes as input the stellar mass, $M_*$, and the subgrid parameter vector, $\mathbf{\theta}$, and it predicts the SMF, $f(M_*)$. The inputs for the gas fraction emulator are the total cluster mass, $M_\text{500c}$ (i.e.\ the mass inside the radius $R_\text{500c}$ within which the mean density is 500 times the critical density), and the subgrid parameters $\mathbf{\theta}$. It outputs the gas fraction as a function of mass, $f_\text{gas,500c}(M_\text{500c})$. 

As discussed in \citet{Schaye2015}, it is generally necessary to recalibrate subgrid parameters for unresolved processes in the ISM when the resolution is changed. A higher-resolution simulation resolves smaller scales and higher gas densities and will therefore, for example, yield different radiative losses and different BH accretion rates (and hence different AGN feedback), which will, in turn, change the structure of the ISM even on scales resolved by the lower-resolution run. To compensate, it is therefore generally necessary to adjust the parameter values when the resolution is modified. This is particularly true if, as is the case here, we demand a very good fit to the calibration data, and if the statistical errors on both the observations and the predictions are small due to the availability of large surveys and large simulation volumes, respectively. A comparison of recalibrated simulations of different resolutions is referred to as a `weak convergence test'. We therefore calibrate each of the three \flamingo\ resolutions separately using their own Latin hypercubes and emulators. For high, intermediate, and low resolution (respectively m8, m9, and m10 in Table~\ref{tab:simulations}) we employ hypercubes consisting of $(100~\Mpc)^3$, $(200~\Mpc)^3$ and $(400~\Mpc)^3$ simulations, respectively. 

We fit the emulator predictions to the data using Markov chain Monte Carlo sampling of the parameter space, accounting for both errors in the data and the emulator uncertainty. We compute the log likelihood separately for the SMF, the X-ray gas fractions and the weak lensing gas fractions (see \S\ref{sec:calibration_data}). These are then combined, giving equal weight to the two types of gas fraction data and equal weight to the combined gas fraction result and the SMF. We use the maximum likelihood values of the subgrid parameters as our fiducial values. 

\subsection{Calibration data} \label{sec:calibration_data}
For the SMF we calibrate to the recent results for $z=0$ from \citet{Driver2022} for the GAMA survey. For the gas fractions of groups and clusters we use X-ray observations at $z\approx 0.1$ compiled from the literature by \citet{FlamingoCal}, which measure the total mass inside $R_\text{500c}$, $M_\text{500c}$, under the assumption of hydrostatic equilibrium, and weak gravitational lensing data at $z\approx 0.3$ from the HSC-XXL survey of \citet{Akino2022}. 

We do not attempt to measure these observables from virtual observations using observational methods, because we lack the resolution to create realistic mock data for galaxies and low-mass clusters. Virtual observations could probably be used to measure gas fractions of clusters sampled with many particles, but more work is needed to see if the simulations are sufficiently realistic for mock observational analyses to be preferable to using observationally inferred gas and total masses. 

If we assume that the bulk of the cluster gas is detectable in X-rays, then there is no ambiguity about the definition of cluster gas fractions, which observations express as the gas mass fraction within $R_\text{500c}$, and which can thus be measured straightforwardly from the simulation output. However, the situation is more murky for stellar masses. Observed stellar masses are typically based on extrapolated S\'ersic fits to surface brightness or inferred mass profiles. Even for stellar mass profiles, we cannot mimic this procedure at our resolution down to the low masses (corresponding to $\sim 10$ stellar particles per object) for which we aim to reproduce the SMF. We therefore choose to define the stellar mass of a galaxy as the stellar mass that is gravitationally bound to the subhalo and contained within a 3D aperture of radius 50~kpc, which is well resolved and which \citet{deGraaff2022} found to yield results close to the masses inferred from virtual observations of the (much higher-resolution) EAGLE simulation. An observational stellar mass bias factor that is discussed below accounts for any systematic offset due to differences in mass definitions. To account for random measurement errors present in the observations, which lead to an Eddington bias (i.e.\ if the SMF is steep, then the number of objects that scatter up into a given observed mass bin strongly exceeds the number that scatter down into that bin, thus flattening the observed slope), we add lognormal scatter of $\sigma(\log_{10}M_{\ast}) =  {\rm min}\left(0.070+0.071z, 0.3\right) \ {\rm dex}$ \citep{Behroozi2019} to the simulation stellar masses before training the emulator.          

We only fit to the data over a limited range of masses. The lower mass limit for the SMF is determined by the resolution limit of the simulation. For high, intermediate, and low resolution the emulator predictions are fit to the data with $M_* > 10^{8.67}$, $10^{9.92}$, and $10^{11.17}\,\Msun$, respectively. The upper mass limit is always $10^{11.5}\,\Msun$, because at higher masses there are significant systematic differences between different data sets and the simulation predictions become highly sensitive to the size of the aperture. For the cluster gas fractions the lower mass limit is always $M_\text{500c} = 10^{13.5}\,\Msun$, while the upper mass limit is determined by the volume of the simulations used for the Latin hypercube. For high, intermediate and low resolution the upper mass limits are, respectively, $10^{13.73}$, $10^{14.36}$, and $10^{14.53}\,\Msun$. 

\subsection{Observational bias factors} \label{sec:biases}
We allow for three types of potential observational biases, which for simplicity we assume to be constants. First, a stellar mass bias factor accounts for possible systematic errors in the observationally inferred stellar masses, e.g.\ because of uncertainty in spectral energy distribution modeling. It shifts the SMF horizontally, i.e.\ along the mass axis. Second, a cosmic variance bias factor accounts for systematic uncertainty in the galaxy number densities due to the possibility that the finite-sized observed volume is unrepresentative. It shifts the SMF vertically, i.e.\ along the number density axis. 

Third, a hydrostatic mass bias factor accounts for the possibility that measurements of total cluster masses inferred from X-ray observations are systematically offset from the true masses because they assume the gas is in hydrostatic equilibrium and neglect non-thermal pressure. Indeed, comparisons between X-ray and weak lensing observations indicate that the former significantly underestimate the total masses \citep[e.g.][]{Hoekstra2015,Eckert2016}, assuming that the weak lensing masses are relatively unbiased. The hydrostatic mass bias shifts the observed cluster gas fraction -- mass relation horizontally, i.e.\ along the mass axis. We neglect the effect on the gas fraction within $R_\text{500c}$. This correction is small, because $R_\text{500c}$ increases if the mass increases, and the gas mass increases if $R_\text{500c}$ increases. Hence, hydrostatic mass bias will only change the gas fraction insofar as the cumulative gas fraction changes between the biased and corrected values of $R_\text{500c}$. Even for a mass bias as large as 25 per cent, $R_\text{500c} \propto M_\text{500c}^{1/3}$ changes by only 10 per cent, which is expected to lead to only small differences in the gas fraction \citep[see e.g.\ fig.~6 of][]{Velliscig2014}. Note that correcting for this small effect would be difficult, because observational studies do not report measurements at the corrected values of $R_\text{500c}$. Our priors for the bias factors are based on results taken from the literature: \citet{Behroozi2019} for the stellar mass bias, \citet{Driver2022} for cosmic variance, and \citet{Hoekstra2015,Eckert2016} for hydrostatic mass bias.

When calibrating the intermediate-resolution simulations, we fit simultaneously for the subgrid parameters and bias factors. Because observational biases should be independent of the resolution of the simulations, we do not re-fit the bias factors for the low- and high-resolution models. We choose to use intermediate resolution to fit the biases because only this resolution probes a large mass range ($\gtrsim 1$~dex) for both types of observables. Similarly, we do not re-fit the biases when we change the model, i.e.\ the cosmology or the subgrid feedback parameters.

The best-fitting (i.e.\ maximum likelihood) stellar mass bias factor correspond to an increase of the observed stellar masses by 0.026~dex and the best-fitting cosmic variance bias corresponds to a change in the observed galaxy number densities by a factor 0.995 (i.e.\ a decrease of 0.5 per cent). The $1\sigma$ posterior confidence intervals for these biases are $0.06 \pm 0.11$~dex and $0.98 \pm 0.06$, respectively, and are thus consistent with the data being unbiased. These can be compared with our adopted Gaussian priors of $\mathcal{N}(\mu,\sigma) = \mathcal{N}(0,0.14)$ and $\mathcal{N}(1,0.06)$, respectively, where $\mu$ is the mean and $\sigma$ the standard deviation. The best-fitting hydrostatic mass bias corresponds to dividing the observed X-ray total masses by a factor 0.743. The evidence for such a bias is strong, with a posterior of $0.74 \pm 0.09$, but fully in line with (and largely determined by) our prior of $\mathcal{N}(0.74,0.10)$ based on the literature. In all the plots shown in this work the observational data has been shifted by the best-fitting bias factors. 

\begin{table}
    \centering
    \caption{Values of the calibrated subgrid parameters (see \S\ref{sec:subgrid} for definitions of the parameters) for the fiducial model for each of the three simulation resolutions and for the feedback variations at intermediate resolution. The low-resolution simulations do not include stellar feedback.}
    \label{tab:parameters}
    \begin{tabular}{llccl}
        \hline
       Prefix & $f_\text{SN}$ & $\vsn$ & $\tagn$ or $\vjet$ & $\betabh$ \\
    & & $(\kms)$ & (K) or $(\kms)$ \\
       \hline
       Fid\_m8            & 0.524 & 259 & $10^{8.07}$ & 0.038 \\
       Fid\_m9             & 0.238 & 562 & $10^{7.95}$ & 0.514 \\
       Fid\_m10             &     0 &   - & $10^{8.29}$ & 0.373 \\
       fgas$+2\sigma$\_m9 & 0.219  & 577 & $10^{7.71}$ & 0.554 \\
       fgas$-2\sigma$\_m9 & 0.206  & 552 & $10^{8.08}$ & 0.497 \\
       fgas$-4\sigma$\_m9 & 0.191  & 532 & $10^{8.21}$ & 0.482 \\
       fgas$-8\sigma$\_m9 & 0.145  & 483 & $10^{8.40}$ & 0.462 \\
       M*$-\sigma$\_m9   & 0.322  & 608 & $10^{8.06}$ & 0.626 \\
       M*$-\sigma$\_fgas-4$\sigma$\_m9 & 0.261 & 557 & $10^{8.27}$ & 0.620 \\
       Jet\_m9                 & 0.166  & 477 & 836 & 0.597 \\
       Jet\_fgas$-4\sigma$\_m9 & 0.176 & 527 & 1995 & 0.439 \\
    \hline
    \end{tabular}
\end{table}

\subsection{Subgrid parameter values}\label{sec:subgridparvalues}
The fiducial subgrid parameter values for all resolutions are listed in Table~\ref{tab:parameters}. These are the maximum likelihood values from table~2 of \citet{FlamingoCal}, which also lists the priors and the posteriors. For intermediate resolution, which is \bahamas\ resolution, we use $f_\text{SN}=0.238$, $\vsn=562~\kms$, $\tagn = 10^{7.95}\,\K$ and $\betabh=0.373$. These differ from the \bahamas\ values\footnote{\bahamas\ used a wind mass loading factor of 2, which corresponds to $f_\text{SN}=0.16$ for their wind velocity of $300~\kms$.} of 0.16, $300~\kms$, $10^{7.8}\,\K$ and 2. However, except for $\betabh=2$, all the \bahamas\ values fall within our $1\sigma$ posteriors. We need a significantly smaller value of $\betabh$, which implies significantly smaller boosts of the Bondi-Hoyle accretion rates. This is despite the fact that we use an order of magnitude smaller BH seed mass than in \bahamas. The difference in $\betabh$ is mostly due to our improvement in the BH repositioning scheme, in particular the absence of a velocity ceiling for particles to be eligible as repositioning targets, discussed in Section~\ref{sec:BHs} and \citet[]{Bahe2022}. 

Compared with intermediate resolution, at high resolution we require about twice as much energy from stellar feedback but about half as large a kick velocity. The increase in energy is probably needed to compensate for the increase in radiative losses due to the higher densities that are resolved at higher resolution. The reduction in the wind velocity likely reflects the fact that at higher resolution we calibrate down to lower galaxy masses and thus lower circular velocities. At high resolution the AGN heating temperature is only 0.12~dex higher than for intermediate resolution, but the slope of the density dependence of the BH accretion rate boost factor is significantly decreased to $\betabh=0.038$, implying almost no boost of the Bondi-Hoyle rate. This reduction is likely due the higher gas densities that can be resolved. The best-fitting values for low resolution ($\tagn = 10^{8.29}\,\K$, $\betabh=0.373$) are not directly comparable to those of the other resolutions because of the absence of stellar feedback, the higher threshold for star formation and the larger BH seed mass (see \S\ref{sec:subgrid}). 

The cosmologies we consider are sufficiently close that it is unnecessary to recalibrate the subgrid model when changing cosmology \citep[e.g.][]{McCarthy2018}. We do, however, change the subgrid model for a series of intermediate-resolution $(1~\Gpc)^3$ runs that vary the stellar and/or AGN feedback. The `M*' and `fgas' models were calibrated analogously to the fiducial models, but after shifting all the observed galaxy stellar masses (for M*) or cluster gas fractions (for fgas). For models M*$-\sigma$ the observed SMF was shifted to 0.14~dex lower stellar masses. For models fgas$+2\sigma$, fgas$-2\sigma$, fgas$-4\sigma$, and fgas$-8\sigma$ the observed cluster gas fraction data points (one point per mass bin) were all shifted by, respectively, $+2$, $-2$, $-4$, $-8$ times the error estimated from bootstrapping in the case of the X-ray data (table~5 of \citealt{FlamingoCal}) or the error on the fit for the weak lensing data from \citet{Akino2022} (as given in \S3.2.2 of \citealt{FlamingoCal}). While even the $2\sigma$ gas fraction variations are formally ruled out at more than $2\sigma$ confidence if the data points are independent, systematic errors are likely correlated between different mass bins. Moreover, we wish to use the feedback variations to obtain conservative estimates of the potential effects of baryonic physics. 

The values of the subgrid parameters for the calibrated feedback variations are listed in Table~\ref{tab:parameters}. These are the maximum likelihood values from table~8 of \citet{FlamingoCal}, which also lists the posteriors. The changes relative to the fiducial model are modest. The main effect of the (0.14~dex) reduction in the observed stellar masses is an increase in $f_\text{SN}$ (by 0.13~dex for the fiducial gas fractions and by 0.14~dex for models fgas$-4\sigma$) though $\vsn$, $\tagn$ and $\betabh$ all increase as well. The differences in the gas fractions are driven mostly by $\tagn$, which changes by 0.13--0.24~dex for each $2\sigma$ shift in the gas fractions, with higher $\tagn$ corresponding to lower $f_\text{gas}$. 

Models Jet and Jet\_fgas$-4\sigma$ use kinetic, jet-like AGN feedback instead of our fiducial thermal AGN feedback implementation. Compared with our fiducial model, model Jet requires slightly weaker stellar feedback and a similar value of $\betabh$. If we convert the fiducial AGN temperature increase of $\tagn = 10^{7.95}\,$K into a velocity (using $2\frac{1}{2}\mu m_\text{p} \vjet^2 = \frac{3}{2} k_\text{B} T$, where $\mu m_\text{p}$ is the mean particle mass) we obtain a value that is just 0.21~dex higher than the adopted jet velocity of $836~\kms$. Compared with model Jet, model Jet\_fgas$-4\sigma$ requires a higher jet velocity and a smaller value of $\betabh$.

\begin{table*}
	\centering
	\caption{Hydrodynamical simulations. The first four lines list the simulations that use the fiducial galaxy formation model and assume the fiducial cosmology, but use different volumes and resolutions. The remaining lines list the model variations, which all use a 1~Gpc box and intermediate resolution. The columns list the simulation identifier (where m8, m9 and m10 indicate $\log_{10}$ of the mean baryonic particle mass and correspond to high, intermediate, and low resolution, respectively; absence of this part implies m9 resolution); the number of standard deviations by which the observed stellar masses are shifted before calibration, $\Delta m_\ast$; the number of standard deviations by which the observed cluster gas fractions are shifted before calibration, $\Delta f_\text{gas}$; the AGN feedback implementation (thermal or jets); the comoving box side length, $L$; the number of baryonic particles, $N_\text{b}$ (which equals the number of CDM particles, $N_\text{CDM})$; the number of neutrino particles, $N_\nu$; the initial mean baryonic particle mass, $m_\text{g}$; the mean CDM particle mass, $m_\text{CDM}$; the Plummer-equivalent comoving gravitational softening length, $\epsilon_\text{com}$; the maximum proper gravitational softening length, $\epsilon_\text{prop}$; and the assumed cosmology which is specified in Table~\ref{tab:cosmologies}.}
	\label{tab:simulations}
	\begin{tabular}{lrrllrrllrrl} 
		\hline
		Identifier & $\Delta m_\ast$ & $\Delta f_\text{gas}$ & AGN & $L$ & $N_\text{b}$ & $N_\nu$ & $m_\text{g}$ & $m_\text{CDM}$ & $\epsilon_\text{com}$ & $\epsilon_\text{prop}$ & Cosmology \\ 
		           & ($\sigma$) & ($\sigma$) && (cGpc) &&& ($\Msun$) & ($\Msun$)  & (ckpc) & (pkpc) \\
		\hline
		L1\_m8             & 0 & 0 & thermal & 1.0 & $3600^3$ & $2000^3$ & $1.34\times 10^8$ & $7.06\times 10^8$    & 11.2  & 2.85 & D3A \\
		L1\_m9             & 0 & 0 & thermal & 1.0 & $1800^3$ & $1000^3$ & $1.07\times 10^9$ & $5.65\times 10^9$    & 22.3  & 5.70 & D3A \\
		L1\_m10              & 0 & 0 & thermal & 1.0 & $900^3$ & $500^3$ & $8.56\times 10^9$ & $4.52\times 10^{10}$ & 44.6 & 11.40 & D3A \\
		L2p8\_m9           & 0 & 0 & thermal &2.8& $5040^3$ & $2800^3$ & $1.07\times 10^9$ & $5.65\times 10^9$    & 22.3  & 5.70 & D3A \\
		fgas$+2\sigma$  & 0 & $+2$ & thermal & 1.0 & $1800^3$ & $1000^3$ & $1.07\times 10^9$ & $5.65\times 10^9$ & 22.3  & 5.70 & D3A \\
		fgas$-2\sigma$  & 0 & $-2$ & thermal & 1.0 & $1800^3$ & $1000^3$ & $1.07\times 10^9$ & $5.65\times 10^9$    & 22.3  & 5.70 & D3A \\
		fgas$-4\sigma$  & 0 & $-4$ & thermal & 1.0 & $1800^3$ & $1000^3$ & $1.07\times 10^9$ & $5.65\times 10^9$    & 22.3  & 5.70 & D3A \\
		fgas$-8\sigma$  & 0 & $-8$ & thermal & 1.0 & $1800^3$ & $1000^3$ & $1.07\times 10^9$ & $5.65\times 10^9$    & 22.3  & 5.70 & D3A \\
		M*$-\sigma$ & $-1$ & 0 & thermal & 1.0 & $1800^3$ & $1000^3$ & $1.07\times 10^9$ & $5.65\times 10^9$    & 22.3  & 5.70 & D3A \\
		M*$-\sigma$\_fgas$-4\sigma$     & $-1$ & $-4$ & thermal & 1.0 & $1800^3$ & $1000^3$ & $1.07\times 10^9$ & $5.65\times 10^9$    & 22.3  & 5.70 & D3A \\
		Jet             & 0 & 0 & jets & 1.0 & $1800^3$ & $1000^3$ & $1.07\times 10^9$ & $5.65\times 10^9$    & 22.3  & 5.70 & D3A \\
		Jet\_fgas$-4\sigma$  & 0 & $-4$ & jets & 1.0 & $1800^3$ & $1000^3$ & $1.07\times 10^9$ & $5.65\times 10^9$    & 22.3  & 5.70 & D3A \\
		Planck          & 0 & 0 & thermal & 1.0 & $1800^3$ & $1000^3$  & $1.07\times 10^9$ & $5.72\times 10^9$    & 22.3  & 5.70 & Planck\\
		PlanckNu0p24Var & 0 & 0 & thermal & 1.0 & $1800^3$ & $1000^3$ & $1.06\times 10^9$ & $5.67\times 10^9$    & 22.3  & 5.70 & PlanckNu0p24Var \\
		PlanckNu0p24Fix & 0 & 0 & thermal & 1.0 & $1800^3$ & $1000^3$ & $1.07\times 10^9$ & $5.62\times 10^9$    & 22.3  & 5.70 & PlanckNu0p24Fix \\
		LS8             & 0 & 0 & thermal & 1.0 & $1800^3$ & $1000^3$ & $1.07\times 10^9$ & $5.65\times 10^9$    & 22.3  & 5.70 & LS8 \\
		\hline
	\end{tabular}
\end{table*}

\begin{table*}
	\centering
	\caption{Gravity-only simulations. The columns list the simulation identifier; the comoving box side length, $L$; the number of CDM particles, $N_\text{CDM}$; the number of neutrino particles, $N_\nu$; the mean CDM particle mass, $m_\text{CDM}$; the Plummer-equivalent comoving gravitational softening length, $\epsilon_\text{com}$; the maximum proper gravitational softening length, $\epsilon_\text{prop}$; and the assumed cosmology which is specified in Table~\ref{tab:cosmologies}. Simulation L1\_m9\_ip\_DMO is identical to  L1\_m9\_DMO except that the phases in the initial conditions were inverted. Note that there are no hydrodynamical counterparts for L5p6\_m10\_DMO, L11p2\_m11\_DMO, PlanckNu0p12Var\_DMO, and L1\_m9\_ip\_DMO.}
	\label{tab:DMO_simulations}
	\begin{tabular}{lrrrlrrl} 
		\hline
		Identifier & $L$ & $N_\text{CDM}$ & $N_\nu$ & $m_\text{CDM}$ & $\epsilon_\text{com}$ & $\epsilon_\text{prop}$ & Cosmology \\ 
		           & (cGpc) &&& ($\Msun$)  & (ckpc) & (pkpc) \\
		\hline

		L1\_m8\_DMO   & 1.0 & $3600^3$ & $2000^3$ & $8.40\times 10^8$    & 11.2  & 2.85 & D3A \\
		L1\_m9\_DMO   & 1.0 & $1800^3$ & $1000^3$ & $6.72\times 10^9$    & 22.3  & 5.70 & D3A \\
		L1\_m10\_DMO  & 1.0 & $900^3$ & $500^3$ & $5.38\times 10^{10}$ & 44.6 & 11.40 & D3A \\
		L2p8\_m9\_DMO & 2.8& $5040^3$ & $2800^3$ & $6.72\times 10^9$    & 22.3  & 5.70 & D3A \\		
		L5p6\_m10\_DMO & 5.6& $5040^3$ & $2800^3$ & $5.38\times 10^{10}$    & 44.6  & 11.40 & D3A \\		
		L11p2\_m11\_DMO & 11.2& $5040^3$ & $2800^3$ & $4.30\times 10^{11}$    & 89.2  & 22.80 & D3A \\	
  Planck\_DMO & 1.0 & $1800^3$ & $1000^3$ & $6.78 \times 10^9$ & 22.3  & 5.70 & Planck \\
		PlanckNu0p12Var\_DMO & 1.0 & $1800^3$ & $1000^3$ & $6.74\times 10^9$ & 22.3  & 5.70 & PlanckNu0p12Var \\
		PlanckNu0p24Var\_DMO & 1.0 & $1800^3$ & $1000^3$ & $6.73\times 10^9$ & 22.3  & 5.70 & PlanckNu0p24Var \\
		PlanckNu0p24Fix\_DMO & 1.0 & $1800^3$ & $1000^3$ & $6.68\times 10^9$ & 22.3  & 5.70 & PlanckNu0p24Fix \\
		LS8\_DMO & 1.0 & $1800^3$ & $1000^3$ & $6.72\times 10^9$ & 22.3  & 5.70 & LS8 \\
		L1\_m9\_ip\_DMO  & 1.0 & $1800^3$ & $1000^3$ & $6.72\times 10^9$ & 22.3  & 5.70 & D3A \\
		\hline
	\end{tabular}
\end{table*}

\section{The simulations} \label{sec:simulations}
The \flamingo\ suite presented here consists of the 16 hydrodynamical simulations listed in Table~\ref{tab:simulations} and the 12 gravity-only simulations listed in Table~\ref{tab:DMO_simulations}. Most of the runs use a $(1~\Gpc)^3$ cubic volume, denoted by `L1' in the simulation identifier, but one run has a volume of $(2.8~\Gpc)^3$ (`L2p8'). The hydrodynamical simulations span three resolution levels (`m10', `m9' and `m8', where the number indicates the rounded logarithm base 10 of the baryonic particle mass\footnote{We use this notation even for DMO simulations, in which case the particle mass can thus be higher than the resolution identifier suggests.}), with the mass (spatial) resolution between consecutive resolutions changing by a factor of 8 (2). Most of our runs are of intermediate resolution (`m9'), which corresponds to an (initial) mean baryonic particle mass of $\approx 1\times 10^9\,\Msun$, a mean cold dark matter particle mass of $\approx 6\times 10^9\,\Msun$, and a maximum proper gravitational softening length of 5.7~kpc. At $z>2.91$ the softening length is held constant in comoving units at 22.3~kpc. All runs use equal numbers of baryonic and dark matter particles, while the number of neutrino particles is a factor $1.8^3$ smaller. Tables~\ref{tab:simulations} and \ref{tab:DMO_simulations} list the parameter values that determine the numerical resolution for all the runs.

The values of the cosmological parameters for our fiducial model are the maximum posterior likelihood values from the Dark Energy Survey year three (DES Y3; \citealt{Abbott2022}) `3×2pt + All Ext.' $\Lambda$CDM cosmology (`D3A' in Table~\ref{tab:cosmologies}). These values assume a spatially flat universe and are based on the combination of constraints from three DES~Y3 two-point correlation functions: cosmic shear, galaxy clustering, and galaxy-galaxy lensing, with constraints from external data from baryon acoustic oscillations (BAO), redshift-space distortions, SnIa, and the Planck observations of the CMB (including CMB lensing), Big Bang nucleosynthesis, and local measurements of the Hubble constant (see \citealt{Abbott2022} for details). Our fiducial cosmology D3A uses the minimum neutrino mass allowed by neutrino oscillation experiments of $\sum m_\nu =0.06$~eV \citep{Esteban2020,Salas2021}, which is consistent with the 95 per cent confidence upper limit of 0.13~eV from DES~Y3. In this model the neutrino contribution is provided by one massive and two massless species.

\flamingo\ includes 4 intermediate-resolution hydrodynamical simulations with the fiducial calibration of the subgrid physics in $(1~\Gpc)^3$ volumes that vary the cosmological parameters. Three of the alternative cosmologies we consider are variations on \citet{Planck2020cosmopars}: their best-fitting $\Lambda$CDM model with the minimum allowed neutrino mass,  $\sum m_\nu = 0.06$~eV (`Planck'); 
a model with a high neutrino mass, $\sum m_\nu = 0.24$~eV, (allowed at 95 per cent confidence by Planck; \citealt{Planck2020cosmopars}) in which the other cosmological parameters take their corresponding best-fitting values from the Planck MCMC chains (`PlanckNu0p24Var'); and a model with the same high neutrino mass, $\sum m_\nu = 0.24$~eV, that keeps all other parameters fixed to the values of model Planck, except for $\Omega_\text{CDM}$ which was reduced in order to keep $\Omega_\text{m}$ constant (`PlanckNu0p24Fix').   Note that for the latter model we fix the primordial power spectrum amplitude, $A_s$, rather than $S_8$. All models with $\sum m_\nu = 0.24$~eV use three massive neutrino species of 0.08~eV. Finally, we include the `lensing cosmology' from \citet{Amom2023} (`LS8'). This model has a lower amplitude of the power spectrum, $S_8 = 0.766$, compared with 0.815 and 0.833 for D3A and Planck, respectively. \citet{Amom2023} show that the lensing cosmology is consistent with observations of galaxy clustering from BOSS DR12 \citep{Reid2016} and galaxy-galaxy lensing from D3A \citep{Abbott2022}, HSC Y1 \citep{Aihara2018} and KiDS-1000 \citep{Kuijken2019} over a wide range of scales, $0.15 - 60~h^{-1}\,\Mpc$, if allowances are made for theoretical uncertainties associated with baryonic feedback and assembly bias. In contrast, they find that the Planck cosmology does not fit the same data on small scales. We note that \citet{Heymans2021} showed that the LS8 model is also consistent with KiDS-1000 cosmic shear measurements. 

All the \flamingo\ cosmologies are spatially flat, $\sum_i\Omega_i = 1$, where the sum is over all components $i$, which includes dark energy, cold dark matter, baryons, massive neutrinos, massless neutrinos, and radiation. All runs assume initial baryonic mass fractions of hydrogen and helium of 0.752 and $1-0.752=0.248$, respectively\footnote{The transfer function used for the initial conditions assumes fractions of 0.754579 and 0.245421, respectively.}. 
The values of the cosmological parameters that vary between runs can be found Table~\ref{tab:cosmologies}.

\begin{table*}
	\centering
	\begin{threeparttable}[b]
	\caption{The values of the cosmological parameters used in different simulations. The columns list the prefix used to indicate the cosmology in the simulation name (note that for brevity the prefix `D3A' that indicates the fiducial cosmology is omitted from the simulation identifiers); the 
 dimensionless Hubble constant, $h$; the total matter density parameter, $\Omega_\text{m}$; the dark energy density parameter, $\Omega_\Lambda$; the
 baryonic matter density parameter, $\Omega_\text{b}$; the sum of the particle masses of the neutrino species, $\sum m_\nu c^2$; the amplitude of
 the primordial matter power spectrum, $A_\text{s}$; the power-law index of the primordial matter power spectrum, $n_\text{s}$; the amplitude of the
 initial power spectrum parametrized as the r.m.s. mass density fluctuation in spheres of radius $8~h^{-1}\,\Mpc$ extrapolated to $z=0$ using linear
 theory, $\sigma_8$; the amplitude of the initial power spectrum parametrized as $S_8\equiv \sigma_8\sqrt{\Omega_\text{m}/0.3}$; the neutrino matter density 
 parameter, $\Omega_\nu \cong \sum m_\nu c^2/(93.14~h^2\,\eV)$. Note that the values of the Hubble and density parameters are given at $z=0$. The values of the parameters that are listed in the
 last three columns have been computed from the other parameters.}
	\label{tab:cosmologies}
	\begin{tabular}{lcccccccccc} 
		\hline
		Prefix & $h$ & $\Omega_\text{m}$ & $\Omega_\Lambda$ & $\Omega_\text{b}$ & $\sum m_\nu c^2$ & $A_\text{s}$ & $n_\text{s}$ & $\sigma_8$ & $S_8$ & $\Omega_\nu$\\
		\hline
		-            & 0.681 & 0.306 & 0.694 & 0.0486 & 0.06~eV & $2.099\times 10^{-9}$ & 0.967 & 0.807 & 0.815 & $1.39\times 10^{-3}$\\
		Planck          & 0.673 & 0.316 & 0.684 & 0.0494 & 0.06~eV & $2.101\times 10^{-9}$ & 0.966 & 0.812 & 0.833 & $1.42\times 10^{-3}$\\
		PlanckNu0p12Var & 0.673 & 0.316 & 0.684 & 0.0496 & 0.12~eV & $2.113\times 10^{-9}$ & 0.967 & 0.800 & 0.821 & $2.85\times 10^{-3}$\\
		PlanckNu0p24Var & 0.662 & 0.328 & 0.672 & 0.0510 & 0.24~eV & $2.109\times 10^{-9}$ & 0.968 & 0.772 & 0.807 & $5.87\times 10^{-3}$\\
		PlanckNu0p24Fix & 0.673 & 0.316 & 0.684 & 0.0494 & 0.24~eV & $2.101\times 10^{-9}$ & 0.966 & 0.769 & 0.789 & $5.69\times 10^{-3}$\\
		LS8             & 0.682 & 0.305 & 0.695 & 0.0473 & 0.06~eV & $1.836 \times 10^{-9}$ & 0.965 & 0.760 & 0.766 & $1.39\times 10^{-3}$\\
		\hline
	\end{tabular}
	\end{threeparttable}
\end{table*}

All runs use the subgrid model described in section~\ref{sec:subgrid}. Up to four subgrid parameters, of which two relate to stellar feedback, one to BH growth and one to AGN feedback, are calibrated to observations of the present-day SMF and low-$z$ cluster gas fractions as described in section~\ref{sec:calibration} and in more detail in \citet{FlamingoCal}. In summary, eight intermediate-resolution, $(1~\Gpc)^3$ volumes vary the subgrid feedback. `Jet' in the simulation name indicates that the AGN feedback is kinetic, jet-like rather than thermal. `M*$-\sigma$' indicates that the observed stellar masses were decreased by the expected systematic error of 0.14~dex before calibration, which mainly results in somewhat stronger stellar feedback. Finally, `fgas$\pm N\sigma$' indicates that for each cluster mass bin the observed cluster gas fraction was shifted by $\pm N$ times the error (see \S\ref{sec:subgridparvalues}). Because one of the main motivations for the model variations is predicting the observational signatures of scenarios that result in larger differences between the LSS in hydrodynamical and DMO simulations, we include more models that vary the gas fractions than the SMF and more models with stronger than with weaker feedback. The values of the calibrated subgrid parameters are listed in Table~\ref{tab:parameters}. 

For each hydrodynamical simulation there is a corresponding gravity-only simulation (postfix `\_DMO' in the simulation name), which uses the same total mass-weighted CDM+baryon perturbations but eliminates the baryon-CDM isocurvature and decaying modes, while leaving the neutrino part untouched. The effective total matter (baryon+CDM) fluid is then discretised using the same number of particles as used for CDM alone (and and also for baryons alone) in the hydrodynamical simulation. This implies that the particle mass is increased by a factor of $(\Omega_\text{CDM} + \Omega_\text{b})/\Omega_\text{CDM}$ relative to the mean mass of CDM particles in the hydrodynamical simulation.

For four DMO simulations we did not run the hydrodynamical counterpart. The intermediate-resolution $(1~\Gpc)^3$ simulation PlanckNu0p12Var\_DMO has a neutrino mass of $\sum m_\nu = 0.12$~eV (allowed at 95 per cent confidence by Planck plus BAO; \citealt{Planck2020cosmopars}) and the other cosmological parameters take their corresponding best-fitting values from the Planck MCMC chains. Simulation L5p6\_m10\_DMO uses $5040^3$ particles in a $(5.6~\Gpc)^3$ box, which corresponds to low resolution ($m_\text{CDM} = 5.38\times 10^{10}\,\Msun$). Simulation L11p2\_m11\_DMO uses the same number of particles, but in a volume of $(11.2~\Gpc)^3$ ($m_\text{CDM} = 4.30\times 10^{11}\,\Msun$). Model L1\_m9\_ip\_DMO is identical to run L1\_m9\_DMO except that the phases were inverted in the initial conditions. Note that all the intermediate-resolution hydrodynamical runs that vary the subgrid physics correspond to the same DMO simulation L1\_m9\_DMO.

Our two most demanding\footnote{L2p8\_m9 and L1\_m8 took 31M and 17M core hours, respectively. These run times include the time spent creating lightcone outputs. These simulations used, respectively, 240 and 120 compute nodes with dual 64-core AMD EPYC 7H12 2.6GHz processors on the DiRAC COSMA8 system in Durham, UK.} simulations are L2p8\_m9, which uses $2\times 5040^3 + 2800^3 \approx 3\times 10^{11}$ (i.e.\ 0.3~trillion) particles, and L1\_m8, which has $2\times 3600^3 + 2000^3 \approx 1\times 10^{11}$ particles. As far as we know, the former uses more particles than any previous cosmological, hydrodynamical simulation that includes radiative cooling and that was run to $z=0$. Its number of particles exceeds that of the similar-resolution \bahamas\ simulations by more than two orders of magnitude. In Fig.~\ref{fig:simulation_sizes} we compared the resolution and box size of \flamingo\ with simulations from the literature.

Before showing some quantitative results, we will first provide a visual impression of the simulations and some of the data products. More visualisations, including videos and interactive sliders, can be found on the \flamingo\ website\footnote{\url{https://flamingo.strw.leidenuniv.nl/}}. We already illustrated the large dynamic range in Fig.~\ref{fig:zoom-in}, which zooms in on a region centered on the most massive halo in the L2p8\_m9 simulation. 

\begin{figure*}
    \centering
    \includegraphics[scale=1.07]{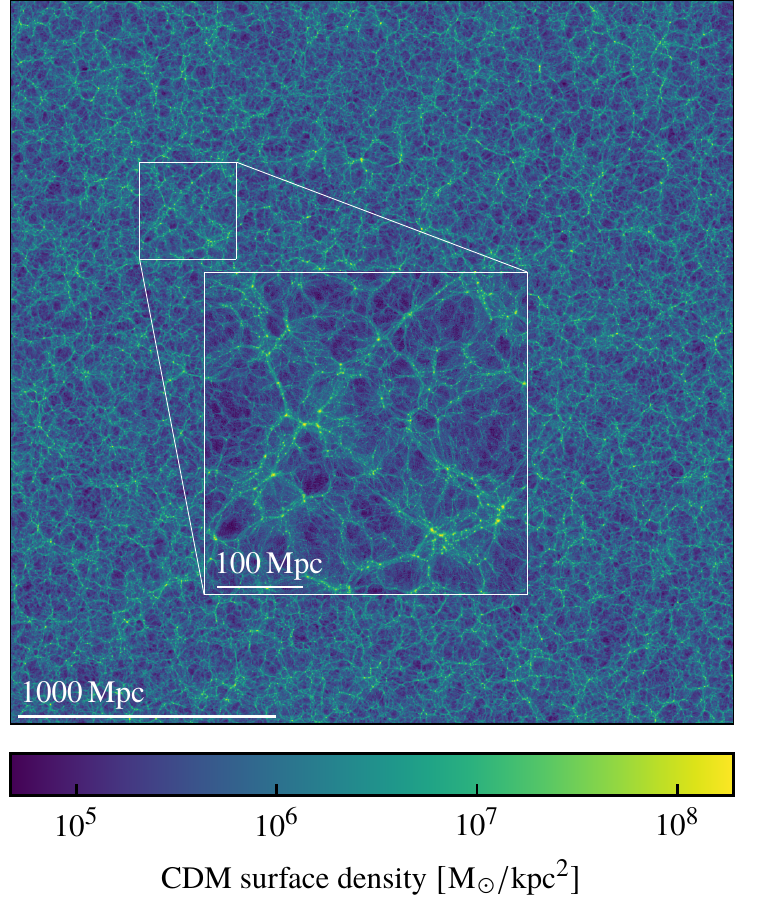}
\includegraphics[scale=1.07]{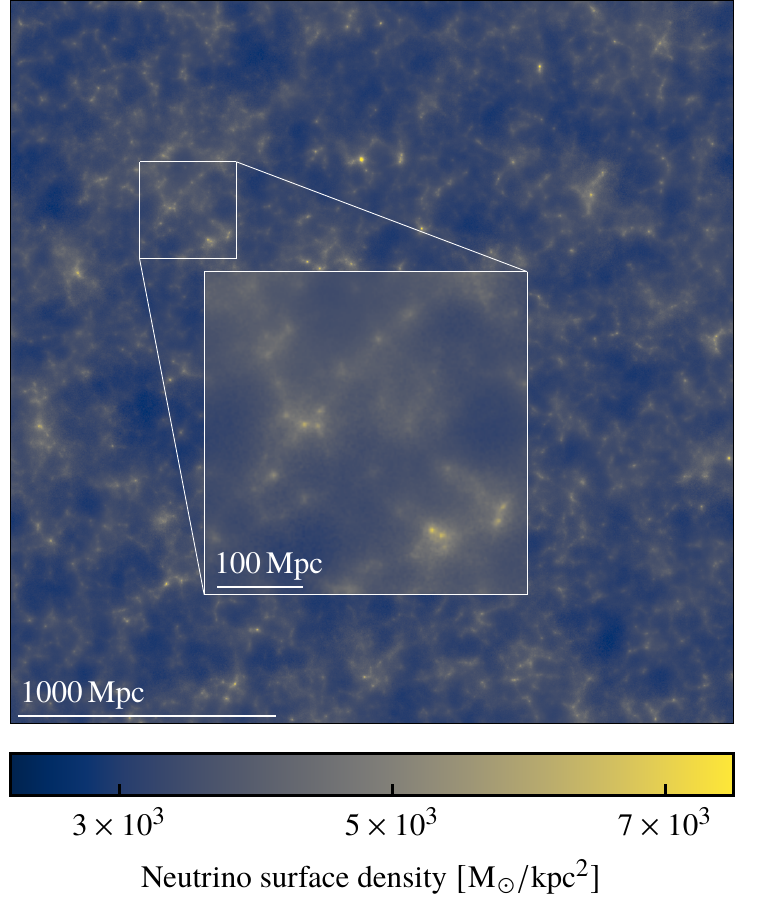}
    \caption{Comparison of the CDM (left panel) and neutrino (right panel) surface density in a 20~Mpc thick slice through the $z=0$ snapshot of the L2p8\_m9 simulation. 
    Note that the dynamic range covered by the color bar is much smaller in the right panel. On scales of $\lesssim 10^2$~Mpc the neutrino distribution is much smoother than that of the CDM.}
    \label{fig:cdm_nu_slice}
\end{figure*}

\begin{figure*}
    \centering
    \includegraphics[width=0.95\textwidth]{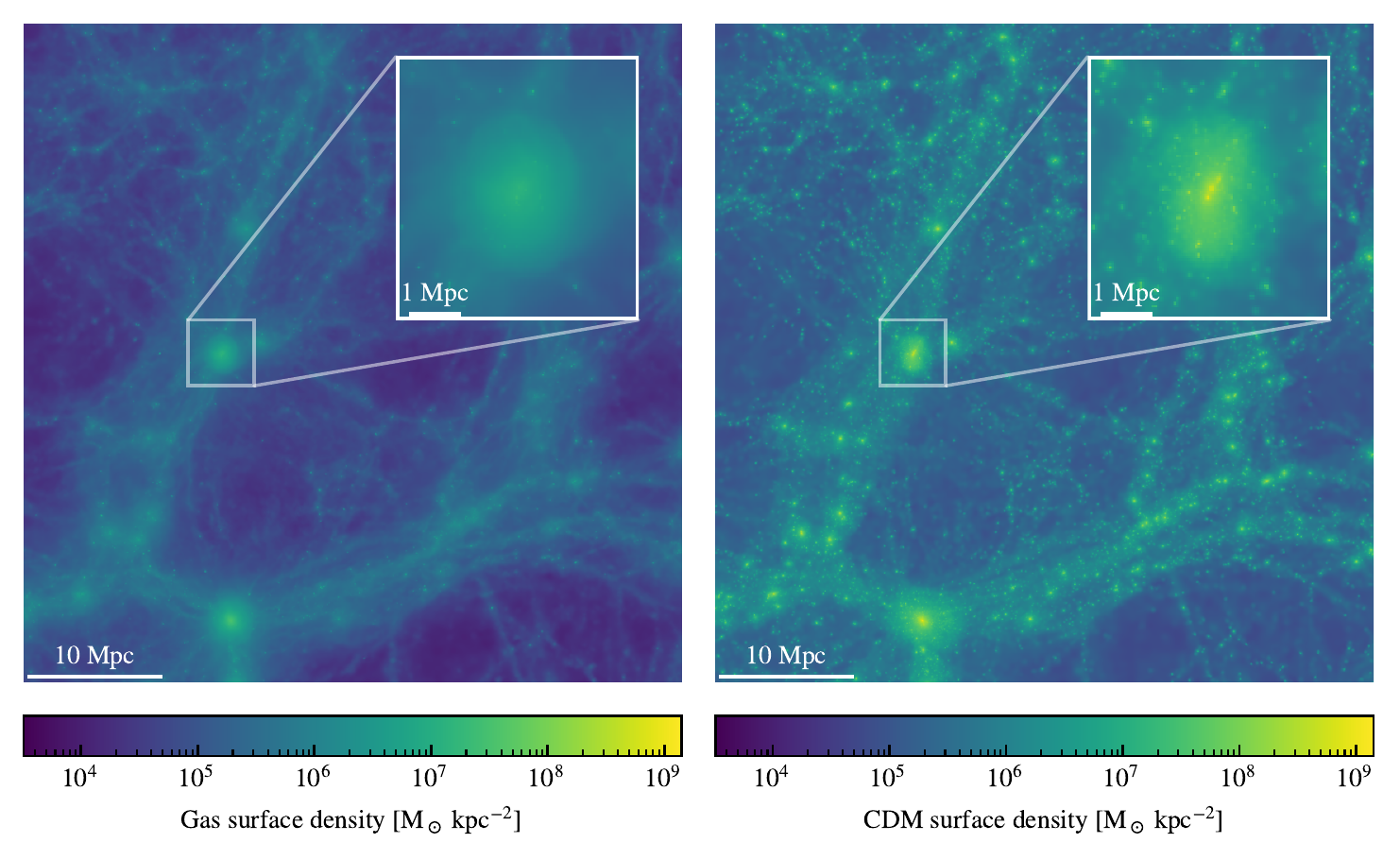}
    \caption{Comparison of the gas (left panel) and CDM (right panel) surface density in a 50x50x20~Mpc slice through the $z=0$ snapshot of the L1\_m8 simulation. The insets zoom in on a halo of total mass $M_\text{200c} = 1.26\times 10^{14}\,\Msun$. Note that the color scale is identical for the two panels. On scales of $\lesssim 1$~Mpc the gas distribution is much smoother than that of the CDM.}
    \label{fig:gas_cdm_slice}
\end{figure*}

Fig.~\ref{fig:cdm_nu_slice} compares the CDM and neutrino distributions in a 20~Mpc thick slice  centered on the most massive halo in the L2p8\_m9 simulation. While neutrinos trace the CDM on very large scales, they are distributed much more smoothly on scales $\lesssim 10^2\,$Mpc. To visualize the difference in the distributions of gas and CDM we have to zoom in. Fig.~\ref{fig:gas_cdm_slice} compares the gas and CDM distributions in a 50x50x20 Mpc slice through simulation L1\_m8, while the inset zooms in further onto a $\sim 10^{14}\,\Msun$ halo. Clearly, on scales $\lesssim 1$~Mpc the gas distribution is much smoother than that of the CDM. Together these two figures illustrate the need for the explicit inclusion of particles representing gas, neutrinos and CDM, and the large dynamic range that is required to simultaneously cover the large- and small-scale differences in the spatial distributions of these species. In addition, there are stellar particles, which trace the CDM better than is the case for gas and neutrinos, and BH particles, which are not shown here. The \flamingo\ website has an interactive slider versions of these figures as well as of other figures.

\begin{figure*}
    \centering
    \includegraphics[width=\textwidth]{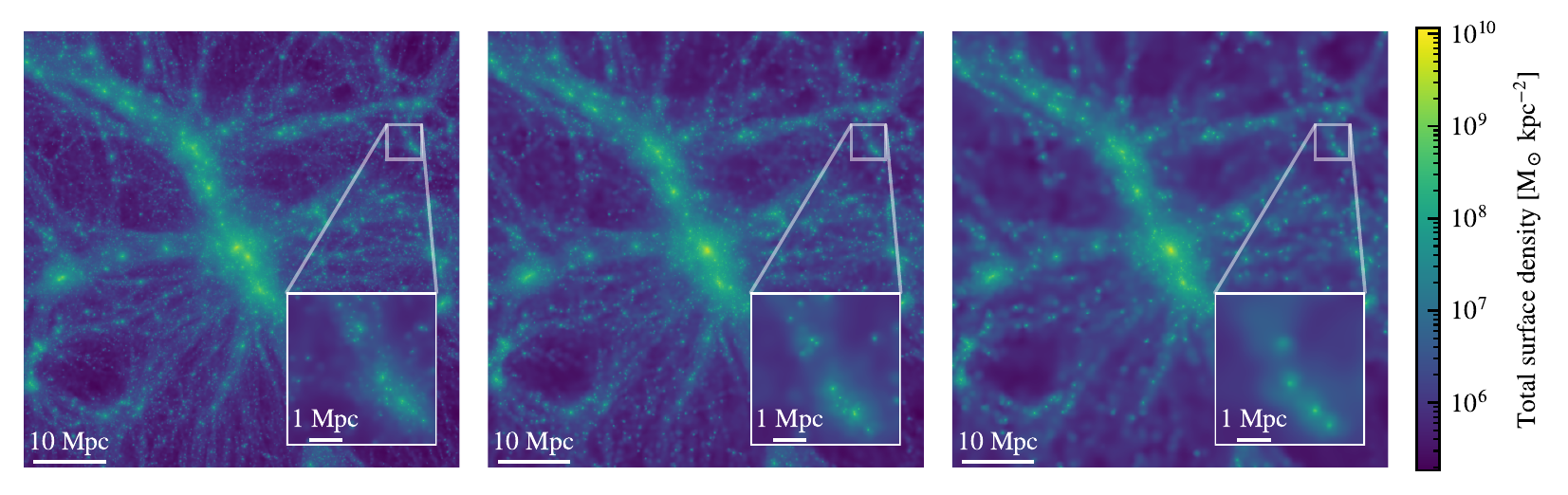}
    \caption{The left, middle and right panels show a 63x63x20 Mpc slice through the $z=0$ snapshot of the L1\_m8, L1\_m9 and L1\_m10 simulations, respectively. The images are centered on a halo of total mass $M_{200c} = 2.25\times{}10^{15}\,\Msun$. From left to right the mass resolution decreases by consecutive factors of 8. The color coding shows the total (i.e.\ cold dark matter plus neutrinos plus gas plus stars) surface density on a logarithmic scale, as indicated by the color bar on the right. The insets zoom into the indicated region and show more clearly that there is structure down to smaller scales in higher-resolution simulations.}
    \label{fig:convergence_images}
\end{figure*}

The three different simulation resolutions are compared in Fig.~\ref{fig:convergence_images}, which shows the same 63x63x20~Mpc region in, from left to right, the L1\_m8, L1\_m9, and L1\_m10 simulations. On large scales the images look identical, but the zooms shown in the insets demonstrate clearly that there is structure down to smaller scales if the simulation resolution is higher. 

Fig.~\ref{fig:lightcone_maps} illustrates the lightcone output for the case of the thermal SZE (see appendix~\ref{sec:lightcones} for a detailed description of the implementation of the lightcone output). The panels show full-sky \healpix\ maps of the dimensionless Compton $y$ parameter,
\begin{equation}
    y(z_1,z_2)  = \int_{z_1}^{z_2} \frac{k_\text{B} T}{m_\text{e} c^2} n_\text{e} \sigma_\text{T} \frac{\dd l}{\dd z}\dd z , \label{eq:Comptony}
\end{equation}
where $k_\text{B}$ is Boltzmann's constant, $m_\text{e}$ is the electron mass, $n_\text{e}$ is the free electron number density, $\dd l$ is the proper distance along the path traveled by the CMB photon, and the redshift limits are indicated below each panel. See appendix~\ref{sec:healpix_map_quantities} for our numerical implementation of the above integral. The bottom right panel of Fig.~\ref{fig:lightcone_maps} shows the contributions of all shells from redshift 0 to 5, while the panels above it show the contributions of smaller redshift intervals, as indicated. The lightcone maps output during the simulation have a redshift width of $\Delta z = 0.05$, but note that thinner shells can be created in post-processing using the particle lightcone output. The \healpix\ maps have $12\times 2^{28}\approx 3\times 10^9$ pixels, corresponding to a maximum size of about 13.46 arcsec. LSS is more clearly visible in the lowest-redshift shells, but this is largely due to the larger angular size of the structures. Zooming in would also reveal LSS in the somewhat higher redshift shells (not shown). 

Besides on-the-fly lightcone output from the perspective of a number of different observers (8 for L2p8 and 2 for the L1 simulations; see appendix~\ref{sec:lightcones}), we save 79 snapshots. They are separated by redshift intervals of $\Delta z=0.05$ below $z=3$, $\Delta z=0.25$ from $z=3$ to 5 and by constant $\Delta\log a$ at higher $z$, yielding outputs at $z=5.5$, 6.04, 6.63, 7.26, 8.70, 9.51, 10.38, and 12.26. The $z=12.26$ snapshot was only saved for the $L>1$~Gpc simulations.

\begin{figure*}
    \centering	\includegraphics[width=2\columnwidth]{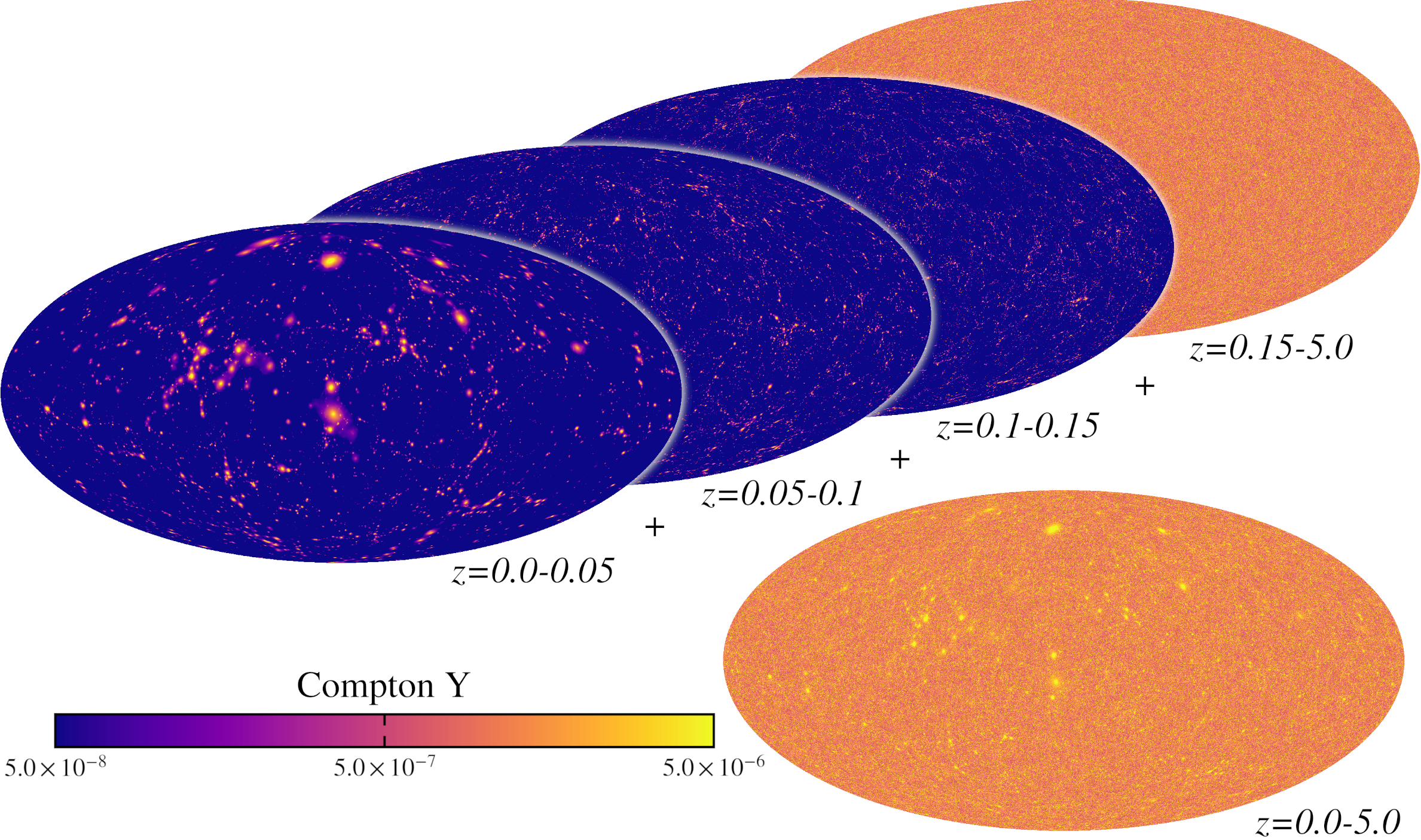}
    \caption{Full sky maps of the thermal Sunyaev-Zel'dovich effect as quantified by the Compton $y$ parameter (eq.~\ref{eq:Comptony}) for different redshift intervals. The first three maps show the shells from $z=0-0.05$, $z=0.05-0.1$ and $z=0.1-0.15$, while the fourth one in the back shows a much larger redshift range, $z=0.15-5.0$, and hence gives higher values. The map on the bottom right shows the total integrated Compton $y$ from $z=0$ to 5, i.e. the sum of the four maps shown above it.}
    \label{fig:lightcone_maps}
\end{figure*}

\section{Comparison with observations used for calibration}\label{sec:comparison_calibration_data}
In this section we compare the simulations to the observables used to calibrate the subgrid model, i.e.\ the $z=0$ SMF (\S\ref{sec:smf}) and $z\approx 0.1-0.3$ gas fractions of low-mass clusters (\S\ref{sec:fgas}). In addition, we show three closely related quantities: the stellar-mass-halo-mass (SMHM) relation (\S\ref{sec:smf}) and cluster stellar and baryon mass fractions (\S\ref{sec:fstarbar}). Because the volume of the flagship runs is 3 orders of magnitudes greater than that of the calibration runs, \flamingo\ enables probing these observables up to much higher masses than used for the calibration, which means that results for massive clusters can, in fact, be considered predictions. Finally, we compare with observations of the relation between stellar and BH mass (\S\ref{sec:BHmass}). Although we did not explicitly calibrate to this last relation, we include it here because we would have adjusted the subgrid AGN feedback efficiency, which we took from \citet{Booth2009}, if the BH masses had been in substantial disagreement with the data.

\begin{figure*}
    \centering
	\includegraphics[width=2\columnwidth]{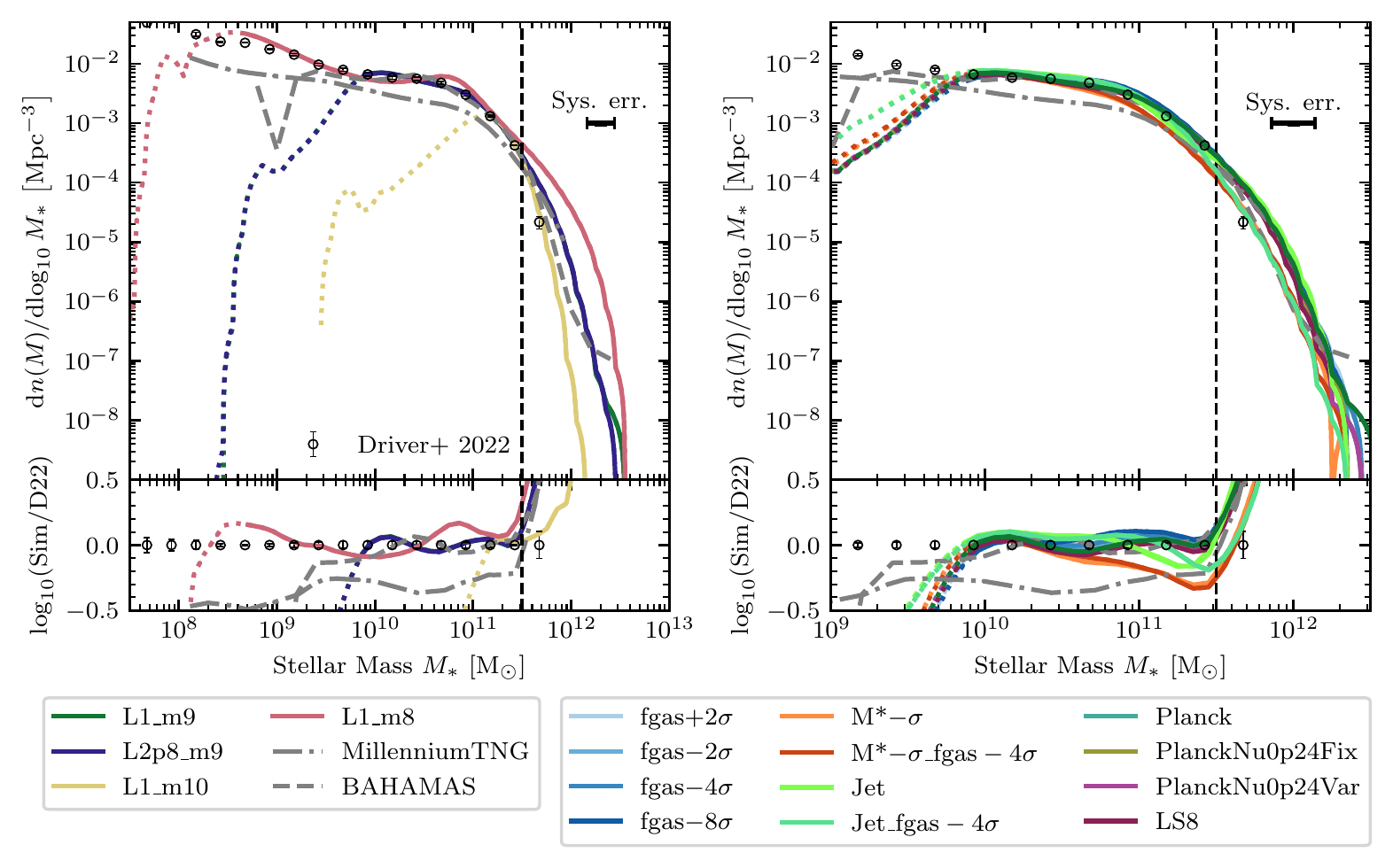}
    \caption{The $z=0$ galaxy stellar mass function for the fiducial galaxy formation model and cosmology with different resolutions and box sizes (left) and the model/cosmology variations in $(1~\Gpc)^3$ volumes at m9 resolution (right; note the different $x$-axis ranges). The simulation results are compared with observations from the GAMA survey from \citet{Driver2022}, shifted by the best-fitting systematic cosmic variance and stellar mass biases, which are however both negligible ($-0.0022$~dex and $+0.026$~dex, respectively). In the bottom panels, the simulation SMFs have been divided by a spline fit through the observations in order to reduce the dynamic range and thus facilitate the comparison with the data. The specifications of the simulations can be found in Tables~\ref{tab:simulations} and \ref{tab:cosmologies}. The parts of the SMFs below the resolution-dependent lower mass limits for the calibration are displayed using dotted line styles. The upper mass limit for the calibration is always the same and indicated by the vertical dashed line in all panels. The error bars labelled `Sys.~err' indicate the $\pm 1\sigma$ systematic errors due to cosmic variance (vertical error bar; negligible) and uncertainty in the stellar mass determination for a fixed IMF (0.14~dex; horizontal error bar). As in all figures, the simulation SMFs include the random lognormal scatter in the stellar mass that is expected to be present in the data (0.3~dex). Except for models M*$-\sigma$, which were calibrated to the observed SMF shifted to 0.14~dex lower masses, all the m10-and m9-resolution models are consistent with the data, while the differences for the m8 resolution are only $\approx 0.1$~dex. For reference, the grey dot-dashed curves show the MTNG simulation, which underestimates the number densities by $\approx 0.3$~dex, and the grey dashed curves show the \bahamas\ simulation, which agrees very well with the \flamingo\ models that have the same resolution (m9).} 
    \label{fig:SMF}
\end{figure*}

\subsection{Galaxy stellar mass function}\label{sec:smf}
The different colored curves in the top left panel of Fig.~\ref{fig:SMF} show the present-day SMFs for the fiducial subgrid model and cosmology for the three different \flamingo\ resolutions in $(1~\Gpc)^3$ volumes and, at intermediate resolution (m9), also for the $(2.8~\Gpc)^3$ run. These are the first four entries in Table~\ref{tab:simulations}. The line style switches from solid to dotted below the minimum galaxy mass used for the calibration for each of the simulations. The vertical dashed line indicates the maximum mass used for the calibration, which is $M_* = 10^{11.5}\,\Msun$ for all resolutions. The data points show the calibration data, i.e.\ the GAMA survey observations from \citet{Driver2022}. The error bar labelled `Sys.~err' in the top-right of the panels shows the expected $1\sigma$ systematic error due to cosmic variance (vertical error bar, too small to see) and systematic uncertainty in the inferred stellar mass ($\pm 0.14$~dex for a fixed IMF; \citealt{Behroozi2019}). The \flamingo\ volumes cover a large dynamic range, which makes it difficult to judge the quantitative agreement with the data. The bottom left panel therefore shows the simulation result divided by a spline fit through the observations. As discussed in \S\ref{sec:calibration_data}, we added 0.3~dex lognormal scatter to the simulated stellar masses to mimic Eddington bias due to random measurement errors. As discussed in \S\ref{sec:biases}, the observed SMF has been shifted by the best-fitting bias factors for cosmic variance and stellar mass, which are, however, far too small to make a visual difference.

The simulations are generally in good agreement with the data over the mass range used for calibration, which extends down to masses corresponding to only $\sim 10$ star particles. For the high-resolution simulation (m8) the agreement is, however, not quite as good as for the lower resolutions. At higher masses than used for the calibration, $M_* \gtrsim 10^{12}\,\Msun$, the results for the different resolutions diverge and only the low-resolution model (m10) appears to agree with the highest mass data point. However, m9 also agrees with the data if we allow for the systematic error on the stellar mass. Indeed, at fixed low number density the stellar masses still only differ at the factor of $\approx 2$ level between the different resolutions, but this corresponds to a diverging difference in number density due to the exponential decline of the SMF at high mass. Moreover, in this mass range the SMF is also sensitive to the assumed Eddington bias, with a smaller random error on the stellar mass resulting in a steeper cut off in the SMF \citep[e.g.][]{Furlong2015}. In addition, above $10^{11}\,\Msun$ the stellar masses become sensitive to the choice of aperture and the treatment of intracluster light \citep[see][]{FlamingoCal}, which implies that comparison to the data requires a more sophisticated analysis of the simulations than our 50~kpc spherical apertures. For these reasons we decided against using these high masses in the calibration\footnote{We note that figures 4 and B2 of \citet{FlamingoCal} indicate that if we had extended the calibration to higher masses, then we could have moved the cut-off of the SMF at m8 resolution to lower masses by using a larger value of $\beta_\text{BH}$ (closer to that used for m9) and a slightly smaller value of $\Delta v_\text{SN}$ without significantly deteriorating the agreement with the rest of SMF and the cluster gas fraction data.}.
The SMFs for the L2.8 (blue) and L1 (green) intermediate-resolution simulations are nearly identical and hence the green curve is invisible except at the highest masses. 

For comparison, the grey dashed and dot-dashed curves show, respectively, the \bahamas\ \citep{McCarthy2017} and MTNG \citep{Pakmor2022} simulations. While the former shows similarly good agreement with the data as \flamingo, the latter strongly underpredicts the SMF. MTNG uses 30~kpc apertures, whereas we use 50~kpc apertures. This difference in aperture is unimportant for $M_* < 10^{11}\,\Msun$, but for higher masses the SMF becomes increasingly sensitive to the aperture, with larger apertures resulting in higher masses. Had we used 30~kpc apertures, then this would have improved the agreement between \flamingo\ and observations at the high-mass end \citep{FlamingoCal}. 

The right panels of Fig.~\ref{fig:SMF} are similar to the left panels, but show the cosmology and feedback variations, which all use intermediate resolution and $L=1~\Gpc$. The SMFs of most models are nearly identical and agree similarly well with the data as the fiducial model. This confirms that re-calibration was unnecessary for the cosmology variations and that models varying only the cluster gas fractions are properly re-calibrated to the SMF. The models using jet-like AGN feedback appear to undershoot the data at $M_* \approx 2\times 10^{11}\,\Msun$, but the difference in number density can easily be accounted for by shifting the masses by the expected systematic error on the observed stellar mass. The only two models that do not fit the data are the ones that are not supposed to: the two M*$-\sigma$ variations, which were calibrated to the observed SMF after decreasing the observed masses by the systematic error of 0.14~dex. Interestingly, these models actually still fit the original, unperturbed data in the mass range where the SMF is flat, $M_* < 10^{10.5}\,\Msun$. 

\begin{figure*}
    \centering
	\includegraphics[width=2\columnwidth]{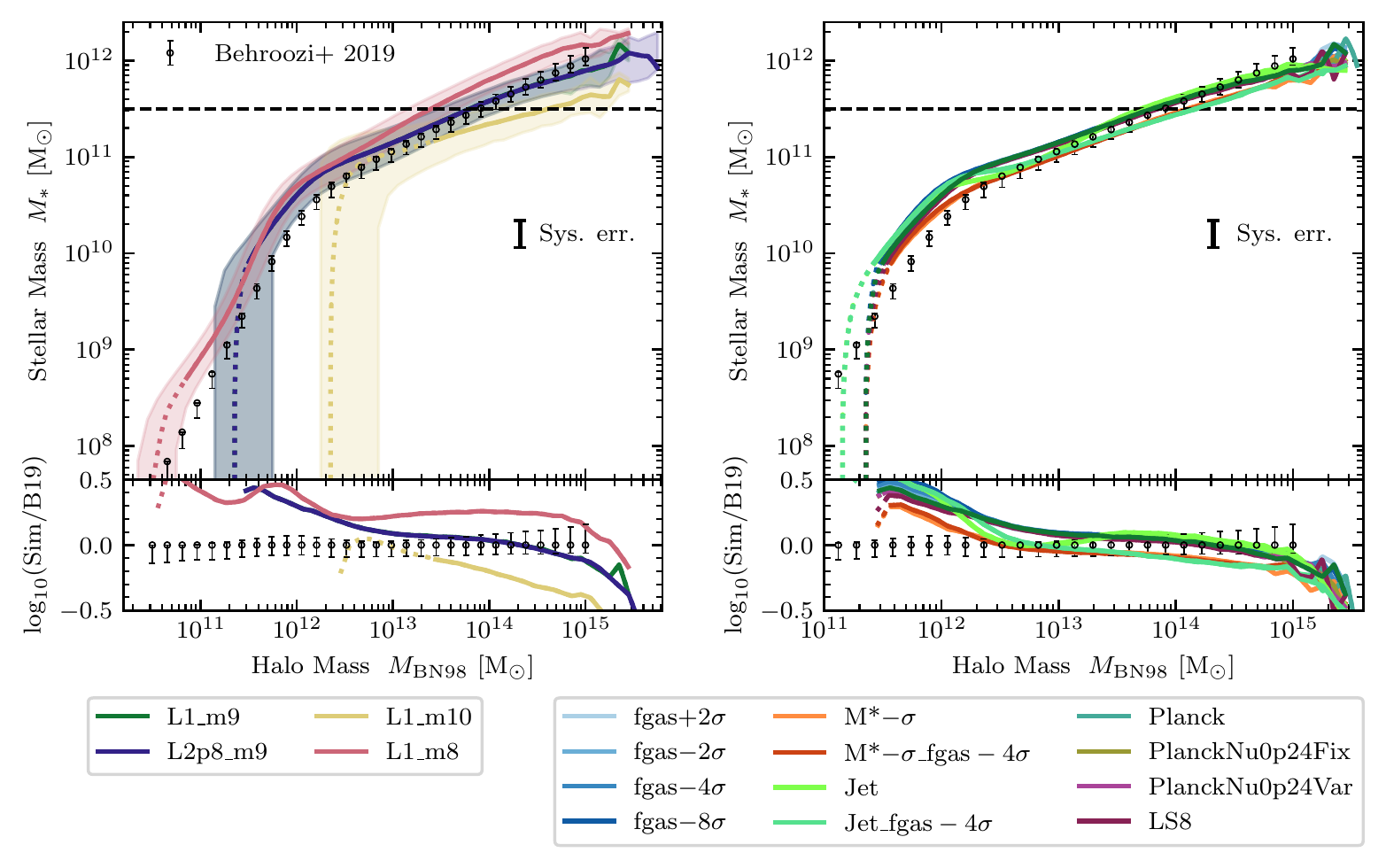}
    \caption{Median stellar mass as a function of halo mass for central galaxies at $z=0$ (solid curves in the top panels). The shaded regions (left panel only) show the 16th to 84th percentile scatter. As in Fig.~\ref{fig:SMF}, the left panels show the simulations using the fiducial galaxy formation model and the fiducial cosmology, while the right panels show the model variations (which all use the same box size and resolution as L1\_m9), as indicated in the legend, and the error bar labelled `Sys.~err' indicates the expected systematic error on the observed galaxy mass for a fixed IMF (0.14~dex). The horizontal dashed lines indicate the upper stellar mass limit for the calibration. Median curves are dotted below the resolution-dependent lower mass calibration limit. For comparison, the data points show the stellar-mass-halo-mass relation inferred by \citet{Behroozi2019} using the semi-empirical model `UniverseMachine'.  The bottom panels show ($\log_{10}$ of) the ratio of the median relations and a spline fit through the data points. For consistency with \citet{Behroozi2019}, we adopted the definition of halo mass from \citet{Bryan1998}. The sharp cut offs at low halo masses are due to resolution effects.}
    \label{fig:SMHM}
\end{figure*}

Fig.~\ref{fig:SMHM} is similar to Fig.~\ref{fig:SMF} but plots the SMHM relation for central galaxies. The curves and shaded regions indicate the median and the 16th - 84th percentile scatter, respectively. The curves become dotted below the resolution-dependent lower mass limits for the calibration while the dashed horizontal line indicates the constant upper mass limit. The sharp downturns at low halo masses are due to the limited resolutions of the simulations. Note that lower percentiles cut off at higher halo masses, which implies that for the lowest halo masses that still have realistic median stellar masses, a significant fraction of the haloes do not form any stars. The data points are from \citet{Behroozi2019}, who used their semi-empirical `UniverseMachine' model to infer the SMHM relation from observations. These data points are model-dependent and should therefore only be used to rule out extreme simulations. The m9 resolutions agree best with UniverseMachine, while m8 and m10 predict, respectively, higher and lower stellar-to-halo mass ratios for $M_* > 10^{11}\,\Msun$. At lower masses the m9 and m8 simulations both predict higher SMHM ratios than UniverseMachine down to their resolution limits. The right panel shows that for halo masses $\lesssim 10^{13}\,\Msun$ the M*$-\sigma$ runs agree better with UniverseMachine than the fiducial model does.  

\begin{figure*}
     \centering
	\includegraphics[width=2\columnwidth]{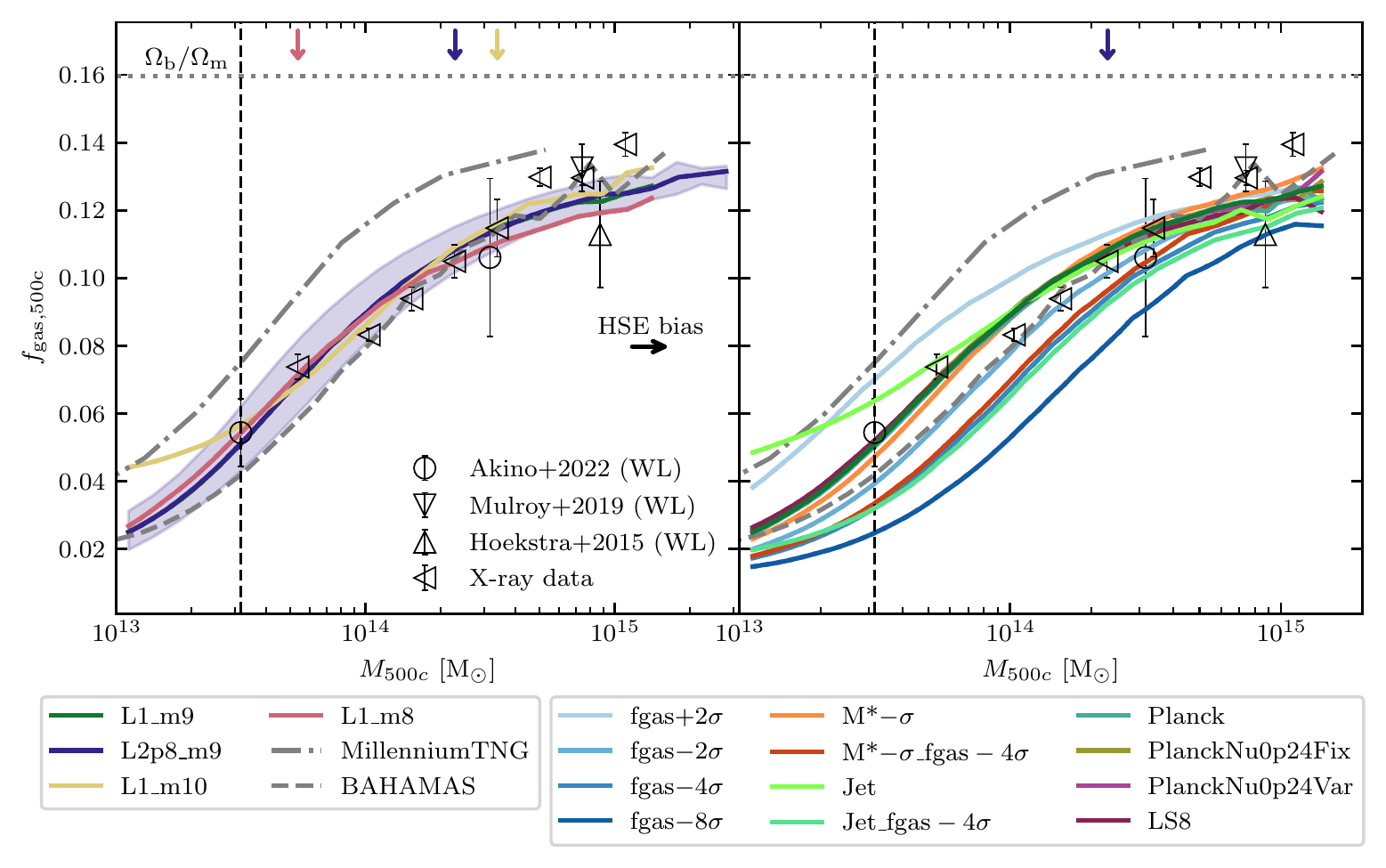}
      \caption{The median gas mass fraction within $R_\text{500c}$ as a function of halo mass ($M_\text{500c}$) for clusters at $z=0.1$. The shaded region in the left panel shows the 16th to 84th percentile scatter for model L2p8\_m9. As in Fig.~\ref{fig:SMF}, the left panel shows the simulations using the fiducial galaxy formation model and the fiducial cosmology, while the right panel shows the model variations, as indicated in the legend. The vertical dashed line indicates the lower (mass limit for the calibration. The resolution-dependent upper mass limits for the calibration are indicated by the coloured downward pointing arrows. The rightward pointing arrow labelled `HSE bias' indicates the correction for hydrostatic bias that has been applied to the observed X-ray data. For reference, the horizontal dotted lines indicate the universal baryon fraction. The simulations are compared with the data used for the calibration: the compilation of $z\approx 0.1$ X-ray data from \citet{FlamingoCal} and the $z\approx 0.3$ weak lensing (plus X-ray) data from \citet{Akino2022}, \citet{Mulroy2019}, and \citet{Hoekstra2015}. For comparison, the grey dot-dashed curves show the MTNG simulation (at $z=0$), which predicts too high gas fractions, and the grey dashed curves show the fiducial \bahamas\ simulation (at $z=0$), which is closest to model fgas$-2\sigma$.}
    \label{fig:fgas}
\end{figure*}

\subsection{Cluster gas fractions} \label{sec:fgas}
Fig.~\ref{fig:fgas} is similar to the top row of the previous figure, but shows the median mass fraction in gas as a function of halo mass for $z=0.1$ clusters of galaxies. The gas fraction, $f_\text{gas,500c}$, and halo mass, $M_\text{500c}$, are measured within $R_\text{500c}$. The dashed vertical line indicates the lower halo mass limit for the calibration, which is always $M_\text{500c} = 10^{13.5}\,\Msun$. The upper mass limits are indicated by the coloured downward pointing arrows. They depend on the resolution, because we used smaller box sizes for the calibration of higher-resolution models (see \S\ref{sec:calibration}). Note that the predictions extend to much higher ($\approx 0.5-1.5$ dex, depending on the resolution) halo masses than the maximum mass used for the calibration. The gas fractions increase monotonically with halo mass, but even at the highest halo masses they fall well short of the universal baryon fraction, $\Omega_\text{b}/\Omega_\text{m}$ (horizontal, dotted line).

The simulations are compared with $z\approx 0.1$ X-ray observations compiled from the literature by \citet{FlamingoCal} and with $z\approx 0.3$ X-ray/weak lensing observations from \citet{Akino2022, Mulroy2019, Hoekstra2015}. The error bar on each X-ray gas fraction data point reflects the error on the median of all the clusters in the mass bin, but it should be noted that the scatter is much greater than the error on the median and that the indicated errors likely underestimate the true uncertainty. The observed X-ray halo masses have been corrected for the best-fitting hydrostatic mass bias (i.e.\ they have been shifted to higher halo masses as indicated by the arrow labelled `HSE bias'). This correction brings them in line with the lensing data, as expected given that our best-fitting bias factor largely reflects the prior that we adopted and that is based on comparisons of X-ray and weak lensing masses in the literature \citep{Hoekstra2015,Eckert2016}. We caution that we have not attempted to account for possible observational selection effects.

The simulations shown in the left panel, which all use the fiducial cosmology and were calibrated to the fiducial data, are in good agreement with the observations, though the gas fractions may be slightly underestimated for high-mass ($M_\text{500c}\sim 10^{15}\,\Msun$) clusters. The fiducial \bahamas\ model (grey dashed curve) predicts gas fractions that are somewhat lower at the low-mass end, but which agree very well with \flamingo\ for $M_\text{500c}> 10^{14}\,\Msun$. In contrast, MTNG (grey dot-dashed curve) predicts gas fractions that are much higher than any of the simulations and that are inconsistent with the data.

In the right panel, where all the m9-resolution model variations are compared, the models that were calibrated to the same data generally fall nearly on top of each other, making them nearly indistinguishable. This confirms that re-calibration was unnecessary for cosmology variations and that model M*$-\sigma$, for which the target SMF was changed while leaving the gas fraction data unchanged, was properly calibrated. As expected, simulation fgas$+2\sigma$ overshoots the fiducial data, while models fgas$-2\sigma$, fgas$-4\sigma$, and fgas$-8\sigma$ undershoot the fiducial data by increasing amounts. Each of these models is, in fact, in good agreement with its own calibration targets. 

Beyond the upper mass limit for the calibration of the m9 resolution, $M_\text{500c} = 10^{14.36}\,\Msun$, the different models converge towards similar values, indicating that the results become less sensitive to the strength of the AGN feedback. Another result of note is that the Jet models yield shallower relations between gas fraction and halo mass. At low mass, below the minimum mass for calibration, model Jet predicts substantially higher gas fractions than the fiducial model. At high mass, above the calibration limit, model Jet\_fgas$-4\sigma$ predicts slightly lower gas fractions than model fgas$-4\sigma$.

\begin{figure*}
    \centering
	\includegraphics[width=2\columnwidth]{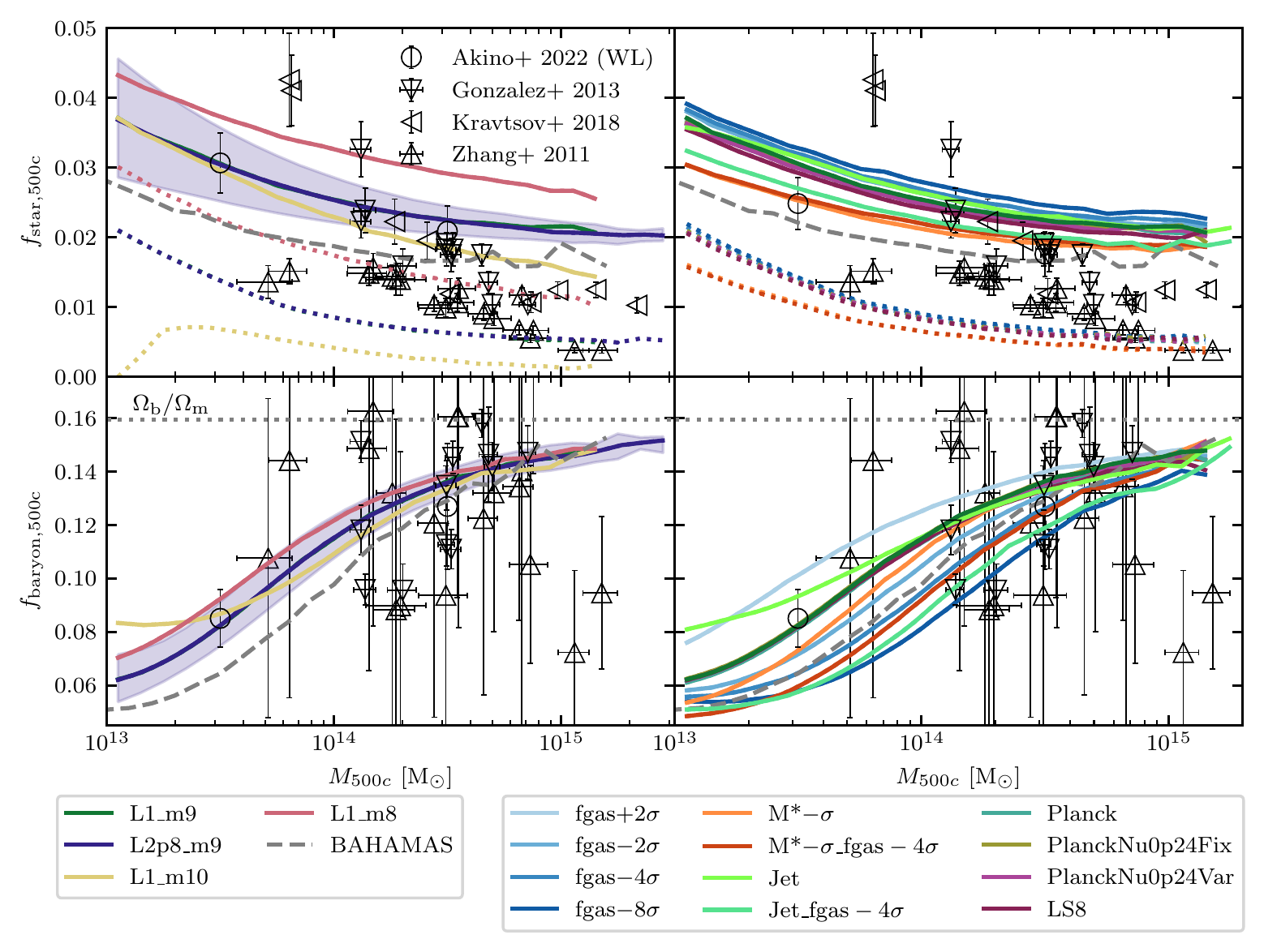}
    \caption{As Fig.~\ref{fig:fgas}, but showing the median stellar (top panels) and baryonic (bottom panels) mass fractions of $z=0.1$ clusters. The solid curves in the top panels, as well as the grey dashed curves showing the \bahamas\ simulation, show the true median total stellar mass fractions, while the dotted curves show the fractions we obtain if we only include the mass that is within our fiducial apertures of 50~kpc (summed over all galaxies). The simulation results are compared with observations from \citet{Zhang2011}, \citet{Gonzalez2013}, \citet{Kravtsov2018}, and \citet{Akino2022}, where the total and brightest cluster galaxy stellar masses from the last study have been increased by factors of 1.15 and 1.30 to account for blue galaxies and intracluster light, respectively. Observed stellar masses have been corrected to our Chabrier IMF following \citet{Chiu2018}. Our fiducial correction for hydrostatic mass bias has been applied to the observed data points. While the star fractions appear higher than observed, they are highly sensitive to the apertures within which the galaxy masses are measured.}
    \label{fig:fstarbar}
\end{figure*}

\subsection{Cluster star and baryon fractions} \label{sec:fstarbar}

Fig.~\ref{fig:fstarbar} shows the median cluster stellar mass fractions as a function of halo mass, both measured within $R_\text{500c}$. From the SMHM relation it follows that the maximum stellar mass used for calibration, $M_* = 10^{11.5}\,\Msun$, corresponds to a halo mass $M_\text{500c}\sim 10^{14}\,\Msun$. Hence, for higher halo masses the stellar masses of the central galaxies were not calibrated. However, what is plotted in Fig.~\ref{fig:fstarbar} is the total stellar mass, which is dominated by satellites and the extended, diffuse component that is traced by the intracluster light and which originates from disrupted satellite galaxies \citep[e.g.][]{Bahe2017Hydrangea,Mitchell2022GasFlowsSMHM}. Hence, the star fraction is predominantly determined by the part of the SMF that accounts for most of the mass, i.e.\ the knee, which we did calibrate to, though the contribution of the central galaxy is not negligible.

The simulations predict star fractions that decrease with halo mass, which is consistent with the observed trend. However, the predicted stellar masses are on the high side for massive clusters. This discrepancy worsens with increasing resolution, which is consistent with the increase in the stellar-to-halo mass ratio with resolution (Fig.~\ref{fig:SMHM}). The low-resolution model agrees well with the data, but this may be for the wrong reason because its SMF already cuts off at $M_*\sim 10^{11}\,\Msun$ due to resolution effects. 

However, it is unclear whether it is the total stellar mass fraction that should be compared with the data. The extended low surface brightness stellar emission around galaxies, including the intracluster light, is difficult to detect. Indeed, that is why we measure the stellar mass in apertures of 50~kpc when we calibrate to the observed SMF. As shown by the dotted curves, including only the stellar mass within the 50~kpc apertures (summed over all member galaxies) results in much lower stellar mass fractions, where the relative difference increases with halo mass. The total and the aperture-limited stellar mass fractions bracket the observed values. We also note that the offsets between different data sets exceed the scatter within a given data set, which suggests that the systematic errors are large compared with both the quoted errors and the intrinsic scatter. Nevertheless, it should be kept in mind that \flamingo\ may overestimate the stellar masses of massive clusters.

The right panel shows that the M*$-\sigma$ models, which were calibrated to the SMF after shifting it to 0.14~dex lower stellar masses, yield lower stellar mass fractions, similar to those in the \bahamas\ simulation (grey, dashed curve). The same holds for Jet\_fgas$-4\sigma$, which Fig.~\ref{fig:SMHM} shows gives similar stellar masses as the M*$-\sigma$ models for the centrals in haloes of mass $\gtrsim 10^{13}\,\Msun$.  

The bottom row of Fig.~\ref{fig:fstarbar} shows the median baryon mass fraction within $R_\text{500c}$ as a function of $M_\text{500c}$, where the baryon fraction is just the sum of the gas and the true total stellar mass fractions that we discussed before. 
The baryon fractions increase with halo mass and at the high-mass end the different models converge to values slightly below the universal baryon fraction. There is a large amount of scatter in the data, much more than is predicted by the simulation, which may indicate that the observational scatter is dominated by measurement or modeling errors. The simulations are in good agreement with the data, with the exception of the $M_\text{500c}\sim 10^{15}\,\Msun$ data from \citet{Zhang2011}. However, as can be seen from Fig.~\ref{fig:fgas}, the baryon fractions measured by \citet{Zhang2011} are much lower, and hence inconsistent with the gas fractions from a variety of studies targeting larger samples of similarly high cluster masses. The \bahamas\ simulation shows a similar trend as \flamingo, but yields slightly lower baryon fractions for $M_\text{500c}\lesssim 10^{14}\,\Msun$. 

\begin{figure*}
    \centering
	\includegraphics[width=2\columnwidth]{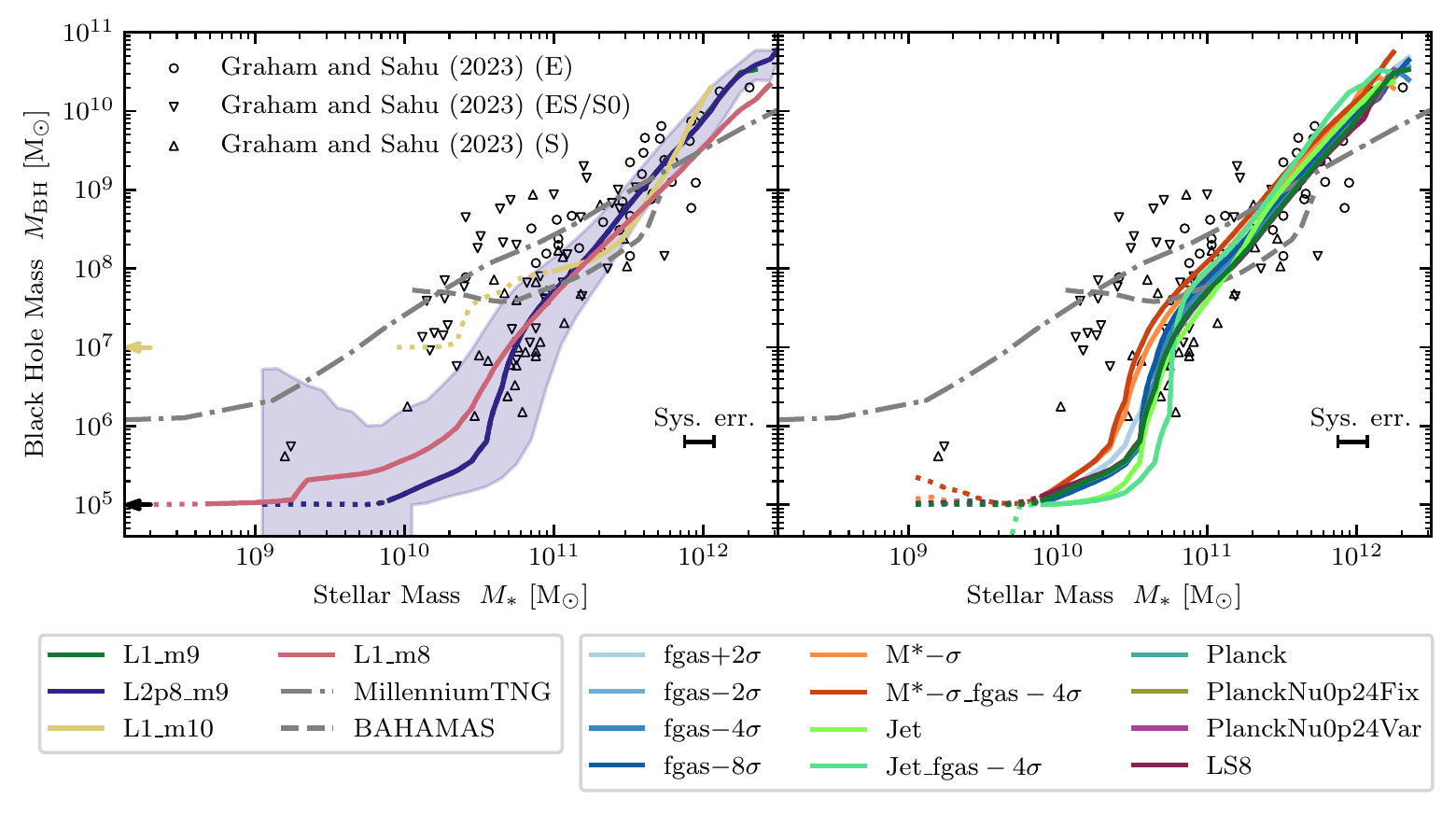}
    \caption{Median mass of the most massive black hole in the galaxy versus the galaxy's stellar mass at $z=0$. The left panel shows the simulations using the fiducial galaxy formation model and cosmology, while the right panel shows the model variations (which all use the same box size and resolution as L1\_m9), as indicated in the legend. The curves become dotted below the minimum stellar mass used for the calibration of the galaxy mass function and stop at the stellar mass corresponding to the initial mass of a single baryonic particle. The shaded region in the left panel shows the 16th to 84th percentile scatter for model L2p8\_m9. The grey dot-dashed and grey dashed curves show the MTNG and \bahamas\ simulations, respectively. The arrows in the left panel indicate the seed BH masses (which are identical for m8 and m9). The simulations are compared with data from \citet{Graham2023} for elliptical (E), spiral (S) and SO galaxies. There is good agreement with the data if we consider that most galaxies with masses $<10^{11}\,\Msun$ are disky.}
    \label{fig:bh}
\end{figure*}

\subsection{Supermassive black holes}\label{sec:BHmass}
Fig.~\ref{fig:bh} shows the median relation between the stellar mass and the mass of the most massive BH of the galaxy. The simulations are compared with observations from \citet{Graham2023} for different galaxy morphologies. At the high-mass end ($M_* \gtrsim 10^{11.5}\,\Msun$) \flamingo\ agrees with the observations for ellipticals, while at the low-mass end ($M_* \lesssim 10^{11}\,\Msun$) there is good agreement with the data for discs. In contrast, the MTNG simulations follows the data for elliptical galaxies even at masses were most galaxies are observed to be disky \citep[e.g.][]{Moffett2016}. 

At low stellar masses the curves asymptote to the BH seed mass, which is $10^5\,\Msun$ for m8 and m9, but $10^7\,\Msun$ for the m10 resolution. As the galaxy (and halo) mass increases, there comes a point when the BH starts to grow rapidly through gas accretion. As can be seen in the figure, this rapid growth phase begins at $M_*\sim 10^{10.5}\,\Msun$ and ends when the BH mass has increased to $\sim 10^8\,\Msun$ and the stellar mass has increased by only a factor of a few to values slightly less than $10^{11}\,\Msun$. This behavior is similar to that in the EAGLE simulation \citep{Schaye2015}, for which \citet{Bower2017} found that the rapid growth phase is triggered when stellar feedback becomes inefficient, which in itself is related to the appearance of a hot, gaseous halo for halo masses $\sim 10^{12}\,\Msun$. The rapid growth phase ends when the BH become sufficiently massive to regulate its own growth via AGN feedback. Beyond this point the relation remains super-linear, which is steeper than observed for elliptical galaxies, but similar to the observations including all galaxies if we account for the fact that low-mass ($M_*< 10^{11}\,\Msun$) galaxies tend to be disky. 

The normalisation of the high-mass end of the relation is sensitive to the assumed efficiency of AGN feedback \citep{Booth2009,Booth2010}. We use a feedback efficiency of $\epsilon_\text{f}=0.15$ and a radiative efficiency of $\epsilon_\text{r}=0.1$, which implies that 1.5 per cent of the accreted rest mass energy is used to heat the gas surrounding the BH. These values are identical to those used by \citet{Booth2009} and were also used in the OWLS \citep{Schaye2010} and EAGLE \citep{Schaye2015} simulations. Although we therefore did not tune the efficiency, \citet{Booth2009} motivated this choice by the desire to fit the normalisation of the BH mass - stellar mass relation, and if the chosen efficiency had resulted in a poor fit then we would have changed it. As shown by \citet{Booth2009,Booth2010}, this would have made hardly any difference for observables other than the BH mass, which can be understood if AGN feedback is self-regulating. We therefore consider the normalisation of the BH mass - galaxy mass relation to be part of the calibration. Moreover, by changing the \citet{Booth2009} boost factor for the BH accretion rate (see \S\ref{sec:BHs}), we have some control over the galaxy mass corresponding to the rapid BH growth phase, which determines the mass above which star formation is quenched.

The right panel of Fig.~\ref{fig:bh} shows that there is little difference between the different models. For models M*$-\sigma$, which were calibrated to produce 0.14~dex lower stellar masses than the fiducial model, the rapid growth phase is shifted to lower stellar masses by about this amount. This indicates that the rapid growth phase is determined mostly by the halo rather than by the stellar mass, which is consistent with the conclusions of \citet{Bower2017}.

\section{Comparison with observations not used for calibration}\label{sec:comparison_other_data}
In the previous section we compared \flamingo\ to observations used for the calibration of the subgrid physics. In this section we compare to selected observables that were not considered in the calibration: the cosmic star formation history (\S\ref{sec:sfh}), the stellar mass dependence of specific star formation rates, passive fractions, stellar metallicities, the sizes of active and passive galaxies (\S\ref{sec:galaxy_properties}), cluster scaling relations (\S\ref{sec:cluster_properties}), and the cross-correlation of thermal SZE and CMB gravitational lensing convergence maps (\S\ref{sec:lss}).

\begin{figure*}
    \centering
	\includegraphics[width=2\columnwidth]{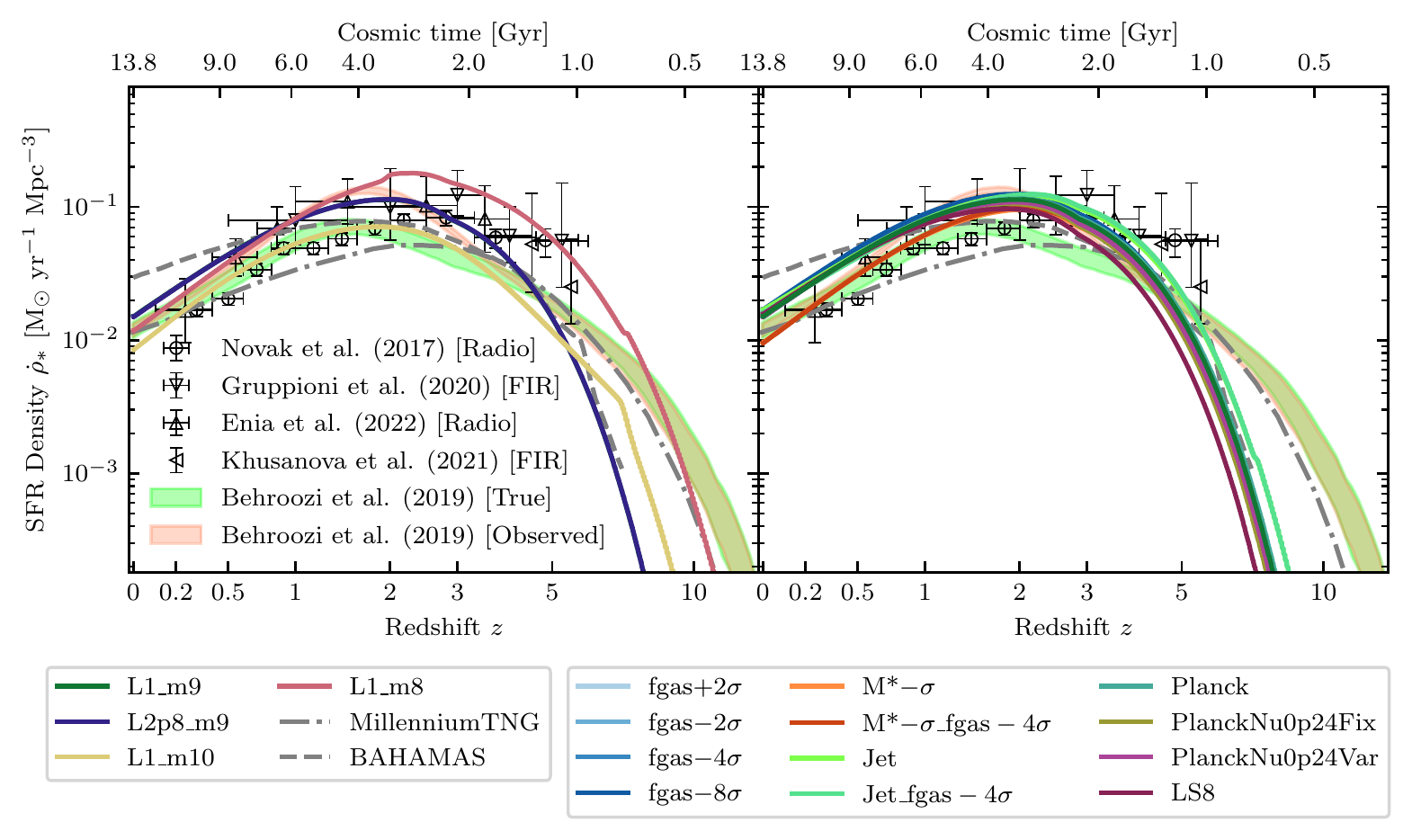}
    \caption{Cosmic SFR density as a function of redshift (bottom axis) and cosmic time (top axis, assuming the fiducial cosmology). As in Fig.~\ref{fig:SMF}, the left panel shows the simulations using the fiducial galaxy formation model and the fiducial cosmology, while the right panel shows the model variations, as indicated in the legend. The predictions are compared with pre-2016 data from the literature compiled by \citet{Behroozi2019}, who provide both the reported measurements and their best estimate of the `true' values after accounting for systematic effects, and more recent radio data from \citet{Novak2017,Enia2022} and far-infrared data from the ALPINE survey \citep{Gruppioni2020,Khusanova2021}. For reference, we also show predictions from \bahamas\ and MTNG. \flamingo\ agrees well with the data for $z<2$, but at higher redshifts the results become increasingly sensitive to the resolution. The differences between the model variations are small, though the models that were calibrated to lower SMFs (M*$-\sigma$) yield lower SFRs at $z<2$.}
    \label{fig:SFH}
\end{figure*}

\subsection{Cosmic star formation history} \label{sec:sfh}
Fig.~\ref{fig:SFH} shows the cosmic SFR density as a function of redshift. The SFR density was computed on-the-fly with high cadence by summing the instantaneous SFRs of all gas particles. As in earlier figures, the left panel shows the models that use the fiducial cosmology and galaxy formation model but which differ in terms of resolution and volume, while the right panel compares model variations in a $(1~\text{Gpc})^3$ volume at m9 resolution. The  L2p8\_m9 and L1\_m9 models lie on top of each other, which implies that the predictions of the L1 models are already converged with box size. The kink at $z=7.8$ is due to photo-heating associated with reionization (see \S\ref{sec:cooling}). Close inspection reveals a much less prominent kink at $z=2.91$, the redshift where we switch from a gravitational softening length that is fixed in comoving units to one that is fixed in physical units. At present we do not have an explanation for the small kink at $z\approx 2$ that is only visible for the high-resolution model and that is absent for smaller volume simulations of the same model (not shown).

The models are compared with a compilation of pre-2016 data from \citet{Behroozi2019} and with a number later FIR and radio surveys. For \citet{Behroozi2019} we show both the originally reported measurements (labelled `Observed') and Behroozi et al.'s best estimate of the intrinsic values after accounting for systematic errors (labelled `True'). All models are in agreement with the data below $z\approx 2$. At higher redshifts the different resolutions start to diverge and eventually fall below the observations. This is expected, we resolve star formation in the haloes that dominate the cosmic SFR up to some resolution-dependent redshift. At very high redshift the low-resolution (m10) simulation yields higher SFRs than the m9 model because it uses a much lower threshold density for star formation (see \S\ref{sec:SF}). 

The right panel shows that for $z<2$ the M*$-\sigma$ models predict lower SFRs. The other models are very close to the fiducial one\footnote{At $z>2$ the Jet models predict slightly higher star formation rate densities, which is due to the fact that the bug affecting star formation in zero metallicity gas, described in footnote~\ref{fn:SF_bug}, was fixed for the Jet models (as well as for the fiducial m8 and m10 simulations).}. \bahamas\ and MTNG predict shallower declines from $z=2$ to $z=0$ than observed and predicted by \flamingo. 

\begin{figure*}
    \centering
	\includegraphics[width=2\columnwidth]{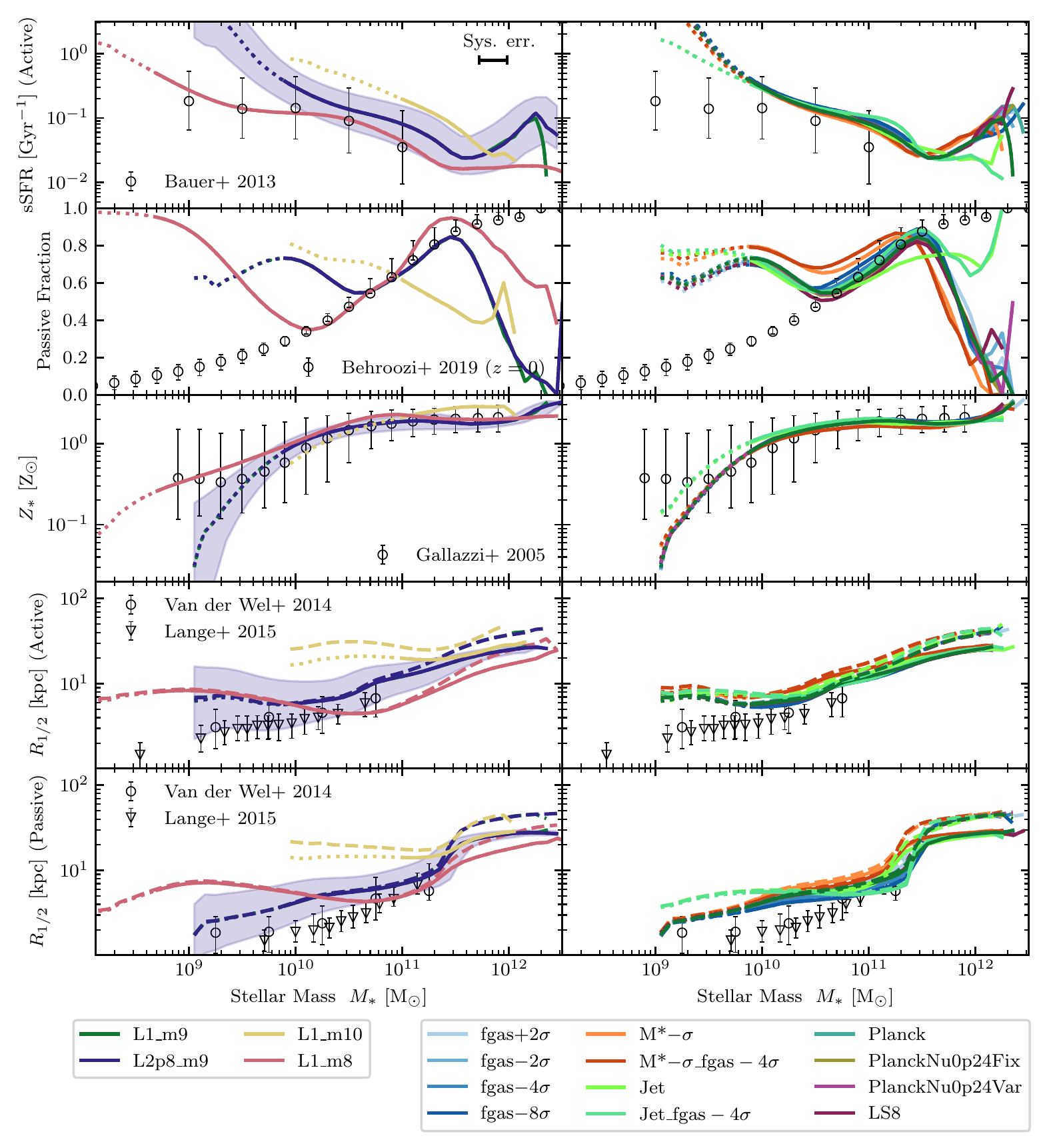}
    \caption{Similar to Fig.~\ref{fig:bh}, but showing a selection of median galaxy properties as a function of stellar mass. All properties are shown at $z=0.1$ with the exception of the passive fractions, which are at $z=0$. From top to bottom the rows show specific SFR of active galaxies, passive fraction (i.e.\ the fraction of galaxies with $\text{sSFR} < 10^{-2}\,\text{Gyr}^{-1}$), stellar metallicity, and projected stellar half-mass radii for active and passive galaxies, respectively. Solid/dotted curves are computed using all particles bound to the subhalo and inside our fiducial 3D apertures of radius 50~kpc, except for half-mass radii which are projected and computed within 2D apertures of radius 50~kpc. The dashed curves in the panels showing galaxy sizes are computed using 2D apertures of radius 100~kpc. As indicated in the legends, the simulations are compared with observed $z\approx 0.1$ sSFRs from \citet{Bauer2013}, $z\approx 0$ passive fractions compiled by \citet{Behroozi2019}, $z\approx 0.1$ stellar metallicities from \citet[][error bars indicate the scatter]{Gallazzi2005}, and $z\approx 0.1$ galaxy sizes from \citet{VanDerWel2014, Lange2015}. Many of the galaxy properties are sensitive to numerical resolution. Over the mass ranges where the simulation results are resolved, which depends on the property and the resolution, the agreement with the data is generally good. However, the passive fractions are underestimated for the highest masses ($M_*\gtrsim 10^{12}\,\Msun$) and the sizes are slightly too large.}
    \label{fig:gal_props}
\end{figure*}

\subsection{Galaxy properties}\label{sec:galaxy_properties}
Although \flamingo\ generally has too low resolution for detailed studies of galaxy structure and evolution, it can provide useful predictions for studies of relatively massive galaxies, particularly for integrated (i.e.\ spatially unresolved) properties. In order to demonstrate the uses and limitations of the simulations when it comes to galaxy properties, we show in Fig.~\ref{fig:gal_props} a number of observables as a function of stellar mass at $z=0$. From top to bottom, the different rows show the median, instantaneous specific star formation rate (sSFR) for active galaxies, i.e.\ galaxies with $\text{sSFR} \ge 10^{-11}\,\yr^{-1}$, the fraction of galaxies that are passive (i.e.\  $\text{sSFR} < 10^{-11}\,\yr^{-1}$), the stellar metallicity\footnote{We show total metal mass fraction in units of the solar abundance $Z_\odot = 0.0134$.}, and the projected stellar half-mass radii of active and passive galaxies. Unless specified otherwise, all quantities except galaxy sizes are computed using all particles bound to the subhalo and within a 3D spherical aperture of radius 50~kpc, while galaxy sizes are computed from all bound particles inside a projected 2D circular aperture of radius 50~kpc. As in earlier figures, the left and right panels compare different resolutions/box sizes and different models, respectively.

The near perfect agreement between L2p8\_m9 and L1\_m9 implies that the galaxy properties are converged with the simulation box size, except at $M_* > 10^{12}\,\Msun$, where the L1\_m9 simulation suffers from small number statistics. Convergence with the numerical resolution is less good, except at high mass. The curves are dotted below the mass limit for the calibration of the SMF, but it is clear that for some properties, e.g.\ passive fractions and sizes, resolution effects kick in at higher masses than this calibration limit. This is hardly surprising given that the SMF was calibrated to masses corresponding to fewer than 10 stellar particles. The upturns of the sSFR and passive fraction towards low masses, as well as the downturn of the metallicity, are all resolution effects. 

The minimum possible nonzero SFR (see \S\ref{sec:SF}), i.e.\ the rate of a single star-forming gas particle with density equal to the star formation threshold, corresponds to a minimum possible nonzero sSFR of $0.4~\text{Gyr}^{-1} (m_\text{g}/10^9\,\Msun) (M_*/10^9\,\Msun)^{-1}$ for the star formation threshold used for the m8 and m9 resolutions. Equating this to the upper limit used to define passive galaxies, $\text{sSFR} = 0.01~\text{Gyr}^{-1}$, yields a critical stellar mass of $M_*=4\times 10^{10}\,\Msun (m_\text{g}/10^9\,\Msun)$. For m10 this is an overestimate, because it uses a much lower star formation threshold. Galaxies with masses lower than this critical value will be active even if they have only a single star-forming gas particle. Galaxies with much lower masses can either have zero sSFR and thus be passive or have a sSFR (far) above the main sequence for star forming galaxies. Hence, to form the right amount of stars in their haloes, such galaxies must be passive most of the time. Therefore, for galaxies with masses below the critical value we expect both the sSFRs of active galaxies and the passive fractions to be too high for purely numerical reasons and this is indeed what we find.  

Resolution effects are also visible for galaxy half-mass radii (solid curves that become dotted below the lower mass limit for the calibration of the SMF). The size-mass relation flattens when the size drops below about three gravitational softening lengths. This flattening is likely caused by spurious collisional heating by dark matter particles rather than by the softening itself, which actually reduces the collisional heating \citep[][]{Ludlow2021,Ludlow2023}. At high mass there is reasonable agreement between the sizes for the different resolutions, though the high-resolution m8 model, which is closest to the data, predicts somewhat smaller values. At the high-mass end the definition of size is however sensitive to the treatment of the extended, diffuse component, as can be seen by comparing to the dashed lines, which show the half-mass radii computed by including all bound particles within 100~kpc circular apertures. Note that at the high-mass end the size-mass relation asymptotes to a size equal to about half the aperture. 

Comparing with the observations, we see that in the resolved mass range (which, for properties other than metallicity, begins at higher masses than the mass limit used for calibration) there is generally good agreement with the observations for sSFR, passive fraction and metallicity. The exception is the passive fraction at $M_*\gtrsim 10^{12}\,\Msun$, where the simulations predict an increasing fraction of active galaxies with increasing mass that is inconsistent with the inference from the semi-empirical model of \citet{Behroozi2019} that we compare with. The decline of the passive fraction and the increase in the sSFR of active galaxies with mass for $M_*\gtrsim 10^{12}\,\Msun$ are less steep at high resolution (m8) than at intermediate resolution (m9). The results from small-volume test runs that we performed suggest this is probably due to the fact that in the m9 simulations the BHs are less well pinned to the halo center because we neglected to exclude the contribution of the BH to the gravitational potential for the purpose of BH repositioning (see \S\ref{sec:BHs}). Galaxy sizes are generally overestimated, except for the high-resolution simulation at high mass ($M_* \gtrsim 10^{11}\,\Msun$). 

The right panels of Fig.~\ref{fig:gal_props} show that the differences between the models is small, though the M*$-\sigma$ models, which were calibrated to a lower SMF, are indeed shifted towards lower stellar masses (i.e.\ to the left in the figure) and to lower metallicities. The only other models predicting significantly different properties are the Jet models, which for $M_* \gg 10^{11}\,\Msun$ yield lower sSFRs for active galaxies and higher passive fractions.\footnote{These differences may partly be due to the fact that the Jet models, as well as the fiducial m8 and m10 simulations, used the improved implementation of BH repositioning discussed in \S\ref{sec:BHs}.}

\begin{figure*}
    \centering
	\includegraphics[width=2\columnwidth, scale=0.8]{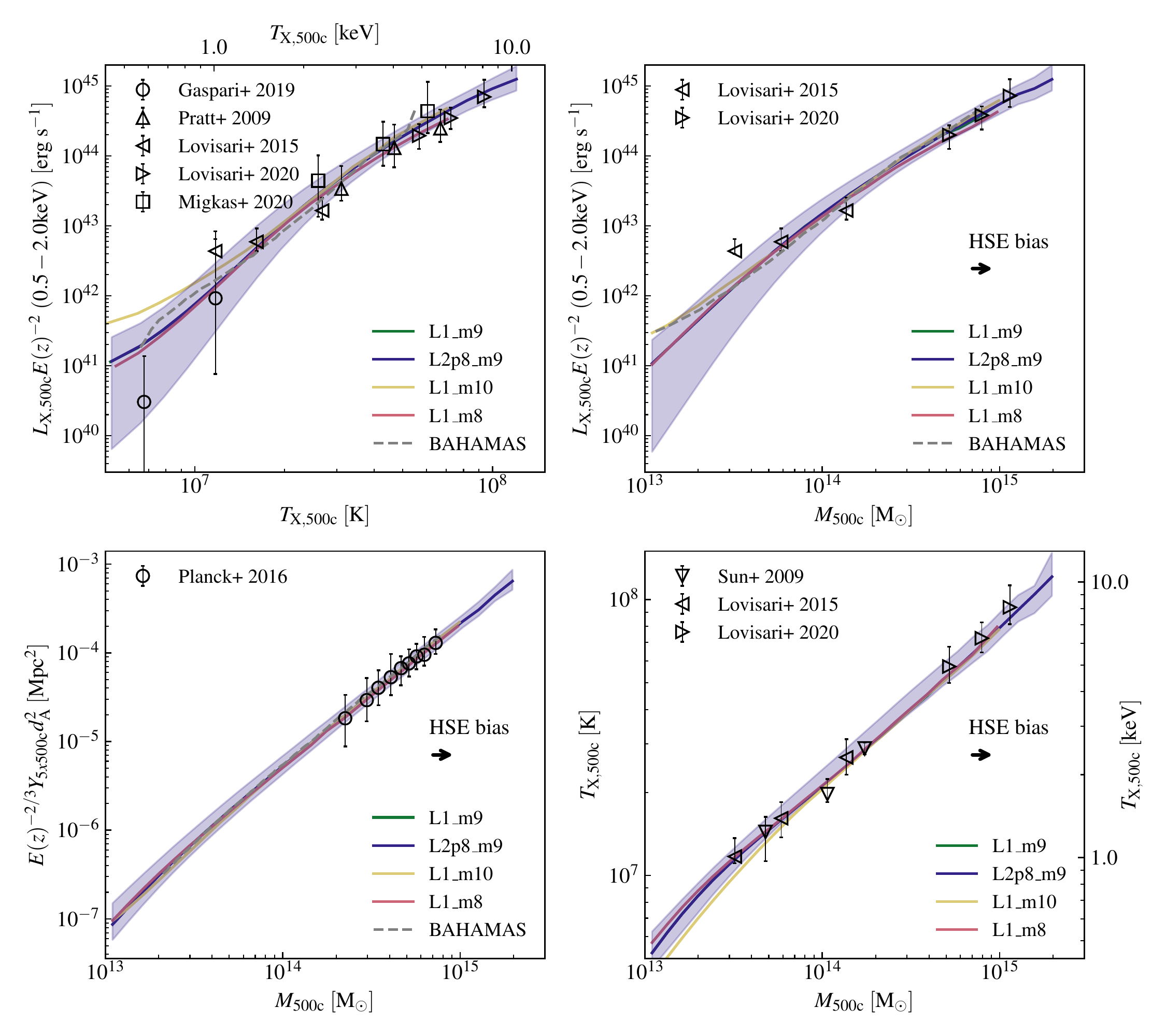}
    \caption{Cluster scaling relations at $z=0$ for the simulations using the fiducial cosmology and the fiducial galaxy formation model but different resolutions and box sizes. The solid lines are the median relations between X-ray luminosity and temperature (top left), X-ray luminosity and halo mass (top right), temperature and halo mass (bottom right) and SZE Compton $Y$ and halo mass (bottom left). Luminosities, temperature and masses are measured within $R_\text{500c}$ while Compton $Y$ is measured within $5R_\text{500c}$. All haloes with $M_\text{500c} \ge 10^{13}\,\Msun$ are included and the luminosity -- temperature relations are shown down to the median luminosity for haloes of this minimum mass. The shaded region indicates the 16th to 84th percentile scatter for L2p8\_m9. The X-ray luminosities are for the 0.5--2~keV band and the temperatures are mass-weighted. As indicated in the legends, data points show X-ray data (medians and 16th-84th percentiles) from \citet[][31 clusters at $0.08 < z < 0.15$]{Pratt2009}, \citet[][23 clusters at $z < 0.035$]{Lovisari2015}, \citet[][120 clusters at $0.059<z<0.546$]{Lovisari2020}, \citet[][85 clusters at $z < 0.04$]{Gaspari2019}, and \citet[][313 clusters at $z < 0.3$]{Migkas2020}, and SZE data for 616 $z<0.25$ clusters from \citet{Planck2016} compiled by \citet{McCarthy2017} (we multiply by $E(z)^{-2/3}$, where $E(z)\equiv H(z)/H_0$, to correct the data to $z=0$ assuming self-similar evolution). Observed X-ray luminosities have been scaled to $z=0$ and the 0.5--2~keV band using \textsc{PIMMS} \citep{Mukai1993}. The arrow labelled `HSE bias' indicates the correction for hydrostatic mass bias that has been applied to the observed X-ray data. The grey dashed lines show the results from the \bahamas\ simulation \citep{McCarthy2017}. The simulations are converged with box size and resolution (except for the low-resolution L1\_m10 at the lowest masses) and the agreement with the data is excellent.} \label{fig:cluster_scalings}
\end{figure*}

\subsection{Cluster scaling relations}\label{sec:cluster_properties}
Modeling galaxy clusters is an important goal of \flamingo. Clusters are widely used for observational cosmology, mainly to measure the halo mass function, and they are also of great interest for studies of the evolution of massive galaxies and the intracluster medium. While the simulations were calibrated to reproduce the observationally inferred cluster gas fractions inside $R_\text{500c}$, this does not guarantee that the X-ray and SZE observations are reproduced, since those can depend on the profiles of the density, temperature, and metallicity as well as the clumpiness and multiphase structure of the gas. Moreover, the observed gas fractions are model dependent and subject to selection effects. We will present detailed studies of the hot gas and its observational signatures elsewhere, but will provide a preview by comparing with some observed cluster scaling relations between integrated thermodynamic properties and halo mass. 

Fig.~\ref{fig:cluster_scalings} compares the predictions of the models using the fiducial calibration and cosmology for the median relations between $z=0$ X-ray luminosity and temperature (top left panel), X-ray luminosity and halo mass (top right panel), temperature and halo mass (bottom right panel), and integrated SZE Compton $Y$ and halo mass (bottom left panel). Here the X-ray luminosities are measured in the 0.5--2~keV band and the temperatures are mass-weighted averages over all gas with $T>10^5\,\K$. As will be detailed in Braspenning et al.\ (in preparation), the particle X-ray luminosities are computed using emissivity tables generated with \textsc{cloudy} \citep[version 17.02]{Ferland2017} and account for the individual elemental abundances of each particle. Hence, the the X-ray luminosities are consistent with the element-by-element radiative cooling rates used during the simulation, which also used \textsc{cloudy} (see \S\ref{sec:cooling}). To avoid artefacts due to the subgrid prescription for AGN feedback, particles that were subject to the injection of AGN feedback energy in the last 15~Myr and whose temperatures are between $10^{-1}\,\Delta T_\text{AGN}$ and $10^{0.3}\,\Delta T_\text{AGN}$, were excluded, but this made negligible difference. Using spectroscopic-like instead of mass-weighted temperatures gives nearly identical median relations, but results in some outliers, particularly if recently heated gas is not excluded (not shown). Luminosities, temperatures and masses are measured inside $R_\text{500c}$.

The cluster Compton $Y$ parameter is the integral of the thermal SZE, $Y = \int y \dd z \dd\Omega$, where $y$ is given by equation~(\ref{eq:Comptony}), $\dd \Omega$ is the solid angle and the integral extends from $z=0$ to the redshift of photon decoupling and over the observed angular aperture. Using $\dd \Omega = \dd l_\theta \dd l_\phi / d_\text{A}^2$, where $d_\text{A}(z)$ is the angular diameter distance of the cluster and $\dd l_\theta$ and $\dd l_\phi$ are the proper sizes in the directions of the observer's spherical polar coordinates $\theta$ and $\phi$, we have $\dd z\dd \Omega = \frac{\dd z}{\dd l}\frac{\dd V}{d_\text{A}^2}$, where $\dd V$ is the proper volume element. Hence, if we ignore contributions from structures in front and behind the cluster and limit the integral to 3D distances $<R$ from the cluster center, we obtain
\begin{equation}
    Y(<R) d_\text{A}^2(z) =  \frac{\sigma_\text{T}}{m_\text{e} c^2} \int_{<R} n_\text{e} k_\text{B} T \dd V.
\end{equation}
We exclude gas recently heated by AGN using the same criteria as mentioned above for X-ray luminosities, but this makes negligible difference. To facilitate comparison with Planck measurements we measure $Y$ within $5R_\text{500c}$. 

The convergence with simulation box size and resolution are both excellent. Even the low-resolution simulation L1\_m10 yields converged X-ray scaling relations for haloes with temperatures $T_\text{X,500c} > 2\times 10^7\,\K$ and halo masses $M_\text{500c} > 10^{13.5}\,\Msun$. For the SZE the L1\_m10 results are even converged down to the lowest masses shown, $M_\text{500c}=10^{13}\,\Msun$. The scatter, which is indicated by the shaded region for L2p8\_m9, increases towards lower masses and is largest for the luminosity. 

\begin{figure*}
	\includegraphics[width=2\columnwidth, scale=0.8]{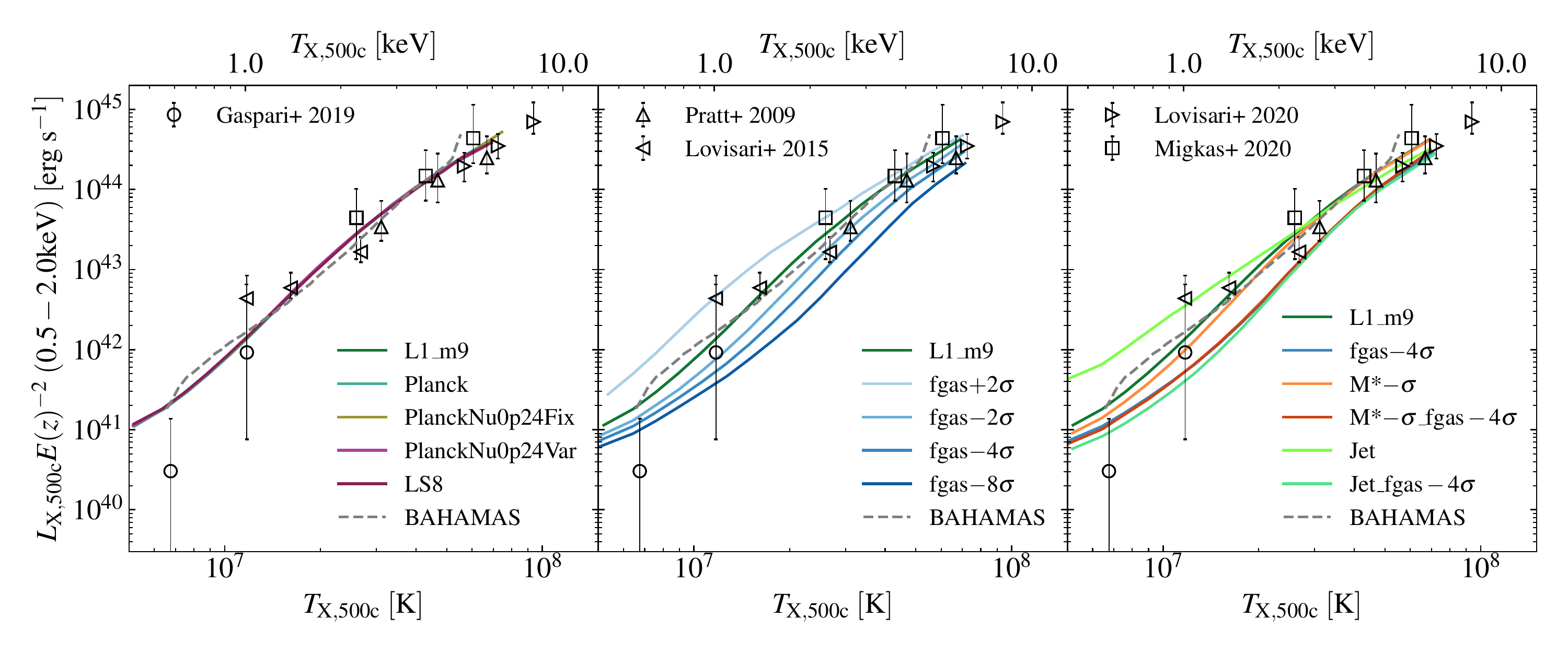}
    \caption{The median cluster X-ray luminosity -- temperature relation at $z=0$ for clusters with mass $M_\text{500c}>10^{13}\,\Msun$. The curves are plotted down to the median temperature for the minimum included halo mass. The X-ray luminosities are for the 0.5--2~keV band and the temperatures are mass-weighted; both are measured within $R_\text{500c}$. The different panels compare simulations with different cosmologies (left), simulations that were calibrated to different gas fractions (middle), and models that were calibrated to different stellar mass functions (right) and simulations that use jet-like AGN feedback (also the right panel). For reference, the fiducial simulation with the same box size and resolution as the models shown, L1\_m9, is repeated in every panel. The data points show the same observations as in the top-left panel of Fig.~\ref{fig:cluster_scalings}. The X-ray luminosity -- temperature relation is sensitive to the gas fraction, but not to the cosmology, stellar mass function or the implementation of AGN feedback (model Jet differs at low temperature, but there its gas fractions are also higher). 
    } \label{fig:L-T}
\end{figure*}

The predictions are compared with a compilation of X-ray observations from the literature (see the legend) and SZE observations of $z<0.25$ clusters from \citet{Planck2016}, where the latter were compiled by \citet{McCarthy2017}. We scaled the X-ray luminosities to the 0.5--2~keV band and $z=0$ using PIMMS\footnote{\url{https://heasarc.gsfc.nasa.gov/docs/software/tools/pimms.html}} \citep{Mukai1993}. The observationally inferred halo masses have been corrected for hydrostatic mass bias by dividing the masses by a factor 0.743 (see \S\ref{sec:biases}), as indicated by the arrows labelled `HSE bias', but the biases on other quantities (as a result of the small change in $R_\text{500c}$) are neglected. The predicted median X-ray luminosity is $\approx 0.5$~dex low compared with the lowest data point from \citet{Lovisari2015}. However, this data point represents only 7 objects, of which the lowest-luminosity one is far below the lower error bar indicating the 16th percentile. Moreover, as discussed by \citet{Lovisari2015}, their low-mass clusters are biased to high luminosities due to Malmquist bias. We conclude that the agreement with the data is excellent for all observables and over the full range spanned by the data. The same holds for the \bahamas\ simulation, which is also shown (grey dashed lines).

In Fig.~\ref{fig:L-T} we compare the X-ray luminosity -- temperature relation for simulations that vary the calibration data for our fiducial cosmology and intermediate resolution. The different cosmologies are indistinguishable (left panel). The simulations that were calibrated to different gas fractions do show large differences, with higher gas fractions yielding a higher luminosity at fixed temperature (middle panel). The fiducial model fits the data best and the most extreme model, fgas$-8\sigma$, is clearly inconsistent with the observations. Varying the SMF at fixed gas fraction has very little effect (right panel). The jet implementation of AGN feedback gives nearly identical results as the fiducial thermal implementation when the models are calibrated to the same gas fractions (right panel). Model Jet does differ from L1\_m9 for $T_\text{X,500c}< 2\times 10^7\,\K$, but this can be attributed to the fact that model Jet has higher gas fractions for $M_\text{500c} < 10^{14}\,\Msun$ (see Fig.~\ref{fig:fgas}), which corresponds to $T_\text{X,500c}\approx 2\times 10^7\,\K$ (see Fig.~\ref{fig:cluster_scalings}). 

In summary, the X-ray and SZE cluster scaling relations are converged and insensitive to the investigated variations in cosmology, to the uncertainty in the SMF and, for a fixed gas fraction, to the implementation of AGN feedback (we showed the model comparison only for the luminosity -- temperature relation, but these conclusions also hold for the other relations shown in Fig.~\ref{fig:cluster_scalings}). The X-ray luminosity -- temperature relation is sensitive to the gas fraction. The model calibrated to the fiducial (i.e.\ unperturbed) gas fractions is in excellent agreement with the data, while the models with gas fractions that differ more disagree more strongly with the observations. The luminosity -- mass relation paints a similar picture, while the other scaling relations are less sensitive to the gas fractions (not shown).

\begin{figure*}
    \includegraphics[width=\columnwidth]{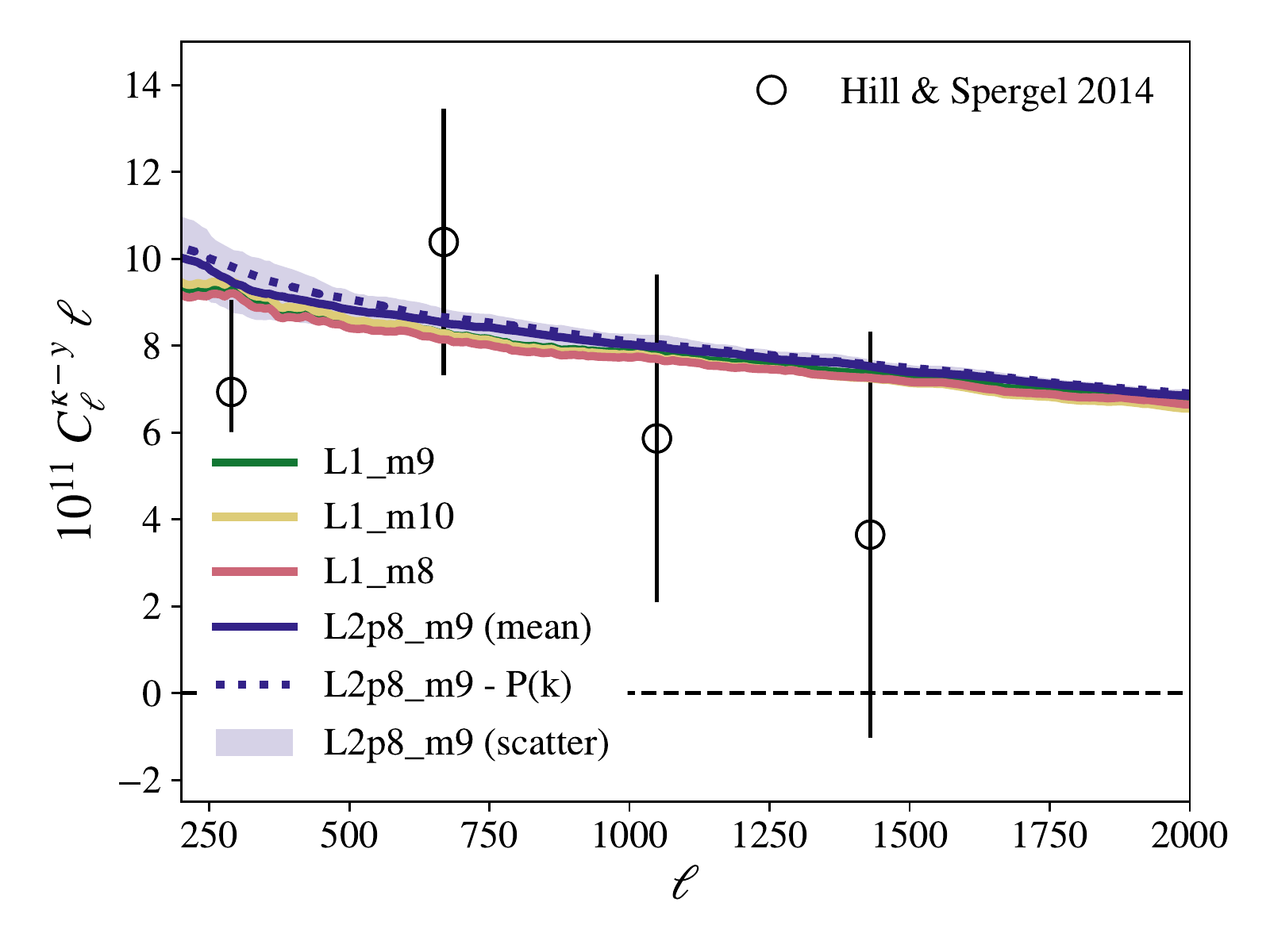}
    \includegraphics[width=\columnwidth]{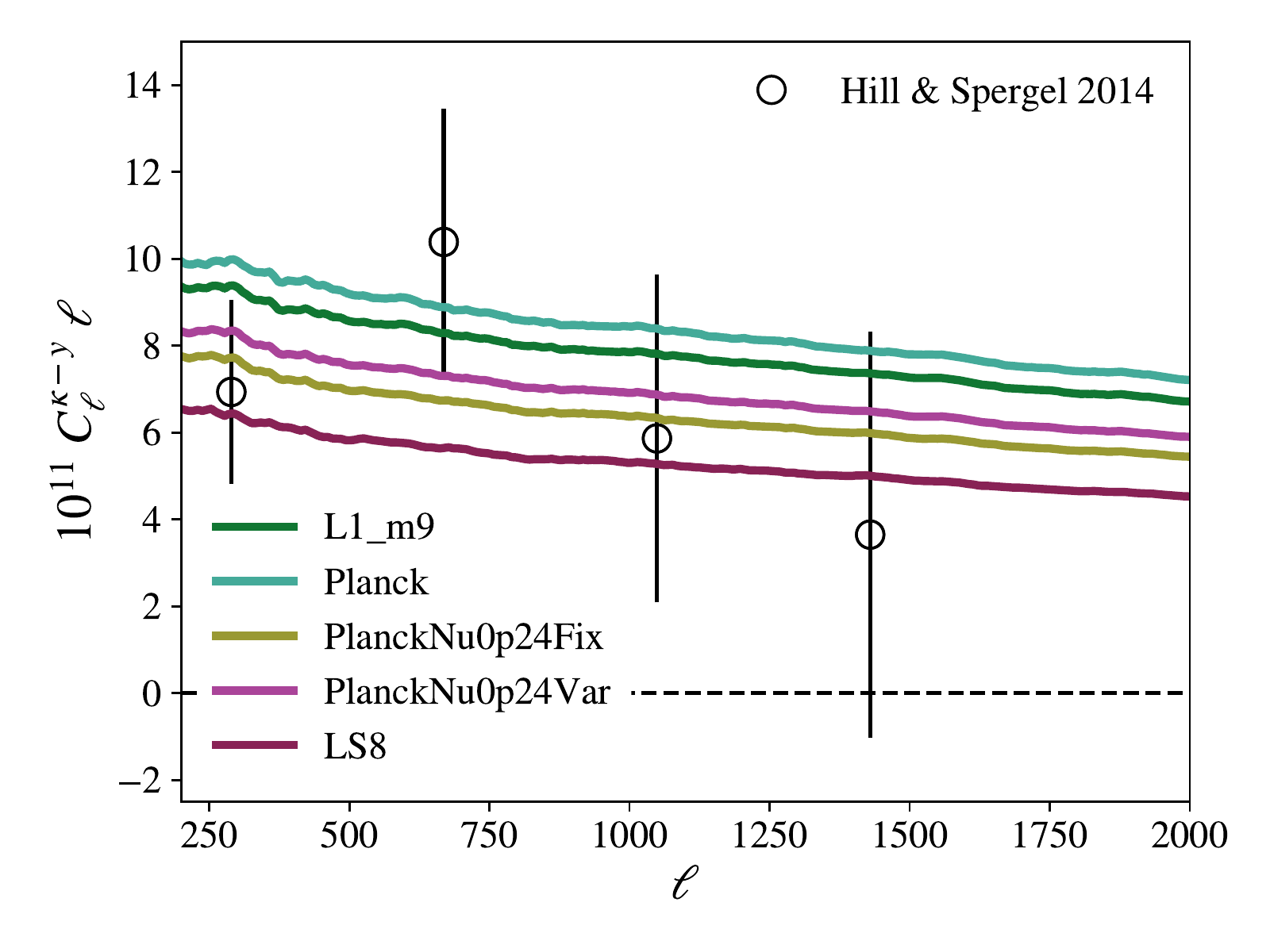}\\
    \includegraphics[width=\columnwidth]{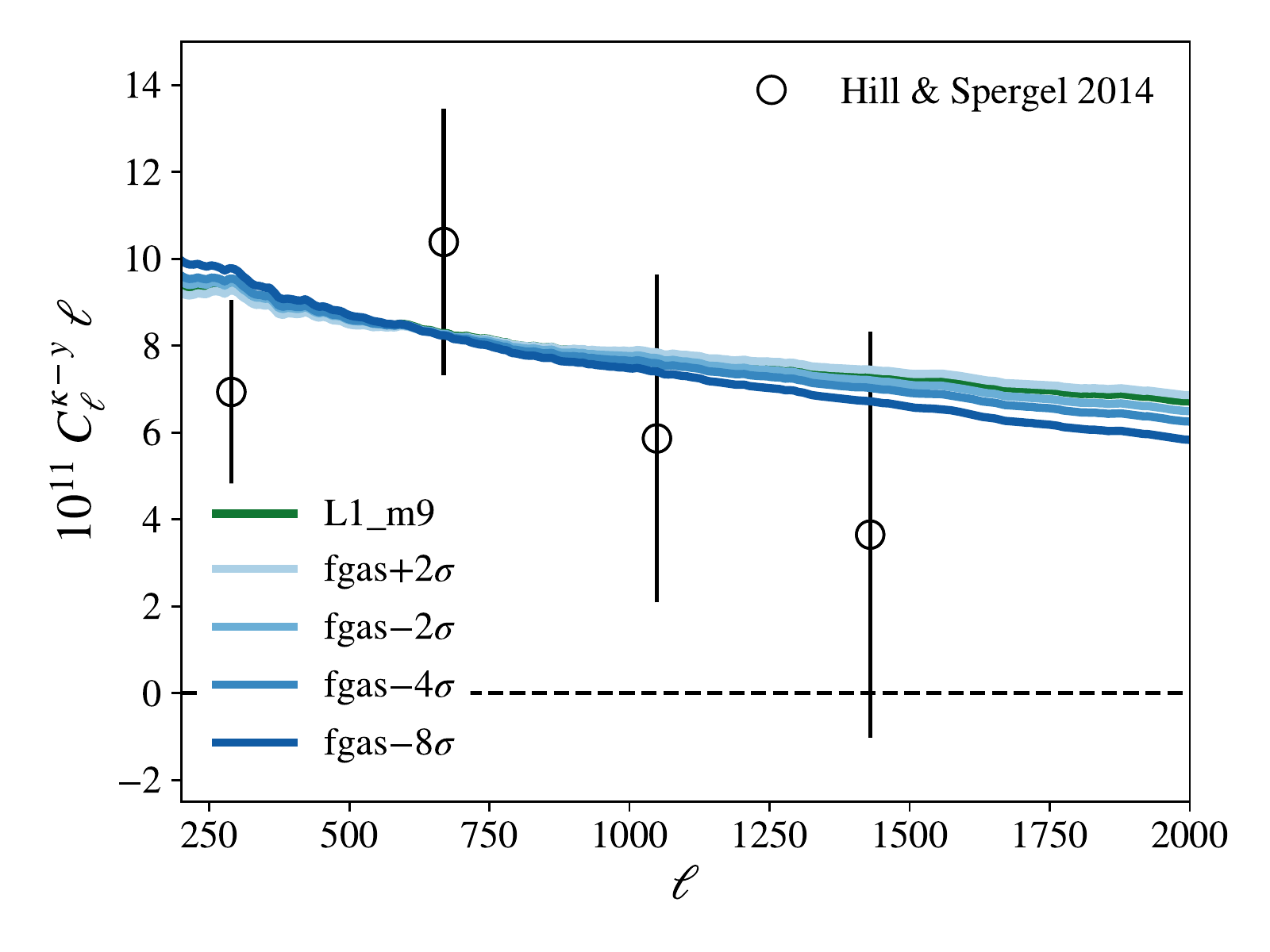}
    \includegraphics[width=\columnwidth]{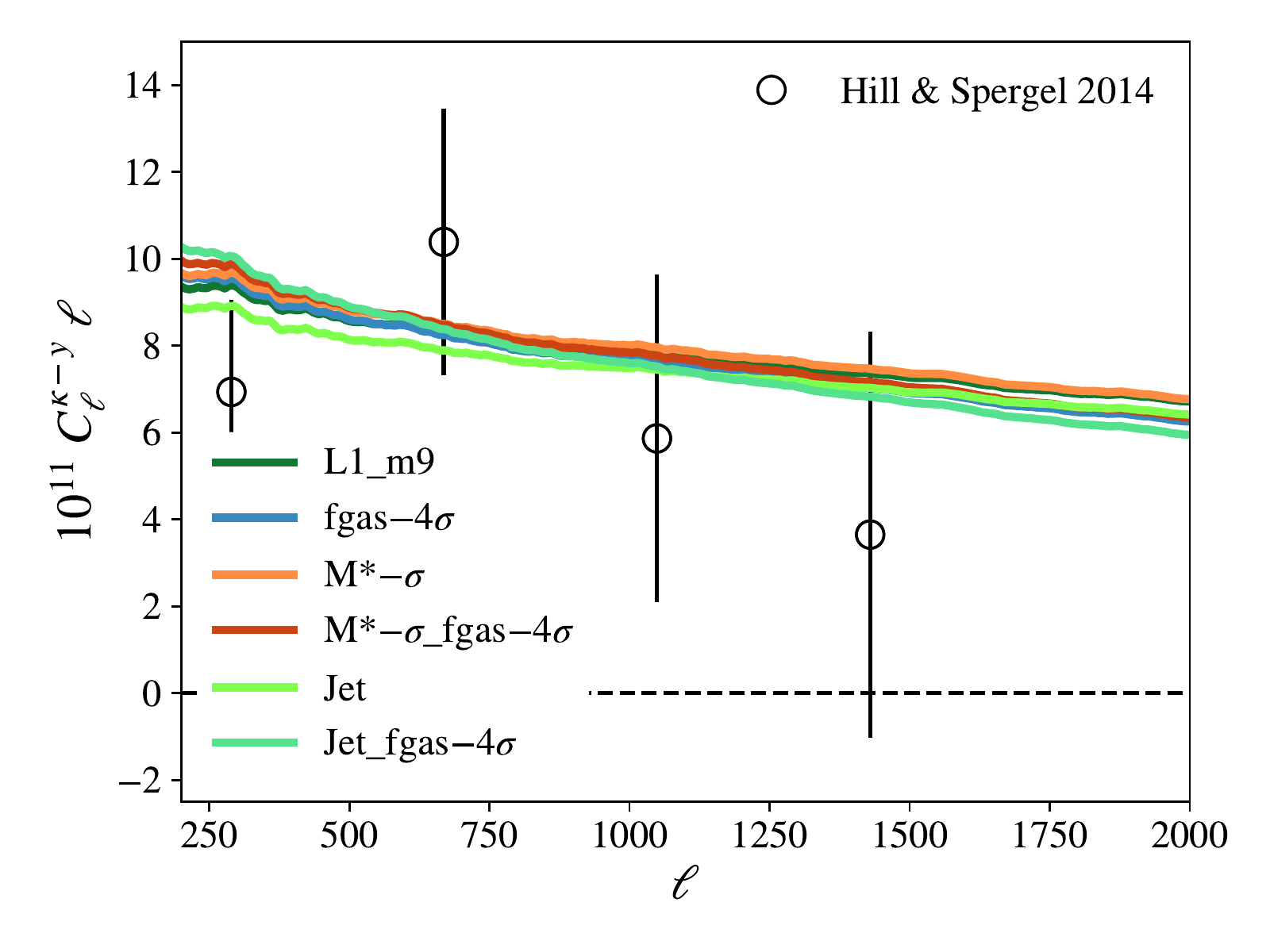}
    \caption{The thermal SZE ($y$) -- CMB lensing convergence ($\kappa$) cross-spectrum.  Thermal SZE and CMB lensing convergence maps are constructed from full-sky lightcone-based \healpix maps (downsampled to $N_{\rm side}=4096$ for the present test) produced on-the-fly while running the simulations.  Cross-spectra are computed using the \namaster pseudo-$C_\ell$ package.  In the top left panel, the shaded region represents the scatter between the 8 independent lightcones (i.e.\ 8 different observer locations in the simulation volume) for the L2p8\_m9 simulation, while the dashed curve represents the cross-spectrum derived by integrating the 3D matter--electron pressure cross-spectrum along the line of sight from 3D power spectra produced on-the-fly by employing the Limber approximation.  The $y$--$\kappa$ cross-spectrum is well converged with the numerical resolution at fixed box size, but comparison of the two intermediate-resolution simulations shows that the 1~Gpc volume misses a small amount of power relative to the 2.8~Gpc simulation (top left panel).  This particular cross-spectrum is more sensitive to variations in cosmology (top right panel) than variations in feedback implementation (bottom panels), though the effect of the latter is also clearly visible.  Comparison to the measurements of \citet{Hill2014} reveals a weak preference for a low $S_8$ cosmology or a high neutrino mass relative to the Planck cosmology.}
    \label{fig:tSZ_CMB_cross}
\end{figure*}

\subsection{Large-scale structure}\label{sec:lss}

The large volumes of the \flamingo\ simulations make them well suited for comparisons to LSS observables such as galaxy clustering, CMB lensing, the SZE, and cosmic shear. Indeed, the ability to make such comparisons was a primary consideration in the design of \flamingo.  To facilitate such comparisons, we have produced on-the-fly full-sky lightcones for multiple observer locations (described in Appendix~\ref{sec:lightcones}), including \healpix maps, and various 3D power spectra of each full simulation volume with a high time cadence.  

In future papers we will explore various measures of the clustering of matter, gas, and galaxies, elucidating the feedback and cosmology dependencies of these quantities, and we will confront the simulations with a wide variety of LSS observables.  Here we present an example of the kind of comparisons that are enabled by the \flamingo~data set.  Specifically, we discuss here the spatial cross-correlation between the thermal SZE signal and the CMB lensing convergence field, $\kappa$.  As this cross-correlation depends on both the clustering of matter and how hot baryons trace the matter field, it is sensitive to both cosmology and feedback variations, making it an interesting test case.  

We first describe the construction of the thermal SZE and CMB lensing maps.  As described in Section \ref{sec:healpix_map_quantities}, we accumulate the Compton $y$ values of individual particles crossing the lightcone onto \healpix maps over fixed intervals in redshift.  To construct a total Compton $y$ map, we only need to sum these maps along the line of sight, which we do back to $z=3$ (which we have verified is sufficient for the SZE-CMB lensing cross power spectrum).  To construct CMB lensing $\kappa$ maps, we follow the method described in \citet{McCarthy2018}, which employs the so-called Born approximation (i.e.\ light ray paths are approximated as straight lines).  In short, for each \healpix total mass map (of which there are 60 per lightcone back to $z=3$), we compute a projected (2D) overdensity map, $\delta(\chi,\bmath{\theta})$.  The maps are then integrated along the line of sight weighted by the CMB lensing kernel to yield the total convergence map 

\begin{eqnarray} 
\kappa(\bmath{\theta}) = \frac{3 \Omega_{\rm m} H_0^2}{2 c^2} \int_0^{\chi(z_{\rm max})} (1+z) \chi(z) \left(1 - \frac{\chi(z)}{\chi(z_{\rm CMB})} \right) \delta(\chi,\bmath{\theta}) {\rm d}\chi 
\label{eq:kappa} 
\end{eqnarray} 

\noindent where $\chi$ is the comoving distance, $z_{\rm max}$ is the maximum redshift of integration (taken to be $z=3$), and $z_{\rm CMB}=1100$ is the surface of last scattering. 

We use the \namaster\footnote{\url{https://namaster.readthedocs.io/en/latest/}} package \citep{Alonso2019} to compute the angular cross-spectrum between the dimensionless scalar (spin-0) quantities $y$ and $\kappa$.  To save computational effort, the \healpix maps here have been downsampled from $N_{\rm side}=16384$ to $N_{\rm side}=4096$, corresponding to an angular resolution of $\approx 0.86$ arcmin, which is sufficient for current measurements.  When computing the cross-spectra, we initially use a multipole moment resolution (bandpower) of $\Delta \ell = 8$ but then employ a Savitzky-Golay filter of order 3 and window size of 25 to further smooth the simulated spectra.  We deconvolve the $N_{\rm side}=4096$ pixel window function from the computed cross-spectra using the \textit{pixwin} function within the \healpix package.  

We compare our map-based cross-spectra with a Limber approximation-based calculation integrating the 3D matter-electron pressure cross-spectrum along the line of sight back to $z=3$,   
\begin{equation} 
    \label{equ:cls} 
    C_\ell^{\kappa-y} = A \int_0^{\chi(z_{\rm max})} \frac{{\rm d}\chi}{\chi} (1+z)^3 \left(1 - \frac{\chi(z)}{\chi(z_{\rm CMB})} \right) \ P_\text{m,e} \left(\frac{\ell+\frac{1}{2}}{\chi}, z(\chi)\right)\, , 
\end{equation} 
where $P_\text{m,e}$ is the 3D matter--electron pressure cross-spectrum and the normalisation factor, $A$, is defined as
\begin{equation}
    A \equiv \frac{3}{2}\left(\frac{H_0}{c}\right)^2\Omega_\mathrm{m}  \frac{\sigma_\mathrm{T}}{m_\mathrm{e}c^2} \, . 
\end{equation}

The derived cross-spectra are compared in Fig.~\ref{fig:tSZ_CMB_cross}.  In the top left panel, we examine the dependence on box size, resolution, and method for computing the cross-spectrum (the fiducial map-based method versus 1D Limber integration), as well as the role of cosmic variance.  At fixed box size (L1), there is excellent convergence between the three different resolutions.  The large 2.8~Gpc run (comparing the mean from the 8 independent lightcones) has slightly more power than the 1~Gpc runs on all scales, which is likely due to a larger number of very rare, massive clusters.  In the context of the 2.8~Gpc run, the Limber 1D calculation agrees well with the mean map-based calculation (compare the dashed curve labelled `P(k)' with the solid blue curve).  The shaded region encapsulates the cone-to-cone scatter from the large run, illustrating that cosmic variance becomes significant at the largest scales shown, $\ell \lesssim 500$.

In the top right panel of Fig.~\ref{fig:tSZ_CMB_cross} we explore the cosmology dependence with fixed baryon physics (the fiducial calibrated model).  Here we see that switching from the Planck cosmology to the lensing LS8 cosmology results in an amplitude reduction of $\approx30\%$.  The impact of massive neutrinos is also clearly discernible, although over this range of scales it would be difficult to discriminate between a variation in $\sum m_\nu$ and a variation in $S_8$, as both mainly alter the amplitude.

In the bottom panels of Fig.~\ref{fig:tSZ_CMB_cross} we explore the impact of varying the feedback on the $y$-$\kappa$ cross-spectrum.  Although the impact of feedback variations is evident, the cross-spectrum is clearly more sensitive to the variations in cosmology we have explored (top right panel) than to the variations in feedback (in spite of the large variations in the latter). The bottom-left panel shows that smaller cluster gas fractions lead to more power for $\ell < 600$, but less power on smaller scales. This is consistent with the findings of \citet{McCarthy2018}, who previously explored this observable using the \bahamas\ simulations, and found that stronger feedback tends to suppress (enhance) the cross-spectrum on small (large) angular scales. This suggests the cross-spectrum probes both the locations of gas ejection from haloes and gas accumulation at larger physical distances from galaxies (i.e.\ where the outflows stall).

Fig.~\ref{fig:tSZ_CMB_cross} also shows the measurements of \citet{Hill2014} who used the CMB lensing and thermal SZE maps from the Planck mission.  As contamination from the clustered infrared background (CIB) is a particular concern for this cross-correlation, the authors devised a novel technique for removing the CIB by making use of the cross-correlation between the high-frequency 857~GHz Planck data and the SZE map.  Nevertheless, \citet{Hurier2015} estimate that the derived cross-spectrum is still likely to be biased high due to residual CIB contamination at the level of $20\%\pm10\%$.  We have therefore rescaled the measurements down by $20\%$.  Furthermore, \citet{Hill2014} actually measure the $y$-$\phi$ cross-spectrum, where $\phi$ is the lensing potential, but this is straightforwardly converted into a $y$-$\kappa$ spectrum via $\phi_\ell = 2 \kappa_\ell/(\ell(\ell+1))$.

Comparing the simulated cross-spectra with the observations, we infer that current data is not sufficiently precise to distinguish between the feedback variations, but that there is a weak preference for a low-$S_8$ (or a high neutrino mass) cosmology, which is consistent with other recent findings in the literature based on LSS observables \citep[e.g.][]{Amom2023}.  However, we expect that forthcoming measurements of this cross-spectrum from the Advanced Atacama Cosmology Telescope \citep{DeBernardis2016} and the Simons Observatory \citep{Ade2019} will yield considerably more precise measurements of this observable, potentially allowing for competitive constraints on both cosmology and astrophysics.

\begin{figure}
	\includegraphics[width=\columnwidth]{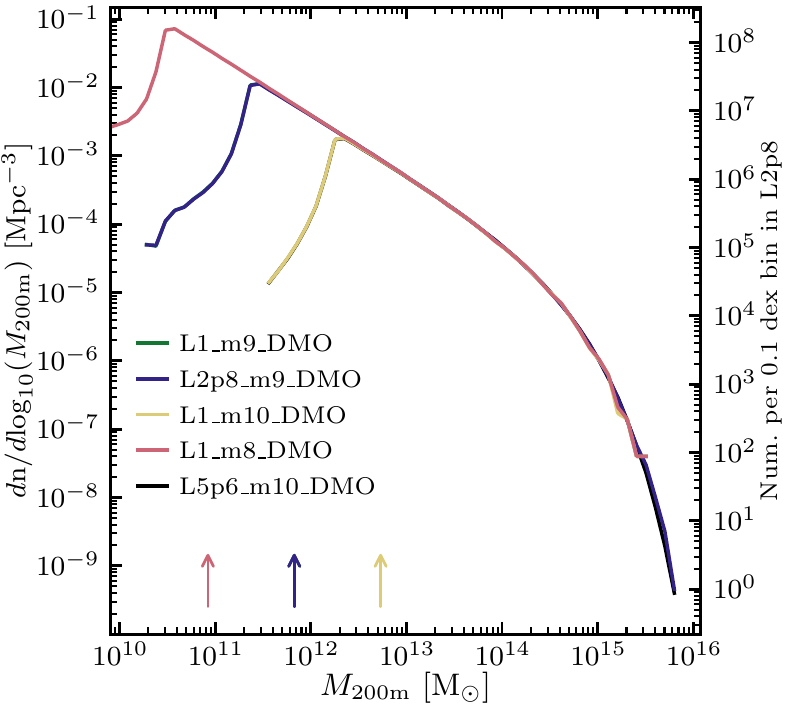}
    \caption{The $z=0$ halo mass function for the DMO simulations using our fiducial D3A cosmology. The left axis indicates the number density, while the right axis shows the number of haloes in the L2p8\_m9\_DMO simulation per halo mass bin of width 0.1~dex. Arrows indicate the halo mass corresponding to 100 dark matter particles for each resolution.}
    \label{fig:HMF_main_runs}
\end{figure}

\begin{figure}
	\includegraphics[width=\columnwidth]{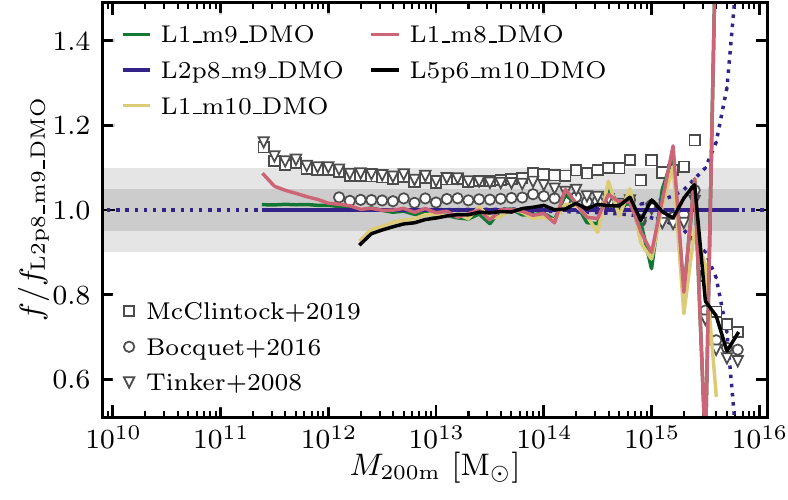}
    ~\\~
    \includegraphics[width=\columnwidth]{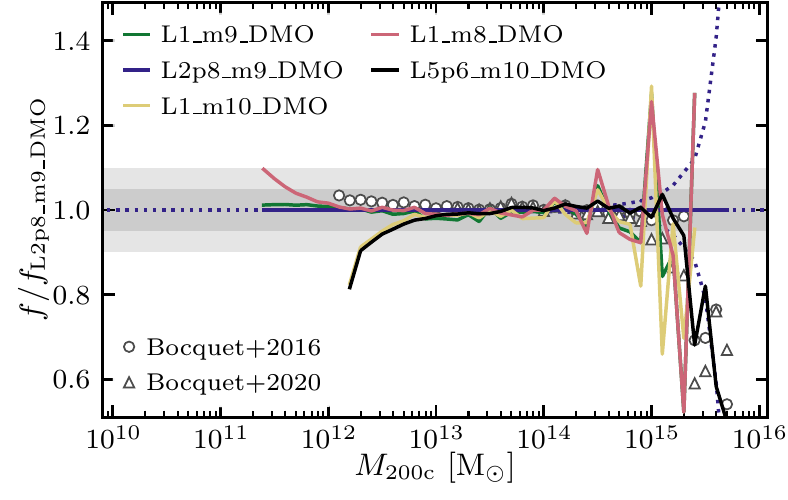}
    \caption{The ratio of the halo mass functions in the DMO simulations using our fiducial D3A cosmology and that for the  L2p8\_m9\_DMO simulation for two different halo mass definitions: $M_\text{200m}$ (top panel) and $M_\text{200c}$ (bottom panel). 
    Grey shaded regions show deviations of $\pm 5\%$ and $\pm 10\%$. Dotted lines show the $1\sigma$ Poisson errors on the counts in each bin for L2p8\_m9\_DMO. For comparison, we show the predictions for our fiducial cosmology from the universal HMF fits from \citet{Tinker2008} (top panel only) and \citet[][based on the DMO Magneticum simulations]{Bocquet2016}, from the Aemulus emulator \citep{McClintock2019} (top panel only) and from the MiraTitan Universe emulator \citep{Bocquet2020} (bottom panel only).
    }
    \label{fig:HMF_ratios_comparison}
\end{figure}

\section{Halo mass function}\label{sec:hmf}
The halo mass function (HMF) forms the basis of a variety of models of LSS, ranging from halo models of the power spectrum to halo occupation distribution models of galaxy clustering \citep[for a recent review see][]{Asgari2023}. Provided we can calibrate the observable-mass relation for clusters, the HMF can also itself be used as a constraint on cosmology. However, baryonic effects are expected to change halo masses and hence the HMF. Radiative cooling and condensation can draw matter in, both baryons and dark matter, while galactic winds driven by star formation and AGN can expel baryons and cause the dark matter distribution to expand. Contraction tends to dominate in the central regions whereas the outer regions of the halo tend to expand relative to a DMO model \citep[e.g.][]{Velliscig2014}. Hence, baryonic effects can both increase and decrease halo masses and this will directly impact the HMF, as has been demonstrated using both hydrodynamical simulations \citep[][]{Stanek2009,Cui2014,Cusworth2014,Velliscig2014,Bocquet2016} and observationally constrained analytic models \citep[e.g.][]{Debackere2021}.  

Before investigating baryonic effects, we first show the HMFs for the DMO simulations. Fig.~\ref{fig:HMF_main_runs} shows the $z=0$ HMF, $f(M) = \frac{\dd n}{\dd \log_{10}M}(M)$, for the DMO \flamingo\ models and the halo mass definition $M= M_\text{200m}$, where  $M_\text{200m}$ is the mass inside $R_\text{200m}$, the radius within which the mean density is 200 times the cosmic mean. The right axis shows the number of haloes per 0.1~dex mass bin in L2p8\_m9\_DMO. 

To facilitate a quantitative comparison of the small, but significant differences, the top panel of Fig.~\ref{fig:HMF_ratios_comparison} shows the ratio of each HMF to that of L2p8\_m9\_DMO. Convergence with the box size roughly follows expectations based on the Poisson errors shown for L2p8\_m9\_DMO (dotted curves). Convergence with numerical resolution is excellent, with systematic deviations smaller than 5 per cent down to haloes of 100 particles, below which the HMF begins to decrease relative to higher-resolution simulations. 

For comparison we also show predictions from the literature for our cosmology (grey open symbols) for $f(M_\text{200m})$ (top panel) and $f(M_\text{200c})$ (bottom panel), where $M_\text{200c}$ is the mass inside $R_\text{200c}$, the radius within which the mean density is 200 times the critical density. We compare with the universal HMF fits from \citet{Tinker2008} (top panel only) and \citet[][based on the DMO Magneticum simulations]{Bocquet2016}, from the Aemulus emulator \citep{McClintock2019} (top panel only) and from the MiraTitan Universe emulator \citep{Bocquet2020} (bottom panel only). Although there is agreement at the per cent level with Magneticum and MiraTitan for $10^{13} \lesssim M_\text{200c}/\Msun \lesssim 10^{14}$, the differences with (and between) predictions from the literature generally far exceed the Poisson errors (which should also be small for the literature studies). The discrepancies are likely dominated by differences between the halo definitions used by different halo finders \citep[e.g.][]{Bocquet2016,Castro2023}. Such large discrepancies are obviously problematic and suggest that a different approach may be needed for cluster counts cosmology \citep[e.g.][]{Debackere2022}.

At $z=0$ the L2p8\_m9 hydrodynamical simulation contains $4.1\times 10^6$, $3.4\times 10^5$ and $4.6\times 10^3$ haloes with mass $M_\text{200m} > 10^{13}$, $> 10^{14}$ and $> 10^{15}\,\Msun$, respectively. These are exceptionally large numbers for a hydrodynamical simulation. The large volume implies that large samples of clusters are available also at higher redshifts. Even at $z=1$ and 2 there are $4.7\times 10^4$ and $1.4\times 10^3$ objects with $M_\text{200m}>10^{14}\,\Msun$. The most massive object at $z=0$, 1, and 2 has a mass of $M_\text{200m} = 6.3\times 10^{15}$, $1.5\times 10^{15}$ and $4.8\times 10^{14}\,\Msun$, respectively. 

\begin{figure*}
    \centering
	\includegraphics[width=0.95\columnwidth]{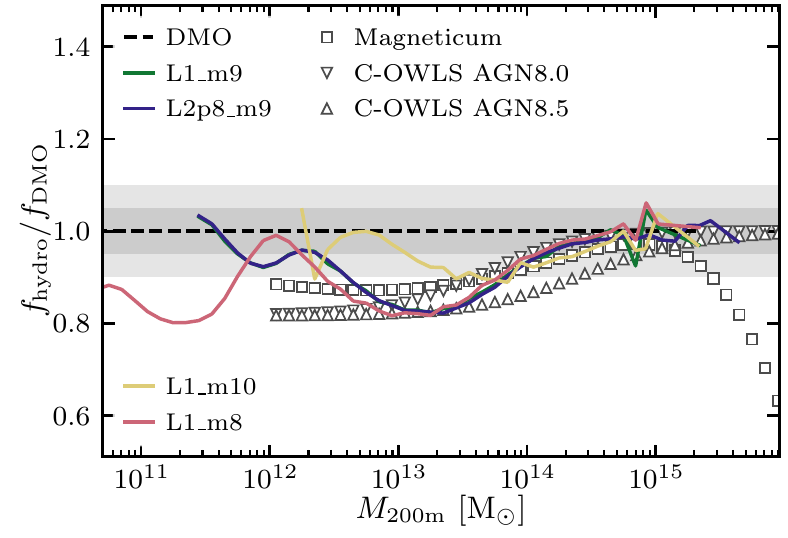}
	~\,~\,~
	\includegraphics[width=0.95\columnwidth]{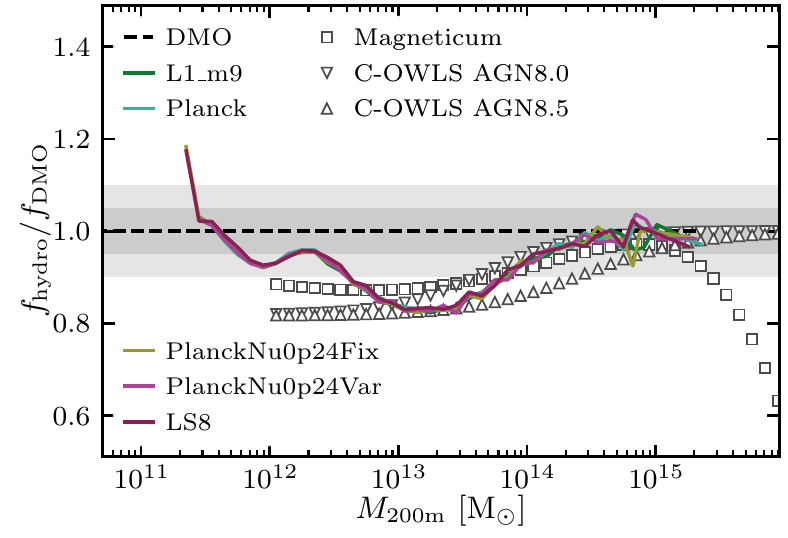}
     ~\\
    \vspace{0.2cm}
    \includegraphics[width=0.95\columnwidth]{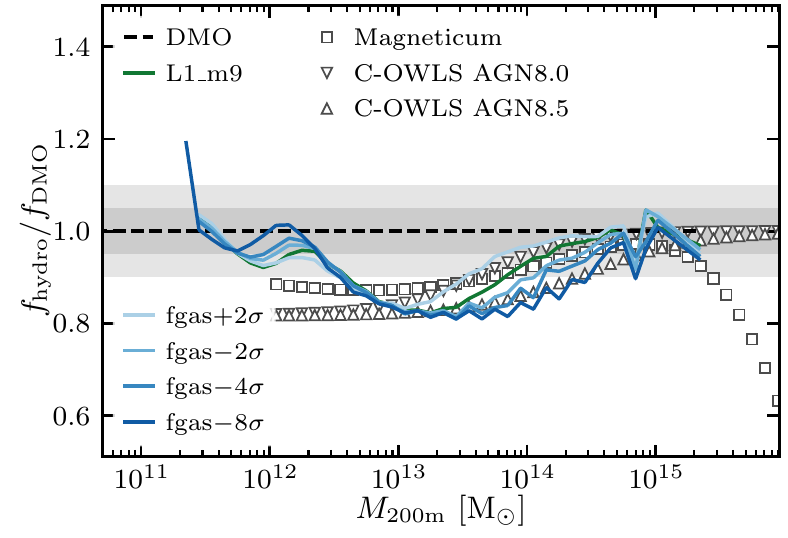}
    ~\,~\,~
    \includegraphics[width=0.95\columnwidth]{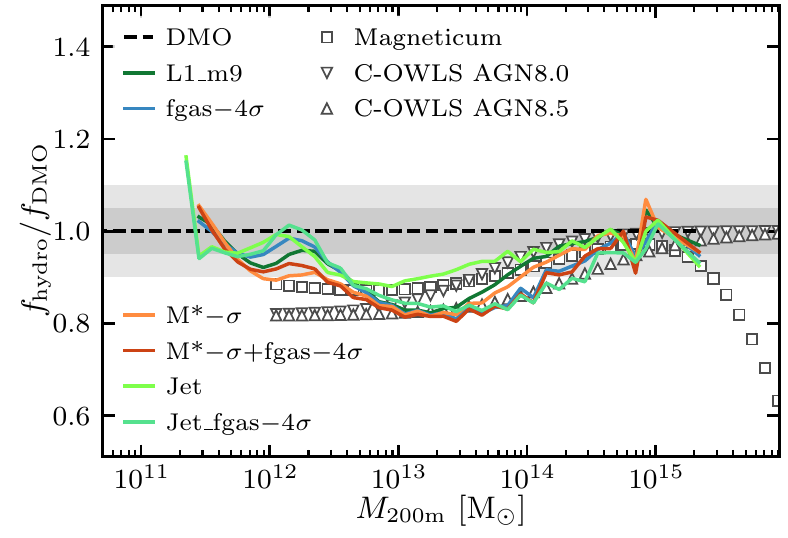}
    \caption{The effect of baryonic physics on the $z=0$ halo mass function. Each curve shows the ratio of the halo mass function in a hydrodynamical simulation and the corresponding DMO model. The different panels compare the models with different box sizes/resolutions (top left), cosmologies (top right), cluster gas fractions (bottom left), SMFs (bottom right), and AGN feedback implementations (also bottom right). For reference, model L1\_m9 is repeated in all panels. Grey shaded regions indicate $\pm 5\%$ and $\pm 10\%$ deviations from L2p8\_m9. For comparison, we also show the results from the Magneticum simulations (grey squares; \citealt{Bocquet2016}) and two cosmo-OWLS models that bracket the observed cluster gas fractions (grey upwards and downwards pointing triangles; \citealt{Velliscig2014}).}
    \label{fig:HMF_ratios}
\end{figure*}

Each curve in Fig.~\ref{fig:HMF_ratios} shows the ratio of the $z=0$ HMF in a hydrodynamical and the corresponding DMO simulation as a function of halo mass, $M_\text{200m}$. The top left panel compares the effects of baryons for the fiducial \flamingo\ simulations. The convergence with box size is excellent, with L1\_m9 and L2p8\_m9 agreeing nearly perfectly up to $M_\text{200m}=10^{15}\,\Msun$, beyond which the smaller volume runs out of haloes. Convergence with resolution is also excellent, at least in the mass range of clusters. L1\_m9 is nearly identical to the high-resolution L1\_m8 for $M_\text{200m} > 10^{13}\,\Msun$. For $11.7 < \log_{10}M_\text{200m}/\Msun < 13.0$ the differences are $< 5$ per cent, but at lower mass the two resolutions diverge. This is expected, because for intermediate resolution (m9) a significant fraction of the haloes with mass $10^{11.7}\,\Msun$ have not formed any stars at all (see Fig.~\ref{fig:SMHM}). 

Focusing for the moment on the high-resolution L1\_m8 simulation, which covers the largest range in halo mass, we see that baryon physics reduces the HMF by more than 5 per cent for both $M_\text{200m}< 7\times 10^{11}\,\Msun$ and $2\times 10^{12}\,\Msun < M_\text{200m} <  1\times 10^{14}\,\Msun$, where the low- and high-mass suppression can be attributed to the ejection of gas by stellar and AGN feedback, respectively. For $10^{12}\,\Msun$ haloes the HMF is nearly unchanged by baryonic effects. This is the halo mass for which stellar feedback becomes inefficient but AGN feedback has not yet had a large impact, resulting in a peak in the ratio $M_*/M_\text{200m}$ (which manifests itself as the inflection in the SMHM relation shown in Fig.~\ref{fig:SMHM}). At very high masses, $M_\text{200m} \gg 10^{14}\,\Msun$ the HMFs in the DMO and hydrodynamical simulations converge for all resolutions. This is consistent with the high baryon fractions of such massive haloes (see the bottom row of Fig.~\ref{fig:fstarbar}), and suggests that even AGN feedback is unable to remove, or keep out, a large fraction of the baryons. The largest deviation, $\approx 18\%$, occurs at $M_\text{200m}=2\times 10^{13}\,\Msun$.

The top right panel of Fig.~\ref{fig:HMF_ratios} shows that the effect of baryons on the HMF is insensitive to cosmology, though it should be noted that we have not varied the universal baryon fraction, which may be expected to have the largest impact, at least without recalibration of the subgrid physics. 

The bottom left panel compares the models that have been calibrated to different cluster gas fractions but to the same SMF. For $M_\text{200m} > 10^{13}\,\Msun$ smaller gas fractions yield a stronger suppression of the HMF, which is unsurprising. For the model with the smallest gas fraction the reduction of the HMF exceeds 5 per cent up to $10^{15}\,\Msun$. 

The bottom right panel shows the remaining models. Comparing the models calibrated to the fiducial SMF with the M*$-\sigma$ models, for which the observed stellar masses have been reduced by the expected systematic error (0.14~dex) before calibration, we see that the latter simulation predicts a 5--10 per cent stronger baryonic suppression of the HMF for $M_\text{200m}\sim 10^{12}\,\Msun$ but almost no difference for higher masses, where stars contribute a smaller fraction of the halo mass. Model Jet predicts a weaker baryonic effect than L1\_m9 for $10^{13} - 10^{14}\,\Msun$  haloes, which is consistent with its higher gas fractions for these masses (Fig.~\ref{fig:fgas}). Models Jet\_fgas$-4\sigma$ and fgas$-4\sigma$ are in close agreement, which is consistent with the good agreement between their gas fractions (Fig.~\ref{fig:fgas}). Hence, it seems that, at least at our relatively low resolution, the effect of AGN feedback on the HMF is insensitive to the subgrid implementation, kinetic jet-like versus isotropic and thermal, provided the models are calibrated to reproduce the same halo gas fractions. The observational uncertainty on this quantity, i.e.\ the difference between the fgas$-2\sigma$ and the fgas$+2\sigma$ models, then implies a $\sim 10$\% uncertainty on the HMF at $10^{14}\,\Msun$. 

Each panel of Fig.~\ref{fig:HMF_ratios} also shows the baryonic suppression of the HMF predicted by the Magneticum simulations \citep{Bocquet2016} and the two cosmo-OWLS simulations that bracket the observed cluster gas fractions \citep{Velliscig2014}. For $M_\text{200m}\gg 10^{13}\,\Msun$ the cosmo-OWLS models indeed bracket the fiducial \flamingo\ model, with cosmo-OWLS model AGN8.0 agreeing very closely for $M_\text{200m}> 10^{14}\,\Msun$, the mass range where this model fits the observed gas fractions well \citep{LeBrun2014}. In the mass range $10^{14}-10^{15}\,\Msun$ Magneticum also agrees very well with fiducial \flamingo, but at higher masses the fit from \citet{Bocquet2016} gives strange results, which is probably due to extrapolation beyond the mass range covered by the simulations. At the low-mass end, $M_\text{200m} \ll 10^{13}\,\Msun$, both cosmo-OWLS and Magneticum predict a stronger suppression of the HMF than any of the \flamingo\ variations. At least for cosmo-OWLS this can be explained by the fact that the simulations strongly underpredict the SMF for $M_*\lesssim 10^{11}\,\Msun$ (see fig.~1 of \citealt{McCarthy2017}), which suggests that feedback is too strong and hence that the gas fractions may also be too low. 
We do not show the MTNG results from \citet{HernandezAguayo2022} because they plot the baryonic suppression as a function of $M_\text{200c}$ rather than $M_\text{200m}$. However, a direct comparison (not shown here) reveals that MTNG predicts a suppression curve with a similar shape but shifted to about a factor of two higher masses. For $\log_{10} M_\text{200c}/\Msun  =14.0-14.5$ MTNG agrees well with model fgas$+2\sigma$, while it predicts even higher $f_\text{hydro}/f_\text{DMO}$ than fgas$+2\sigma$ for $\log_{10} M_\text{200c}/\Msun =14.5-15.0$. These differences are qualitatively consistent with the differences in the gas fractions measured at $R_\text{500c}$ shown in Fig.~\ref{fig:fgas}.

We close this section by noting that failing to account for baryonic effects of the magnitude found here would significantly bias cosmological parameters inferred from cluster counts \citep{Castro2021,Debackere2021}. 

\begin{figure}
    \centering
    \includegraphics{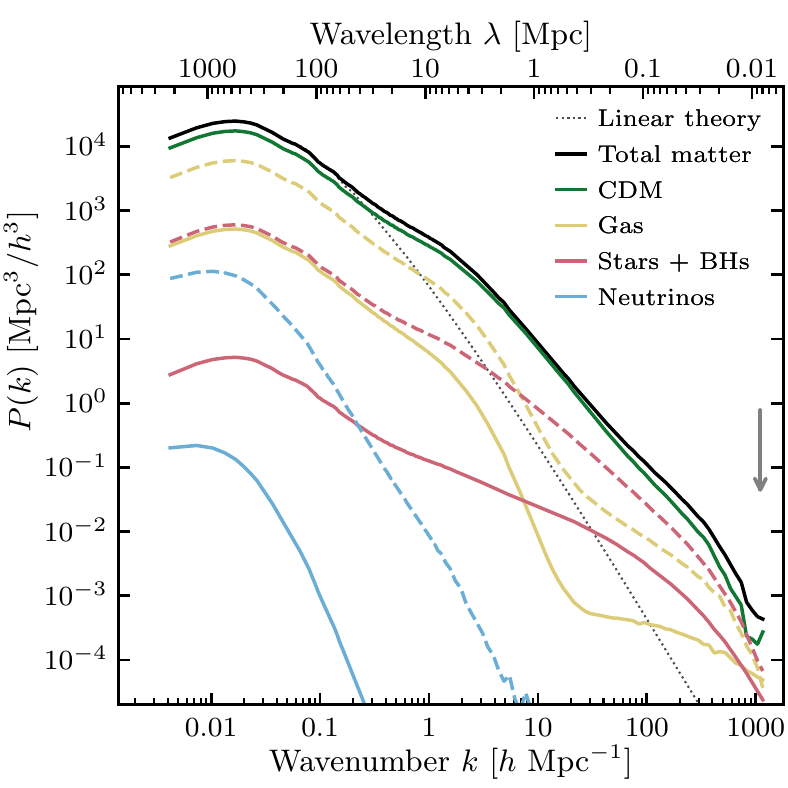}
    \caption{Matter power spectra as a function of wavelength $\lambda$ (top axis; in units of Mpc) and wavenumber $k=2\pi/\lambda$ (bottom axis; in units of $h\,\Mpc^{-1}$) for model L2p8\_m9 at $z=0$. The solid black curve shows the total matter power spectrum, while the dotted curve is the linear theory prediction, which strongly underestimates the power for $k> 0.1~h\,\Mpc^{-1}$. Coloured lines show the contributions from individual species (see the legend), where solid lines indicate auto-spectra and dashed lines show cross-spectra with CDM. The downward arrow 
    indicates the gravitational softening length. The importance of different species varies with scale. Gas and stars provide the dominant non-CDM contributions for $k< 10$ and $k>10~h\,\Mpc^{-1}$, respectively. Neutrinos are always much less important and only non-negligible on very large scales.}
    \label{fig:power_spectrum_by_species}
\end{figure}

\section{Matter power spectra}\label{sec:power_spectra}
Obtaining precision measurements of the total matter power spectrum is a key goal of observational surveys. Baryons modify the power spectrum in various ways. Before recombination, BAO imprint wiggles on very large scales, $\sim 10^2\,\Mpc$, that are well understood theoretically and can be modelled accurately using CMB Boltzmann codes. However, non-linear effects are not completely negligible for the observational signature of BAO in the low-$z$ galaxy distribution \citep[e.g.][]{Eisenstein2007} and simulations like \flamingo\ may be of some help in modelling them. On smaller scales baryonic effects on the power spectrum are much larger and much more uncertain. Outflows driven by stellar and AGN feedback suppress the power on intermediate scales, while gas cooling boosts the power on small scales \citep[e.g.][]{VanDaalen2011}. For current and upcoming surveys the former is much more important than the latter. Modelling the effect of galaxy formation on the power spectrum is one of the motivations for the \flamingo\ project. Because the baryonic suppression is thought to depend mainly on the gas fractions in clusters \citep[e.g.][]{Semboloni2011,Semboloni2013,Schneider2019,VanDaalen2020,Arico2021,Salcido2023}, \flamingo\ was calibrated to reproduce this observable. 

The solid, black curve in Fig.~\ref{fig:power_spectrum_by_species} shows the $z=0$ total matter power spectrum in the fiducial L2p8\_m9 simulation. The total matter power spectrum is given by $P(k)= \left < \left |\hat{\delta}(k)\right |^2\right >$, where $\hat{\delta}$ is the Fourier transform of the density contrast, $\delta = (\rho-\bar{\rho})/\bar{\rho}$, with $\rho$ the total density (i.e.\ CDM+gas+star+BHs+massive neutrinos) and $\bar{\rho}$ the cosmic mean density. Comparison with the dotted, black line, which shows the linear theory prediction, demonstrates that modelling non-linear effects is essential for wavenumbers $k> 0.1~h\,\Mpc^{-1}$ ($\lambda < 10^2~\,\Mpc$). 

The colored lines show the contributions of CDM (green), gas (yellow), stars plus BHs (red), and neutrinos (blue). Solid lines indicate the auto-spectra, i.e.\ $\left < \left |\hat{\delta_i}(k)\right |^2\right >$, where $\delta_i=(\rho_i-\bar{\rho})/\bar{\rho}$ and $i$ indicates the species. Dashed lines show the cross-spectra with CDM, $\left <\hat{\delta_i}\hat{\delta}_\text{CDM}^* + \hat{\delta}_\text{CDM}\hat{\delta_i}^* \right >$, which dominate over the auto-spectra. For $k\ll 0.1~h\,\Mpc^{-1}$ the baryons trace the CDM, but the relative contribution of neutrinos declines visibly with $k$ over all the scales probed by the simulation. For $k\sim 1$ to $10^2\,h\,\Mpc^{-1}$ the contribution of gas is suppressed. \citet{Debackere2020} used an observationally constrained halo model to show that this suppression is mainly due to ejection of gas from low-mass ($M_\text{500c}/\Msun = 10^{13}-10^{14}$) clusters for $k\lesssim 10~h\,\Mpc^{-1}$ and from groups ($M_\text{500c}/\Msun = 10^{12}-10^{13}$) for $k\gtrsim 10~h\,\Mpc^{-1}$. At very small scales ($k\sim 10^3\,h\,\Mpc^{-1}$) gas boosts the power spectrum because dissipation allows it to condense into galaxies. Because only cold and dense gas turns into stars, the stellar component (BHs contribute negligibly) boosts the small-scale power and for $k\gg 10~h\,\Mpc^{-1}$ this effect dominates over the suppression due to gas expulsion. Note that the CDM contribution shown in the figure is taken from the hydrodynamical simulations. Its shape differs from that in the corresponding DMO simulation (not shown) because of the gravitational back reaction of the baryons on the CDM \citep{VanDaalen2011,VanDaalen2020}. 

\begin{figure*}
    \centering
	\includegraphics[width=0.95\columnwidth]{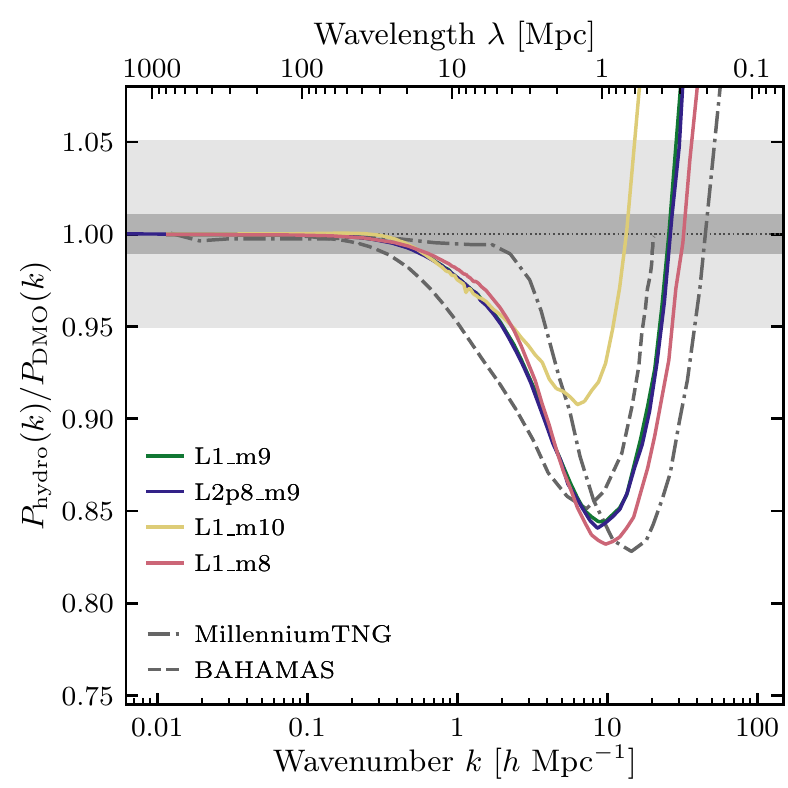}
	~\,~\,~
	\includegraphics[width=0.95\columnwidth]{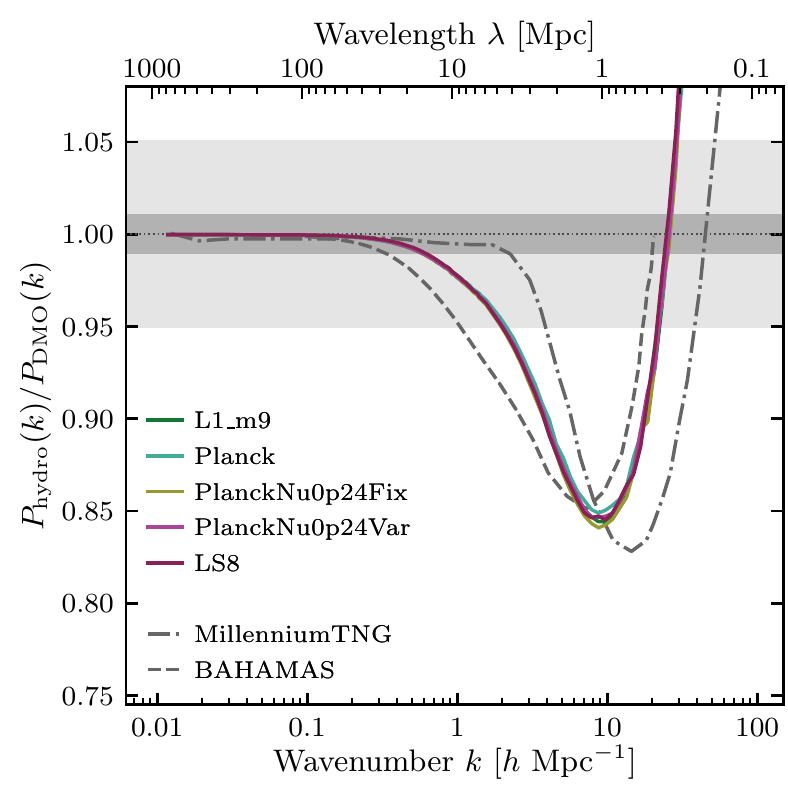}
    	\includegraphics[width=0.95\columnwidth]{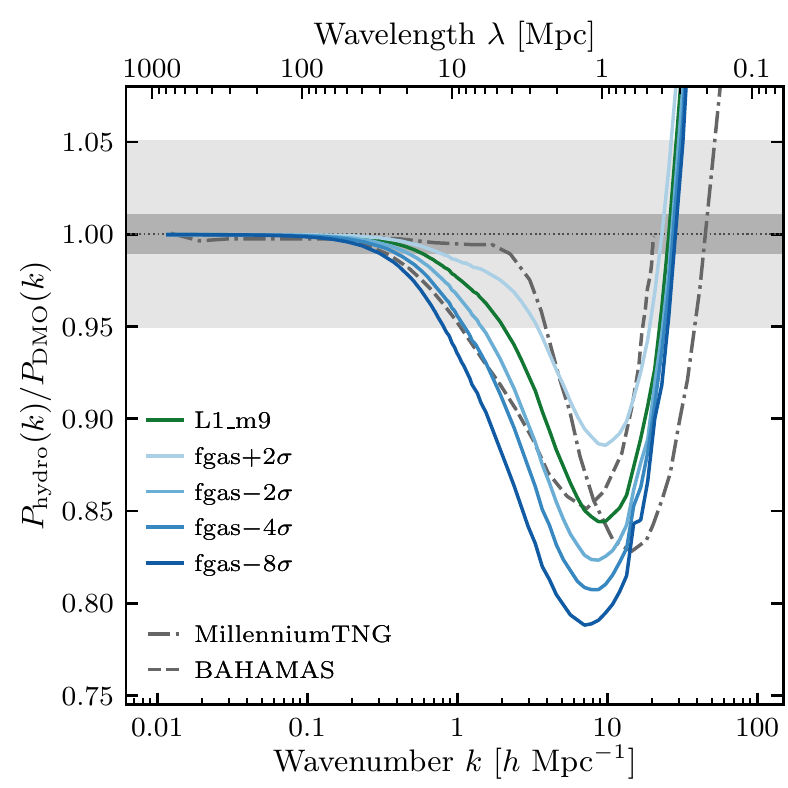}
    	~\,~\,~
     	\includegraphics[width=0.95\columnwidth]{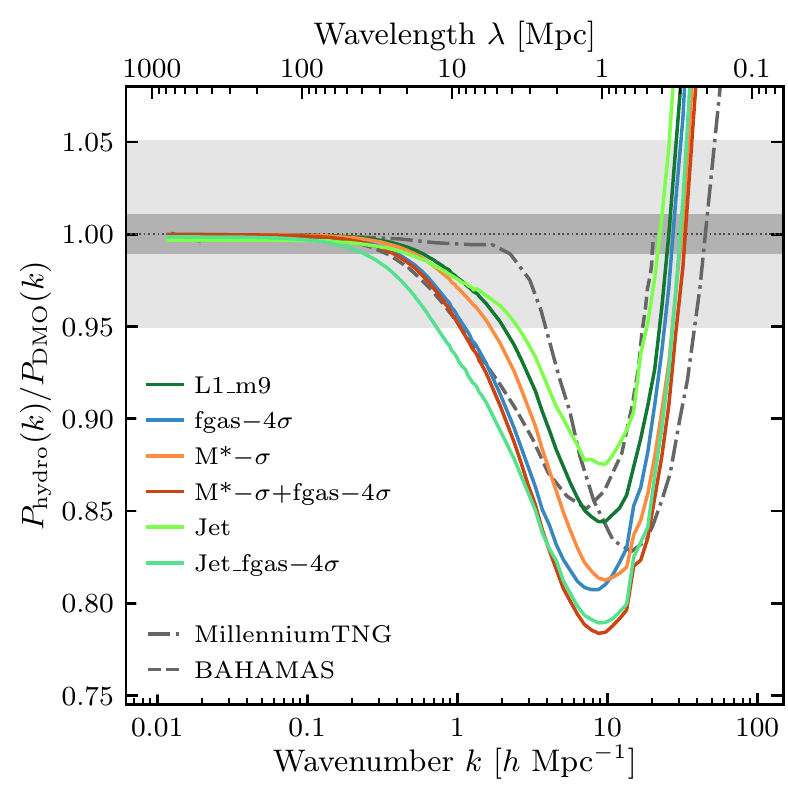}
    \caption{The baryonic suppression (or boost) factor for the total matter power spectrum, i.e.\ the ratio of the total matter power spectra in the hydrodynamical and corresponding DMO simulation, at $z=0$ as a function of wavenumber (bottom axis) or wavelength (top axis). The different panels compare the models with different box sizes/resolutions (top left), cosmologies (top right), cluster gas fractions (bottom left), SMFs (bottom right), and AGN feedback implementations (also bottom right). For reference, model L1\_m9, \bahamas\ and MTNG are repeated in all panels. Dark and light grey shaded regions indicate deviations from DMO of $\pm 1\%$ and $\pm 5\%$, respectively.}
    \label{fig:power_spectrum_ratios}
\end{figure*}

The total baryonic suppression/boost factor, i.e.\ the ratio of the total matter power spectrum in the hydrodynamical and the corresponding DMO simulations, is shown in Fig.~\ref{fig:power_spectrum_ratios}. Clockwise starting from the top left, the different panels compare the fiducial models, the cosmology variations using the fiducial galaxy formation model, the models using jet-like AGN feedback as well as models calibrated to different SMFs, and the models calibrated to different gas fractions. Each panel clearly shows that baryons trace the total matter on large scales (i.e.\ $P_\text{hydro}/P_\text{DMO}\approx 1$), suppress the power on intermediate scales ($P_\text{hydro}/P_\text{DMO}< 1$) and boost the power on small scales ($P_\text{hydro}/P_\text{DMO}> 1$).

The top-left panel demonstrates that the baryonic effects are converged with the box size, because the results for L1\_m9 and L2p8\_m9 fall on top of each other. Convergence with the resolution is excellent on large scales, but L1\_m10 and L1\_m9 deviate by more than 1 per cent from the high-resolution L1\_m8 simulation for $k> 3$ and $k>20~h\,\Mpc^{-1}$, respectively. 

For comparison, the grey curves show the MTNG and fiducial \bahamas\ simulations. MTNG predicts much weaker baryonic effects, which is consistent with its cluster gas fractions being too high (see Fig.~\ref{fig:fgas}). \bahamas\ predicts significantly stronger baryonic suppression than fiducial \flamingo, even though its cluster gas fractions are only slightly lower (Fig.~\ref{fig:fgas}). The difference in star fractions is a bit larger (Fig.~\ref{fig:fstarbar}), but that does not explain the fact that the difference in power suppression increases towards large scales, particularly since the \bahamas\ gas and star fractions agree better with \flamingo\ for more massive clusters, whose contribution to the power spectrum increases towards larger scales. The halo models of \citet{Debackere2020} suggest that the fact that for $k\sim 1~h\,\Mpc^{-1}$ \bahamas\ predicts a stronger suppression indicates that the gas is ejected to larger distances than in \flamingo. 

The top-right panel shows that the baryonic effect is insensitive to the cosmology, as had already been shown by \citet{VanDaalen2011,Mummery2017,VanDaalen2020}. However, we note that none of these papers nor \flamingo\ tests a wide range of cosmological parameters. In particular, none explore large variations in the cosmic baryon fraction, which might be expected to show the strongest interaction with the galaxy formation physics. 

The bottom-left panel compares the models calibrated to the same SMF but to different cluster gas fractions. Clearly, the smaller the gas fraction, and hence the baryon fraction, the stronger the baryonic suppression. Since the models were calibrated to the same SMF, they converge on small scales where stars dominate over gas. The fgas$-2\sigma$ and fgas$+2\sigma$ models could be taken as upper and lower limits on the magnitude of the baryonic suppression, given that they were calibrated to gas fractions that were shifted by $\pm 2$ times the estimated error on the gas fractions \citep[see][]{FlamingoCal}. However, comparison with the \bahamas\ prediction shows that this would be naive, because depending on $k$, \bahamas\ predicts a suppression as strong as in fgas$-4\sigma$ or even fgas$-8\sigma$, even though the \bahamas\ gas fractions are much closer to the fiducial \flamingo\ model than to these fgas variations (Fig.~\ref{fig:fgas}). This again suggests that the gas (and star) fractions inside $R_\text{500c}$ do not fully constrain the baryonic suppression. 

Indeed, the bottom-right panel of Fig.~\ref{fig:power_spectrum_ratios} shows that the Jet models predict a different shape for the scale dependence, with relatively stronger suppression on larger scales. Hence, for a fixed baryon fraction in clusters, there is a residual dependence on the implementation of AGN feedback, as anticipated by \citet{Debackere2020} who showed that the distance out to which the gas distribution is modified is also important. However, the differences between the fiducial and jet-like AGN feedback at fixed gas fraction, as well as those between fiducial \flamingo\ and \bahamas, are small in terms of percentage points. While the relative differences in the baryonic suppression factors are large for $k<1~h\,\Mpc^{-1}$, on such large scales $P_\text{hydro}$ is only a few per cent smaller than $P_\text{DMO}$. Finally, the comparison of the different M* variations at fixed gas fraction suggests that uncertainties in the SMF only become important for $k\gtrsim 10~h\,\Mpc^{-1}$.

\begin{figure}
    \centering
	\includegraphics[width=1.0\columnwidth]{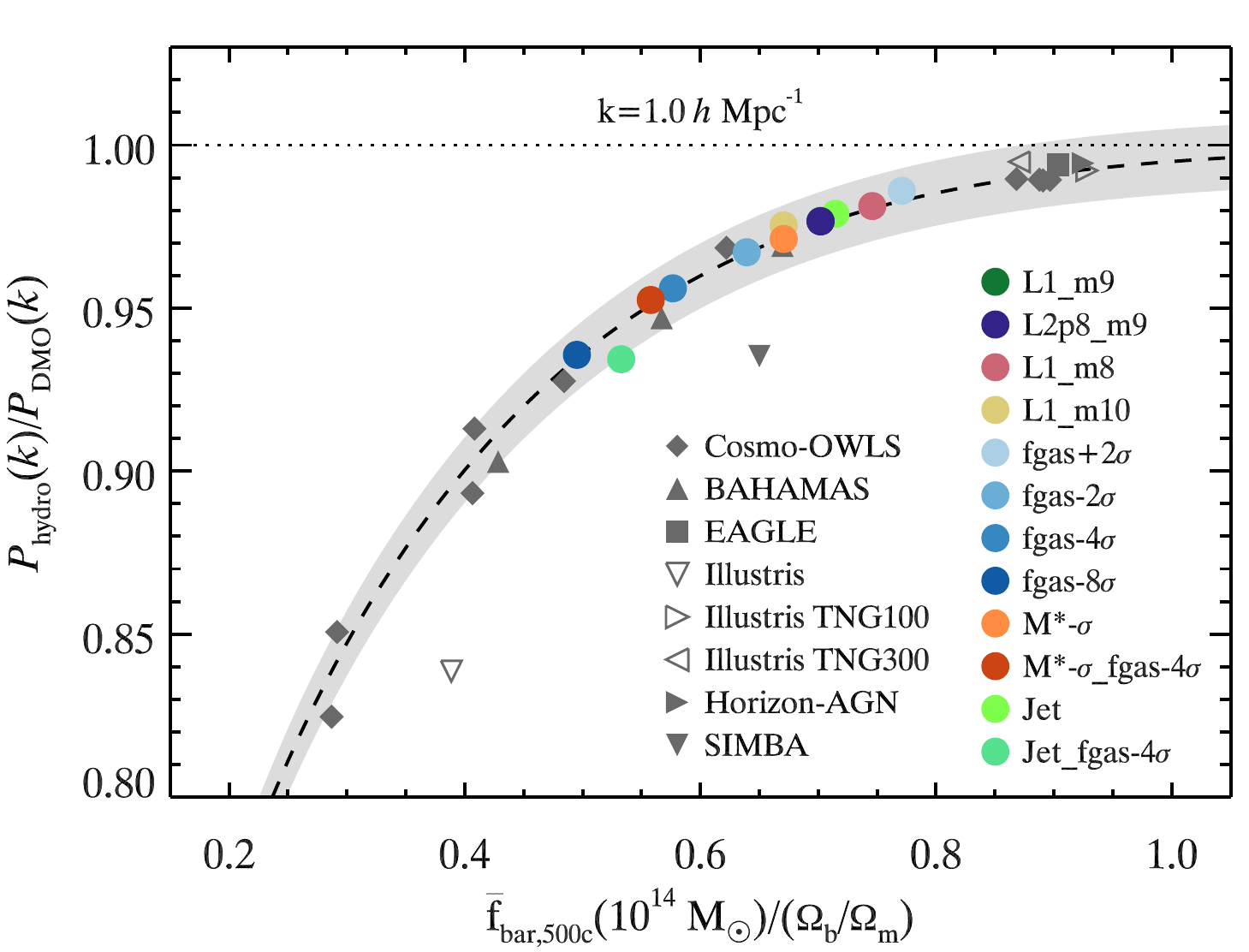}
 \caption{The ratio of the $z=0$ total matter power spectrum of hydrodynamical and DMO simulations, $P_\text{hydro}/P_\text{DMO}$, at wavenumber $k=1.0~h\,\Mpc^{-1}$ as a function of the mean baryon fraction within $R_\text{500c}$ in haloes of mass $M_\text{500c}=10^{14}\,\Msun$ for different \flamingo\ simulations and simulations from the literature: cosmo-OWLS \citep{LeBrun2014}, fiducial \bahamas\ \citep{McCarthy2017}, EAGLE \citep{Schaye2015}, Illustris \citep{Vogelsberger2014}, IllustrisTNG \citep{Springel2018}, Horizon-AGN \citep{Dubois2014}, and SIMBA \citep{Dave2019}. MillenniumTNG uses the same model as TNG300 and should therefore give nearly identical results in this plot. Note that the L2p8\_m9 point falls on top of the L1\_m9 point. The \flamingo\ cosmology variations are not shown as they would also be indistinguishable from L2p8\_m9 (see the top-right panel of Fig.~\ref{fig:power_spectrum_ratios}). The dashed line shows the relation between these quantities that \citet{VanDaalen2020} fit to the cosmo-OWLS and \bahamas\ simulations. The \flamingo\ simulations are consistent with the relation established for previous simulations to within $\left |\Delta P\right |/P \sim 10^{-2}$ (grey shaded region). }
    \label{fig:VD20_relation}
\end{figure}

Fig.~\ref{fig:VD20_relation} shows the effect of baryons on the $z=0$ matter power spectrum at a fixed scale of $k=1.0~h\,\Mpc^{-1}$ as a function of the mean baryon fraction within $R_\text{500c}$ in clusters of mass $M_\text{500c}=10^{14}\,\Msun$. This is similar to fig.~16 of \citet{VanDaalen2020}, who showed that the latter is a remarkably good predictor of the baryonic suppression for $k\le 1~h\,\Mpc^{-1}$. Each data point represents a different simulation. The dashed curve shows the fit of \citet{VanDaalen2020} to the cosmo-OWLS \citep{LeBrun2014} and \bahamas\ simulations with box sizes of at least $400~\Mpc/h$. The coloured points correspond to the \flamingo\ simulations shown in Fig.~\ref{fig:power_spectrum_ratios}, except for the cosmology variations which would have fallen on top of L2p8\_m9 and L1\_m9. The grey points are for simulations taken from the literature. 

Note that the widely used galaxy formation simulations EAGLE \citep{Schaye2015}, Illustris-TNG \citep{Pillepich2018TNGmethod}, and Horizon-AGN \citep{Dubois2014} predict very small suppression factors because their cluster baryon fractions are larger than observed. While we do not have the baryon fractions for MTNG, we expect it to give nearly the same small baryonic suppression as TNG300 because it uses the same galaxy formation model and \citet{Pakmor2022} show that TNG300 and MTNG predict the same baryonic suppression of the power spectrum. Some of the cosmo-OWLS models as well as Illustris \citep{Vogelsberger2014} predict much stronger baryonic suppression factors than \flamingo\ because they predict cluster baryon fractions that are too small.

Given the DMO power spectrum and the mean baryon fraction at $M_\text{500c}=10^{14}\,\Msun$, the \citet{VanDaalen2020} relation predicts the matter power spectrum of nearly all hydrodynamical simulations to $\sim 1$ per cent accuracy (grey shaded region). We find that the same holds for $k=0.5$ and $2.0~h\,\Mpc^{-1}$, but that deviations (and resolution effects) are larger for $k=10~ h\,\Mpc^{-1}$ (not shown). \citet{Salcido2023} use a suite of \bahamas-like simulations to show that the suppression of the power spectrum on these smaller scales correlates strongly with the baryon fraction in lower-mass haloes. Note that the relative error on the baryonic suppression, i.e.\ the deviation of the points from the model relative to the value of $P_\text{hydro}/P_\text{DMO}$ plotted along the $y$-axis, can be much larger than 1 per cent, particularly when the baryonic correction is small. However, it is the absolute error on $P_\text{hydro}/P_\text{DMO}$ that needs to be small to enable robust measurements of cosmological parameters. 

Even though none of the models used by \citet{VanDaalen2020} used jet-like AGN feedback, the Jet models do follow the relation, although Jet\_fgas$-4\sigma$ is the most deviant of the \flamingo\ models. Much more discrepant are Illustris and SIMBA \citep{Dave2019}, but we note that this may in part be due to their relatively small box sizes ($\sim 10^2\,\Mpc$), which will cause them to underestimate the contribution of massive haloes to the power spectrum, which are less affected by baryonic physics. If the difference cannot be explained by the simulation volume, then, based on the results of \citet{Debackere2020}, it is likely that AGN feedback affects baryons out to larger distances than in the simulations that do follow the \citet{VanDaalen2020} relation.

\section{Summary}\label{sec:conclusions}
Observational cosmology based on measurements of the growth of large-scale structure (LSS), such as cosmic shear, galaxy-galaxy and cosmic microwave background (CMB) lensing, galaxy clustering, and Sunyaev-Zel'dovich Effect (SZE)/X-ray observations of hot gas, is increasingly limited by the accuracy of theoretical predictions. The models that are compared with the data are nearly always based on dark matter only (DMO) simulations, though some allow for marginalization over expected baryonic effects associated with galaxy formation. However, baryonic effects become increasingly important as observations target smaller scales, and may be considerably more complex than is assumed in the corrections applied to DMO simulations. Hydrodynamical simulations can in principle help resolve this issue, but they tend to have volumes that are too small to study LSS, they often do not reproduce the relevant observables, and/or they do not include model variations that cover the relevant parameter space. The \flamingo\ project aims to address these shortcomings. 

\flamingo\ consists of a suite of new large-volume cosmological, hydrodynamical simulations. There are three different resolutions, corresponding to baryonic particle masses of $1.3\times 10^8\,\Msun$ (referred to as `high' or `m8' resolution), $1.1\times 10^9\,\Msun$ (intermediate/m9 resolution) and $8.6\times 10^9\,\Msun$ (low/m10 resolution), where the latter is used only for convergence testing. The flagship runs are the $(1~\Gpc)^3$ high-resolution L1\_m8 and the $(2.8~\Gpc)^3$ intermediate-resolution L2p8\_m9 simulations. The latter follows $2.8\times 10^{11}$ particles, which makes it the largest hydrodynamical simulation ever run to $z=0$. Importantly, the \flamingo\ suite contains 12 additional simulations at m9 resolution in the 1~Gpc box that vary the cosmology and galaxy formation physics (see Tables~\ref{tab:simulations} and \ref{tab:cosmologies} for an overview of all the hydrodynamical simulations). In addition, there is a DMO counterpart to each hydrodynamical run, plus some additional DMO simulations, including a 5.6~Gpc and a 11.2~Gpc box (see Table~\ref{tab:DMO_simulations}). Besides regular snapshot outputs, lightcone output is generated on-the-fly from the perspective of a number of different observers (8 for L2p8 and 2 for the L1 simulations).

The simulations are performed with the \swift\ code \citep{SWIFT_release} using the \sphenix\ SPH scheme \citep{Borrow2022sphenix}. The simulations include neutrino particles using the new $\delta f$ method of \citet{Elbers2021}. The initial conditions include separate fluids for CDM, baryons and neutrinos, and discreteness errors are suppressed by perturbing particle masses rather than displacing particles \citep{Hahn2020,Hahn2021,Elbers2022a}. The simulations include subgrid models for unresolved physical processes whose importance is widely accepted. Radiative cooling is calculated element-by-element while accounting for self-shielding \citep{Ploeckinger2020}. Star formation is implemented using a pressure law that reproduces the observed Kennicutt-Schmidt law \citep{Schaye2008}. Stellar mass loss is tracked element-by-element for winds from AGB and massive stars as well as core collapse and Type Ia supernovae \citep{Wiersma2009Chemo,Schaye2015}. Energy feedback from star formation is injected stochastically and isotropically in kinetic form \citep{DallaVecchia2008,Chaikin2022isotropic,Chaikin2023}. BHs are seeded, repositioned down the gravitational potential gradient, and can grow through mergers and gas accretion \citep{Springel2005BHs,Booth2009,Bahe2022}. Most models use thermal, isotropic AGN feedback \citep{Booth2009}, but two models instead use jet-like kinetic AGN feedback \citep{Husko2022}. 

The subgrid models for BH accretion and for stellar and AGN feedback are calibrated to the observed $z=0$ SMF, gas mass fractions inside $R_\text{500c}$ for clusters at $z\approx 0.1-0.3$ from a combination of X-ray and weak lensing data, and the $z=0$ relation between BH mass and stellar mass. Contrary to common practice, the calibration (to the SMF and cluster gas fractions) is not performed by trial and error, but using machine learning \citep{FlamingoCal}. In particular, for each resolution and each observable a Gaussian process emulator is trained on a 32-node Latin hypercube consisting of simulations with the target resolution but much smaller volumes (which limits the maximum cluster mass used for the calibration to $\log_{10} M_\text{500c}/\Msun = 13.7$, 14.4, and 14.5 for m8, m9 and m10, respectively). Four subgrid parameters are varied: the amount of supernova energy, the target velocity for kinetic stellar feedback, the AGN heating temperature or jet velocity, and the density dependence of the BH accretion rate (at m10 resolution we do not need stellar feedback and vary only the two BH parameters). 

Another novelty is that the calibration accounts for expected observational errors and biases. We impose random errors on the simulated stellar masses to account for Eddington bias. During the calibration of the fiducial intermediate-resolution model we fit for systematic errors in the SMF due to cosmic variance, bias in the inferred stellar mass, and for hydrostatic mass bias in the cluster gas fractions inferred from X-ray observations. The best-fitting bias factors, which are negligible for cosmic variance and stellar mass, and which is consistent with the literature for the hydrostatic mass bias, are then applied to the calibration data for all resolutions and models.

The emulators are not only used to design simulations that reproduce the observations, but also to create models in which the SMF and/or cluster gas fractions are shifted to higher/lower values. This allows us to specify model variations in terms of the number of $\sigma$ by which they deviate from the calibration data, which is more intuitive and useful than specifying simulations solely by the values of subgrid parameters that are not directly observable. \flamingo\ includes four models in which cluster gas fractions are varied (by $+2$, $-2$, $-4$ and $-8\sigma$, respectively) while keeping the SMF unchanged, one model in which the SMF is reduced by decreasing the stellar masses by the expected systematic error (0.14~dex; \citealt{Behroozi2019}) while keeping gas fractions fixed, and two models that simultaneously vary the gas fractions and the SMF. Finally, two models use jet-like AGN feedback rather than the fiducial isotropic and thermal feedback, one of which is calibrated to the fiducial data and one to gas fractions shifted down by $4\sigma$. Comparison of these last two models with the corresponding fiducial ones enables estimates of the uncertainty due to differences in the implementation of AGN feedback for a common calibration. 

The flagship runs and galaxy formation variations all assume the cosmological parameters from the Dark Energy Survey year three (3x2pt plus external constraints; \citealt{Abbott2022}) for a spatially flat universe and the minimum allowed summed neutrino mass of $\sum m_\nu c^2 = 0.06~\eV$ (one massive and two massless species). In addition, \flamingo\ includes three runs based on the \citet{Planck2020cosmopars} cosmology, one with $\sum m_\nu c^2 = 0.06~\eV$ and two with $\sum m_\nu c^2 = 0.24~\eV$. Finally, one model is motivated by the preference of many LSS surveys for a lower amplitude of the power spectrum than inferred from the CMB \citep{Amom2023}. 

The fiducial models reproduce the calibration data, i.e.\ the $z=0$ SMF (down to $\log_{10} M_*/\Msun = 8.7$, 9.9 and 11.2 for m8, m9, and m10, respectively; Fig.~\ref{fig:SMF}), the gas fractions of $z\approx 0.1-0.3$ clusters (Fig.~\ref{fig:fgas}) and the $z=0$ BH mass-stellar mass relation (Fig.~\ref{fig:bh}), within the mass ranges for which the results are converged with the resolution and box size. The same holds for all cosmology variations using the fiducial galaxy formation parameters, which implies that the changes in cosmology did not necessitate re-calibration. Similarly, the galaxy formation variations calibrated to perturbed data yield good fits to their own calibration targets. An exception is the SMF at $M_* \gtrsim 10^{12}\,\Msun$ where we find large differences between the different simulations and with the calibration data. However, in this regime the mass is sensitive to the aperture within which it is measured (we apply a 50~kpc 3D aperture to the simulations) and the treatment of the intracluster light and we therefore ignored masses exceeding $10^{11.5}\,\Msun$ when calibrating to the observed SMF. This systematic uncertainty also complicates comparison with the observed stellar mass fractions in clusters (Fig.~\ref{fig:fstarbar}). For cluster gas fractions the agreement with the data extends to higher masses than considered during the calibration, e.g.\ to more than an order of magnitude higher cluster masses for the high-resolution model.

Although the resolution of the \flamingo\ simulations is too low for detailed studies of galaxy structure and evolution, except perhaps for massive galaxies, we did compare the simulations to a number of observables that characterize galaxy properties. The simulations reproduce the observed cosmic star formation history (Fig.~\ref{fig:SFH}). In the stellar mass range for which we find convergence (which, depending on the property, can begin at higher masses than for the SMF) there is generally good agreement with the observations for sSFR, passive fraction and metallicity. The exception is the passive fraction at $M_*\gtrsim 10^{12}\,\Msun$, where the simulations predict an increasing fraction of active galaxies with increasing mass that is not observed. The Jet models look better in this respect, though they also do not quench star formation sufficiently in very massive galaxies. As expected given the relatively low resolution, galaxy sizes are generally overestimated, except for the high-resolution simulation at high mass ($M_* \gtrsim 10^{11}\,\Msun$) (see Fig.~\ref{fig:gal_props}). 

Thanks to its large volumes, the \flamingo\ simulations provide extremely large numbers of galaxy clusters. At $z=0$, simulation L2p8\_m9 contains $4.1\times 10^6$, $3.4\times 10^5$ and $4.6\times 10^3$ haloes with mass $M_\text{200m} > 10^{13}$, $10^{14}$ and $10^{15}\,\Msun$, respectively. Even at $z=1$ and 2 there are, respectively, $4.7\times 10^4$ and $1.4\times 10^3$ objects with $M_\text{200m}>10^{14}\,\Msun$.

We compared the simulation predictions to observations of a number of low-redshift cluster scaling relations, namely the relations between X-ray luminosity and temperature, X-ray luminosity and halo mass, X-ray temperature and halo mass, and SZE Compton Y parameter and halo mass (Fig.~\ref{fig:cluster_scalings}). The simulation predictions are converged and the fiducial model is in excellent agreement with the data. The X-ray relations are sensitive to the gas fractions (Fig.~\ref{fig:L-T}), but not to the investigated variations in the cosmology, SMF, and AGN feedback implementation. More detailed comparisons, including profiles, will be presented in a future study. 

As a first test of the predicted large-scale distributions of matter and hot gas, we investigated the cross-correlation of the thermal SZE and CMB lensing convergence signals. The simulation predictions are converged, but on large scales cosmic variance becomes significant, as evidenced by the scatter between the past lightcones of different observers. The results differ more for the cosmology than for the galaxy formation variations. Higher cluster gas fractions result in an increase of the cross spectrum for $\ell > 600$ but a decrease on larger scales. We found good agreement with the data, which are slightly better fit by the low $S_8$ and high neutrino mass cosmologies (Fig.~\ref{fig:tSZ_CMB_cross}).

We provided two examples of applications relevant for observational cosmology: the suppression due to baryonic effects of the halo mass function and the matter power spectrum, both at $z=0$. Except at masses $M_\text{200m}\sim 10^{12}\,\Msun$, where galaxy formation is most efficient, the HMF in the hydro simulations is suppressed relative to that in the corresponding DMO simulation. For low-mass clusters ($M_\text{200m}\sim 10^{13}\,\Msun$) the mass function is reduced by $\approx 20$ per cent. At higher masses the baryonic effects are weaker and for the models with the fiducial gas fractions the difference in the mass function is smaller than 5 per cent for $M_\text{200m}\sim 10^{15}\,\Msun$ (Fig.~\ref{fig:HMF_ratios}). In the mass range of clusters, the main factor determining the suppression is the gas fraction. The observational uncertainty on this quantity translates into a $\sim 10$ per cent uncertainty on the HMF at $10^{14}\,\Msun$. 

The matter power spectrum is suppressed for $1 \lesssim k\lesssim 10~h\,\Mpc^{-1}$, mainly because gas is distributed more smoothly than CDM on these scales, and boosted for $k \gtrsim 10^2\,h\,\Mpc^{-1}$, mainly due to stars (Fig.~\ref{fig:power_spectrum_by_species}). The reduction in power peaks at $k\sim 10~h\,\Mpc^{-1}$, where it exceeds 10 per cent in all models, and remains greater than 1 per cent down to at least $k=1~h\,\Mpc^{-1}$ even for our model with too high gas fractions (Fig.~\ref{fig:power_spectrum_ratios}). On large scales the baryonic suppression in the fiducial model is smaller than for the fiducial \bahamas\ simulation, but all these simulations follow the \citet{VanDaalen2020} relation between the reduction in power for $k=1~h\,\Mpc^{-1}$ and the mean baryon fraction in haloes of mass $M_\text{500c}=10^{14}\,\Msun$ (Fig.~\ref{fig:VD20_relation}).

Together with \citet{FlamingoCal}, where we describe the calibration of the model using Gaussian process emulation, this paper serves to document the methods used for the \flamingo\ suite of simulations. In addition, we have provided an overview of basic results for galaxies and clusters, and a preview of some of its applications to observational cosmology. Upcoming papers will investigate cluster selection effects, thermodynamic profiles of clusters, galaxy clustering, the effect of massive neutrinos on LSS, and the consistency between LSS and the primary CMB (the so-called $S_8$ tension). There are many more potential applications, including, for example, the validation and improvement of methods to correct DMO simulations for baryonic effects. Thanks to the availability of multiple resolutions and box sizes, and the enormous data volume, we anticipate that the simulations may also prove useful for machine learning applications. 

In the future we intend to use the \flamingo\ galaxy formation model and calibration strategy to run a new suite of simulations that will enable emulation of LSS observables as a function of both cosmological parameters and baryonic effects. This will enable further investigation of the interplay between baryonic effects and cosmology. Moreover, it will allow the application of observational constraints, such as the cluster gas fractions used in this work, during the inference of cosmological parameters from LSS data. This approach has the potential to reduce the uncertainty in the magnitude of baryonic effects and their impact on precision cosmology.

Some additional information as well as visualizations can be found on the \flamingo\ website.\footnote{\url{https://flamingo.strw.leidenuniv.nl/}}

\section*{Acknowledgements}
We are grateful to Zorry Belcheva, Jeger Broxterman, Yanling Chen, Shaun Cole, Jonah Conley, Lorenzo Filipello, Owen Hagedooren, Yi Kang, Elia Pizzati, Alex Smith, Evelyn van der Kamp, Liz van der Kamp, Tim van der Vuurst, Falco Verhoef, Zhen Xiang, Kun Xu, Jiaqi Yang, and Yun-Hao Zhang who helped analyze early \flamingo\ results. We are grateful to the late Richard Bower for convincing us of the utility of emulation for the calibration of subgrid prescriptions. We also gratefully acknowledge discussions with Shaun Cole, Scott Kay, Sylvia Ploeckinger, Alex Richings, and Sam Stafford, as well as help with \vr from Pascal Elahi, Claudia Lagos, and Rodrigo Tobar. We thank the members of the Virgo Consortium for their contributions to joint DiRAC computing time applications, in particular Baojiu Li for his coordination, and the COSMA support team, in particular Alastair Basden and Peter Draper, for accommodating our computing needs and for their help with running our simulations.
The research in this paper made use of the \textsc{SwiftSimIO} open source tools \citep{Borrow2020}. This work used the DiRAC@Durham facility managed by the Institute for Computational Cosmology on behalf of the STFC DiRAC HPC Facility (\url{www.dirac.ac.uk}). The equipment was funded by BEIS capital funding via STFC capital grants ST/K00042X/1, ST/P002293/1, ST/R002371/1 and ST/S002502/1, Durham University and STFC operations grant ST/R000832/1. DiRAC is part of the National e-Infrastructure. EC acknowledges support from the European Union’s Horizon 2020 research and innovation programme under the Marie Skłodowska-Curie grant agreement No 860744 (BiD4BESt), ARJ, CSF, and CGL from the STFC consolidated grants ST/T000244/1 and ST/X001075/1, FH from STFC studentship ST/P006744/1. This project has received funding from the Netherlands Organization for Scientific Research (NWO) through research programme Athena 184.034.002 and Veni grant 639.041.751, from the Swiss National Science Foundation (SNSF) under funding reference 200021\_213076, and from the European Research Council (ERC) under the European Union's Horizon 2020 research and innovation programme (grant agreements No 769130 and GA 786910).

\section*{Data Availability}

The data supporting the plots within this article are available on reasonable request to the corresponding author. A public version of the \swift\ code \citep{SWIFT_ascl} is available at \url{http://www.swiftsim.com}. The swift-emulator framework used for the calibration is publicly available, see \citet{Kugel2022}. The \flamingo\ simulation data will eventually be made publicly available, though we note that the data volume (several petabytes) may prohibit us from simply placing the raw data on a server. In the meantime, people interested in using the simulations are encouraged to contact the corresponding author.



\bibliographystyle{mnras}
\bibliography{main} 




\appendix

\section{Lightcone data}\label{sec:lightcones}
For this project we added functionality to the \swift\ simulation code to output particles as they cross the past (full-sky) lightcone(s) of one or more observers in the
simulation volume as well as spherical \healpix\ maps for user specified quantities and redshift intervals.

In the $L=1~\Gpc$ simulation boxes we place two observers at coordinates $(L/4, L/4, L/4)$ and $(-L/4, -L/4, -L/4)$ relative to the centre of the box, where $L$ is the simulation box size. In the larger boxes we place eight observers at coordinates $(\pm L/4, \pm L/4, \pm L/4)$. Table~\ref{tab:lightcone_redshift_limits} lists the number of observers placed in each simulation.

\begin{table*}
	\centering
	\caption{Number of observer positions and the maximum redshifts at which lightcone particle and \healpix\ map outputs are generated in each simulation. From left to right the particle types are dark matter (DM), neutrinos, black holes (BH), stars and gas. Filtered gas particles are those satisfying the density and temperature criteria described in appendix~\ref{sec:lightcones_observer_pos}. In simulations with two lightcones the individual particles are only output for the first lightcone.}
	\label{tab:lightcone_redshift_limits}
	\begin{tabular}{lcrrrrrrrr}
		\hline
                                                 &                   &\multicolumn{6}{l}{Maximum redshift for particle output} & \multicolumn{2}{l}{Maximum redshift for \healpix maps} \\
		Identifier                       & No. of lightcones & DM     & Neutrino & BH    & Stars     & Filtered gas & All gas & First lightcone & Other lightcones \\
		\hline
		L1\_m8                           & 2                 & 0.25   & 0.25    & 15     & 0.5       & 0.5     & 0.25         & 3.0       & 0.5     \\
		L1\_m9                           & 2                 & 0.25   & 0.25    & 15     & 0.5       & 0.5     & 0.25         & 3.0       & 0.5     \\
		L1\_m10                          & 2                 & 0.25   & 0.25    & 15     & 0.5       & 0.5     & 0.25         & 3.0       & 0.5     \\
		L2p8\_m9                         & 8                 & 0.78   & 0.78    & 15     & 0.78      & 5.0     & 0.78         & 5.0       & 5.0     \\
		fgas$+2\sigma$                   & 2                 & -      & -       & 15     & -         & 0.5     & 0.25         & 3.0       & 0.5     \\
		fgas$-2\sigma$                   & 2                 & -      & -       & 15     & -         & 0.5     & 0.25         & 3.0       & 0.5     \\
		fgas$-4\sigma$                   & 2                 & -      & -       & 15     & -         & 0.5     & 0.25         & 3.0       & 0.5     \\
		fgas$-8\sigma$                   & 2                 & -      & -       & 15     & -         & 0.5     & 0.25         & 3.0       & 0.5     \\
		M $-1\sigma$                     & 2                 & -      & -       & 15     & -         & 0.5     & 0.25         & 3.0       & 0.5     \\
		M $-1\sigma$\_fgas$-4\sigma$     & 2                 & -      & -       & 15     & -         & 0.5     & 0.25         & 3.0       & 0.5     \\
		Jet                              & 2                 & -      & -       & 15     & -         & 0.5     & 0.25         & 3.0       & 0.5     \\
		Jet\_fgas$-4\sigma$              & 2                 & -      & -       & 15     & -         & 0.5     & 0.25         & 3.0       & 0.5     \\
		Planck                           & 2                 & -      & -       & 15     & -         & -       & -            & 3.0       & 0.5     \\
		PlanckNu0p24Var                  & 2                 & -      & -       & 15     & -         & -       & -            & 3.0       & 0.5     \\
		PlanckNu0p24Fix                  & 2                 & -      & -       & 15     & -         & -       & -            & 3.0       & 0.5     \\
		LS8                              & 2                 & -      & -       & 15     & -         & -       & -            & 3.0       & 0.5     \\
		\hline
		L1\_m8\_DMO                      & 2                 & 0.25   & 0.25    & -      & -         & -       & -            & 3.0       & 0.5     \\
		L1\_m9\_DMO                      & 2                 & 0.25   & 0.25    & -      & -         & -       & -            & 3.0       & 0.5     \\
		L1\_m10\_DMO                     & 2                 & 0.25   & 0.25    & -      & -         & -       & -            & 3.0       & 0.5     \\
		L2p8\_m9\_DMO                    & 8                 & 0.78   & 0.78    & -      & -         & -       & -            & 5.0       & 5.0     \\
		L5p6\_m10\_DMO                   & 8                 & 0.78   & 0.78    & -      & -         & -       & -            & 25.0       & 25.0     \\
		L11p2\_m11\_DMO                   & 8                 & -    & -    & -      & -         & -       & -            & 30.0       & 30.0     \\
		Planck\_DMO                      & 2                 & 0.25   & 0.25    & -      & -         & -       & -            & 3.0       & 0.5     \\
		PlanckNu0p12Var\_DMO             & 2                 & 0.25   & 0.25    & -      & -         & -       & -            & 3.0       & 0.5     \\
		PlanckNu0p24Var\_DMO             & 2                 & 0.25   & 0.25    & -      & -         & -       & -            & 3.0       & 0.5     \\
		PlanckNu0p24Fix\_DMO             & 2                 & 0.25   & 0.25    & -      & -         & -       & -            & 3.0       & 0.5     \\
		LS8\_DMO                         & 2                 & 0.25   & 0.25    & -      & -         & -       & -            & 3.0       & 0.5     \\
		L1\_m9\_ip\_DMO                  & 2                 & 0.25   & 0.25    & -      & -         & -       & -            & 3.0       & 0.5     \\
		\hline
	\end{tabular}
\end{table*}

\subsection{Particle data} \label{sec:particle_lightcones}

\subsubsection{Implementation}

The position of each observer and the redshift range over which
lightcone particle output will be generated are specified in the
simulation parameter file. At each time step we compute the earliest and
latest times between which particles might be drifted during this
time step and the corresponding comoving distances. This defines a
shell around the observer in which particles might cross the lightcone as a result of drift operations carried out during this time step. An
additional boundary layer is added to the inside of the shell
to account for particles that move during the time step. The thickness
of this boundary is computed by assuming that all particles travel at
less than the speed of light.

Since the simulations employ periodic boundary conditions, we need to output any periodic copy of a particle which crosses the
observer's lightcone. We therefore generate a list of all periodic
copies of the simulation volume that overlap the shell around the
observer. Then, whenever a particle is drifted during the time step, we
iterate over the periodic copies in the list and check whether that
periodic copy of the particle crossed the observer's lightcone during
the drift operation\footnote{As an additional optimization, we take advantage of the way that \swift\ stores particles in a cubic grid of cells. Before the particles in a particular cell are drifted, we take the list of periodic replications of the volume computed at the start of the time step and find the
subset of those replications in which particles in the current cell may
cross the lightcone. Then, when drifting a particle, we only iterate
over this subset of replications rather than over the full list.}. If so, the particle's position is interpolated to the redshift at which it crossed the lightcone and the particle is added to a buffer. At the end of each step we check whether the size of the
buffer exceeds a specified threshold size and if it does, we write out the particles.

\subsubsection{Redshift limits}\label{sec:lightcones_observer_pos}

Table~\ref{tab:lightcone_redshift_limits} gives the maximum redshifts at which we output particles of each type for each simulation. There are two redshift thresholds for the output of gas particles crossing the lightcone. All gas particles are output at redshifts less than the limit shown in the `All gas' column in the table. Gas particles which have both temperature $T > 10^{5}\,\K$ and hydrogen number density $\nH > 10^{-6}(1+z)^4\,\cm^{-3}$ are output at redshifts below the limit shown in the `Filtered gas' column. These temperature and density thresholds were chosen based on experimentation with X-ray observables.

The redshift limits for dark matter, neutrino and unfiltered gas particles of $z=0.25$ and $z=0.78$ approximately correspond to the simulation box size in the 1~Gpc and 2.8~Gpc boxes, respectively. Gas particles that contribute significantly to X-ray (and SZE) observations are stored up to $z=0.5$ because the output is not prohibitively large and it enables the generation of multi-frequency X-ray emission lightcones in post-processing. BH particles are stored at redshifts below $z=15$ because the output size is modest and they are useful for the construction of halo lightcone catalogues in post-processing.

\subsection{\healpix\ maps}\label{sec:healpix_maps}

Lightcone particle outputs rapidly grow in size as the upper redshift limit is increased and can become impractical to store. We therefore also implement a scheme to store spherical maps of arbitrary quantities on the lightcone with user specified angular resolution and redshift bins.

The observer's past lightcone is split into a set of concentric spherical shells in comoving distance. For each shell we create one full sky \healpix\ \citep{Gorski2005} map for each quantity to be recorded. Whenever a particle is found to have crossed the lightcone according to the criteria above, we determine which shell it lies in at the time of crossing and accumulate the particle's contributions to the \healpix\ maps for that shell\footnote{In practice it is not necessary to store the maps for all of the shells simultaneously. Each map is allocated and set to zero when the simulation first reaches the time corresponding to the outer edge of the shell. The shell is written to disk and the memory is deallocated once all particles have been drifted to times later than the time corresponding to the inner edge of the shell. This means that the code will usually have \healpix\ maps for 1-2 shells in memory at any time, regardless of the total number of shells used.}.

The shell radii are specified in terms of redshift. From redshifts $z=0$ to 3 we use shells of thickness $\Delta z=0.05$. From $z=3$ to 5 we use shells of thickness $\Delta z=0.25$. We have a total of 68 shells to $z=5$. Between $z=5$ and 25 we use the shell boundaries $z=5.50$, 6.04, 6.63, 7.26, 7.95, 8.70, 9.51, 10.38, 12.26, 15.00, 20.00 and 25.00. In the 2.8, 5.6 and 11.2~Gpc boxes we produce \healpix maps to $z=5$, 25 and 30, respectively, for all 8 observer positions. In the 1~Gpc boxes we produce \healpix maps to $z=3$ for the first observer and to $z=0.5$ for the second observer.

We set the \healpix map resolution parameter, $N_\text{side} = 16384$, which gives a maximum pixel radius of 13.46 arcseconds and $12*16384^2=3,221,225,472$ pixels in each full sky map. We note that the number of pixels exceeds the size of a signed 32-bit integer ($2^{31}$), which means that some software packages for post-processing \healpix\ maps do not work for the full-resolution maps. For that reason we have also created maps that are down-sampled to a factor 4 fewer pixels.  

\subsubsection{Smoothing the \healpix\ maps}\label{sec:lightcones_healpix_smoothing}

Since the gas particles have associated smoothing lengths, quantities derived from the gas can be smoothed onto the \healpix\ maps. When a gas particle crosses the lightcone, we compute its angular smoothing length:
\begin{equation}
\label{eq:lightcone_projected_smoothing_length}
\theta_\text{h} = \arctan(h/r),
\end{equation}
where $h$ is the particle's SPH smoothing length and $r$ is the distance from the observer at which the particle crossed the lightcone. The particle will update all \healpix pixels with centres within an angular search radius $\theta_\text{s}=\gamma\theta_\text{h}$, where $\gamma$ is the radius at which the SPH smoothing kernel falls to zero in units of the smoothing length.

Given the resolution parameter, $N_\text{side}$, of a \healpix\ map, it is possible to compute the maximum angular radius of any pixel in the map. If the angular search radius, $\theta_\text{s}$, of a particle is smaller than this maximum radius then the particle's full contribution to the map is accumulated to the pixel which contains the particle's position on the sky and no smoothing is done. Otherwise we compute a weighting factor for each pixel within the search radius and distribute the contribution from the particle between pixels in proportion to their weights.

The weighting factors are determined by integrating the 3D SPH kernel over one dimension to produce a 2D projected smoothing kernel, using equation (30) of \cite{Price2007}:
\begin{equation}
\label{eq:lightcone_kernel_projection}
F(q_{xy}) = \int_{-\sqrt{R^2-q^{2}_{xy}}}^{\sqrt{R^2-q^{2}_{xy}}} W(q) \dd q_{z},
\end{equation}
where $q^2 = q^2_{xy} + q^2_{z}$, $R = h\gamma$ is the radius at which the kernel reaches zero, and, in our case, $W(q)$ is the Wendland C2 kernel.

\subsubsection{Parallelisation scheme}

In order to allow the generation of maps with high angular resolution and large numbers of pixels, the maps are distributed over all compute nodes involved in running the simulation. The maps are stored using the \healpix\ ring pixel ordering scheme and each MPI rank is assigned a contiguous range of pixel indices, corresponding to some range in latitude on the sky. When a particle is drifted and found to cross the lightcone, it is added to a buffer. At the end of the time step a copy of each buffered particle is sent to each MPI rank which contains pixels which may be updated by that particle and the affected parts of the map are updated.

\begin{table}
    \centering
    \caption{Overview of \healpix\ maps produced on-the-fly.}
   \begin{tabular}{lc}
 \hline
Quantity & SPH-smoothed?\\
\hline
Total mass &  N \\
Dark matter mass & N \\
Gas mass unsmoothed & N \\
Gas mass smoothed & Y \\
Stellar mass & N \\
Neutrino mass & N\\
Star formation rate & N\\
0.2--2.3~keV X-ray emission & Y\\
2.3--8.0~keV X-ray emission & Y\\
0.5--2.0~keV X-ray emission & Y\\
Thermal SZE & Y\\
Kinematic SZE & Y \\
Dispersion measure & Y \\
 \hline
    \end{tabular}
    \label{tab:healpix_maps}
\end{table}

\subsubsection{Quantities stored in \healpix\ maps} \label{sec:healpix_map_quantities}

Table~\ref{tab:healpix_maps} lists all the \healpix\ maps that are computed and saved on-the-fly. We compute \healpix\ maps of the total mass of particles of all types, i.e.\ dark matter, gas, star, BH, and neutrino particles, that cross the lightcone in each shell. In addition, we produce mass maps for each individual particle species except BHs. Most of these maps are for collisionless particles and hence not smoothed; the mass of the particle is simply accumulated to the pixel containing its position on the sky. For the gas mass we however compute both unsmoothed and SPH-smoothed maps, where the latter uses the method described in section~\ref{sec:lightcones_healpix_smoothing}.

We compute maps of the star formation rate in each shell. Each time a gas particle crosses the lightcone its associated star formation rate, given by equation~(\ref{eq:SFlaw}), is accumulated to the pixel containing the particle's position on the sky. This map is not smoothed.

We compute smoothed maps of X-ray energy and photon flux in three observer-frame bands: eROSITA 0.2--2.3~keV, eROSITA 2.3--8.0~keV, ROSAT 0.5--2.0~keV.
In order to avoid artefacts due to the specific subgrid implementation of AGN feedback, gas particles are only allowed to contribute to the X-ray maps if their temperature is not significantly and directly affected by recent AGN feedback. Specifically, if a particle crossing the lightcone has undergone direct AGN heating within the last 15~Myr and if its temperature is between $10^{-1}{\Delta}T_\text{AGN}$ and $10^{0.3}{\Delta}T_\text{AGN}$, then it is excluded from the calculation. The particle X-ray luminosities are computed using emissivity tables that depend on the gas density, temperature, the individual elemental abundances, and redshift. The tables are generated using \textsc{cloudy} \citep[version 17.02]{Ferland2017} and hence are consistent with the tables from \citet{Ploeckinger2020} used for radiative cooling during the simulation. Full details of the generation of the X-ray emission will be provided in Braspenning et al.\ (in preparation). 

We construct smoothed maps of the Compton $y$ parameter of the thermal SZE. When a gas particle crosses the lightcone we accumulate the dimensionless quantity
\begin{equation}
\Delta y = \frac{\sigma_\text{T} k_\text{B}}{m_\text{e}c^2}  \frac{m_\text{g} n_\text{e} T}{\Omega_\text{pixel}^2 d_\text{A}^2 \rho}
\end{equation}
to the map, where $m_\text{g}$ is the particle's mass, $\Omega_\text{pixel}$ is the solid angle of a \healpix pixel and $d_\text{A}$ is the angular diameter distance to the observer. As for the X-ray emission, gas particles whose temperatures are affected by recent, direct AGN feedback are excluded.

We also construct smoothed maps of the Doppler $b$ parameter of the kinematic SZE. When a gas particle crosses the lightcone the dimensionless quantity which ought to be accumulated to the map is:
\begin{equation}
\Delta b = \frac{n_\text{e} m_\text{g} \sigma_\text{T} v_\text{r}}{\Omega_\text{pixel}^2 d_\text{A}^2 \rho c} ,
\end{equation}
where $v_\text{r}$ is the particle's radial velocity relative to the observer. Due to a bug in our implementation, an extra factor of $a$ was introduced. This has been approximately corrected by dividing each map by the expansion factor at the shell mid point. Note that if necessary the maps can be corrected precisely using the particle lightcone outputs, at least for the observers and redshifts for which such data was stored (see Table~\ref{tab:lightcone_redshift_limits}). Again, particles that have recently received AGN feedback energy are excluded using the same criteria as described for X-ray emission.

Finally, we compute a smoothed map of the dispersion measure (DM). Each time a gas particle crosses the lightcone the following quantity ought to be accumulated to the \healpix map for the appropriate shell:
\begin{equation}
\Delta \text{DM} = \frac{n_\text{e} m_\text{g} a}{\Omega_\text{pixel}^2 d_\text{A}^2 \rho} ,
\end{equation}
where $a$ is the expansion factor at which the particle crossed the lightcone. However, due to a bug, the $a$ factor was missing in our implementation and so we have approximately corrected the dispersion measure maps by multiplying each map by the expansion factor at the mid point of the shell. Note that if necessary the maps can be corrected precisely using the particle lightcone outputs, at least for the observers and redshifts for which such data was stored (see Table~\ref{tab:lightcone_redshift_limits}). Gas particles recently heated by AGN are excluded from the dispersion measure maps in the same way as for the X-ray maps.
 
\begin{table*}
	\centering
	\caption{\panphasia\ descriptors for different box sizes. Simulation variations in the same volume use the same descriptor.}
	\label{tab:phase_descriptors}
	\begin{tabular}{rl}
		\hline
		$L$ (Gpc) & Descriptor \\
		\hline
		1.0 & \verb+[Panph6,L18,(56034,71400,250000),S1,KK1025,CH3774755196,Flamingo_Gpc1]+\\
		2.8 & \verb+[Panph6,L15,(27965,4226,16598),S3,KK1025,CH2241377117,Flamingo_2800]+\\
		5.6 & \verb+[Panph6,L14,(1069,8462,10972),S3,KK1025,CH1863120676,Flamingo_5600]+ \\
     11.2 & \verb+[Panph6,L13,(17113,34063,27542),S3,KK1025,CH1329212371,Flamingo_11200]+ \\
		\hline
	\end{tabular}
\end{table*}

\section{The choice of the linear phases for the initial conditions}\label{sec:panphasia}

The Gaussian phases for all the \flamingo\ volumes use a newer version of the Panphasia hierarchical Gaussian White Noise field than was described and published in \citet{Jenkins2013}. There are two main changes: (i) a larger set of polynomials, completed to sixth order, is used instead of the S8 scheme; (ii) a faster and more flexible pseudo-random generator, ThreeFry4x64 \citep{Salmon2011}, is used instead of a multiple linear congruence generator.

 The text descriptors in Table~\ref{tab:phase_descriptors} specify the phases for the different \flamingo\ volumes and, in principle, define the phases for all future possible zoom simulations of these volumes.  The public version of \monofonIC{} \citep{Hahn2020} and the version of \monofonIC{} we used to create the Flamingo initial conditions, are both able to use the new Panphasia field descriptors to set up the phases.

  The initial conditions for the \flamingo\ simulations represent a significant advance in accuracy for cosmological initial conditions with CDM, baryons and neutrinos. To our knowledge, there is currently no code that is able to make zoom initial conditions to the same degree of accuracy as our cosmological volumes when including all three of these components. DMO zoom initial conditions can be generated for the \flamingo\ volumes with the latest version of the IC\_Gen code \citep{Jenkins2013}.

\bsp	
\label{lastpage}
\end{document}